\documentclass[twocolumn,hyperpdf,amsmath,amssymb,aps,prd,10pt,superscriptaddress,nofootinbib,noeprint,preprintnumbers,floatfix]{revtex4-2}

\usepackage{orcidlink}
\usepackage{graphicx, color}
\usepackage{enumitem,kantlipsum}
\usepackage[letterspace=-10]{microtype} 

\usepackage{bm, amsmath, amsfonts, amssymb,xfrac}
\usepackage{multirow, tabularx, dcolumn, soul}
\usepackage{mathtools, leftidx, braket, slashed, cancel, bigdelim, booktabs}
\usepackage{blkarray}
\usepackage[figures]{rotating}

\usepackage{tikz}
\usetikzlibrary[plotmarks]
\usepackage{anyfontsize}

\usepackage[utf8]{inputenc} 
\usepackage{hyperref}
\newcommand\bef{\begin{figure}}
\newcommand\eef[1]{\label{fg:#1}\end{figure}}
\newcommand\beq{\begin{equation}}
\newcommand\eeq[1]{\label{#1}\end{equation}}
\newcommand\beqa{\begin{eqnarray}}

\newcommand\bet{\begin{table}}
\newcommand\eet[1]{\label{tb:#1}\end{table}}

\definecolor{link_blue}{RGB}{51,102,204}

\hypersetup{%
pdftitle = {title},
pdfsubject = {},
pdfkeywords = {},
colorlinks = {true},
filecolor = {black},
linkcolor = {link_blue},
menucolor = {black},
citecolor = {link_blue},
urlcolor = {link_blue},
}{}

\begin{document}

\title{Estimating energy levels from lattice QCD correlation functions using a transfer matrix formalism}
\author{Debsubhra Chakraborty~\orcidlink{0000-0001-5815-4182}}
\email{debsubhra.chakraborty@tifr.res.in}
\author{Dhruv Sood~\orcidlink{0009-0000-5555-9723}}
%
\author{Archana Radhakrishnan~\orcidlink{0000-0001-9357-1360}}
%
\author{Nilmani Mathur~\orcidlink{0000-0003-2422-7317}}
\email{nilmani@theory.tifr.res.in}
\affiliation{Department of Theoretical Physics, Tata Institute of Fundamental Research, \\ Homi Bhabha Road, Mumbai 400005, India }
%


\begin{abstract}
\noindent
We present an efficient method for extracting energy levels from lattice QCD correlation functions by computing the eigenvalues of the transfer matrix associated with the lattice QCD Hamiltonian. While mathematically and numerically equivalent to the recently introduced Lanczos procedure~\cite{Wagman:2024rid}, our approach introduces a novel prescription for removing spurious eigenvalues using a kernel density estimator (KDE) and Gaussian-convoluted histogram method. This strategy yields a robust and stable estimate of the energy spectrum, outperforming the Cullum-Willoughby filtering technique in efficiency. In addition, we detail how this method can be applied to extract overlap factors from two-point correlation functions, as well as matrix elements from three-point functions with a current insertion. Furthermore, we extend the methodology to accommodate correlation matrices constructed from a variational basis of operators, with its Block formulation.
We demonstrate the efficacy of this framework by computing the two lowest energy levels for a broad range of hadrons, including several nuclei. Although the signal-to-noise ratio is not significantly improved, the extracted energy levels are found to be more reliable than those obtained with conventional techniques. Within a given statistical ensemble, the proposed method effectively captures both statistical uncertainties and systematic errors, including those arising from the choice of fitting window, making it a robust and practical tool for lattice QCD analysis.

\end{abstract}

\maketitle

\section{Introduction}
Extraction of energy levels from two-point correlation functions is an essential component in most lattice QCD studies. The methods involved in that process need to be reliable and efficient not only for determining the energy spectra accurately but also for probing and understanding the internal structures of hadrons, particularly for any precision study. 
In general, by exploiting the asymptotic decay of two-point correlation functions,
\( C(\tau) \sim e^{-E_n \tau} \), one can extract the lowest-lying energy level
\( E_0 \) that couples to a given interpolating operator at sufficiently large
temporal separations from the source (\( \tau \rightarrow \infty \)).
 Over the years, better methodologies for extracting energy levels have been developed within the framework of the generalized eigenvalue problem (GEVP) \cite{MICHAEL198558, Luscher:1990ck}. These approaches use a large variational basis of interpolating fields to estimate the energy levels more reliably. All these methods have produced remarkable results, particularly for stable hadrons \cite{doi:10.1126/science.1163233, HPQCD:2003rsu}, achieving per-mille accuracy in many cases and even predicting subatomic particles before their experimental discovery \cite{PhysRevD.86.094510,Brown:2014ena,Mathur:2018epb}. 
More recently, these methods have also been extensively applied to study resonance properties and scattering phenomena \cite{RevModPhys.90.025001}. However, a significant challenge in these procedures remains the signal-to-noise-ratio (SNR) problem in Euclidean correlation functions, as pointed out by Parisi \cite{PARISI1984203} and Lepage \cite{Lepage89}. This issue, intrinsic to lattice QCD calculations, becomes more pronounced for systems involving multi-hadron states, and becomes a hindrance particularly for nuclei \cite{PhysRevD.87.114512,PhysRevD.90.034503,PhysRevD.93.094507,Chakraborty:2024oym}.
Overcoming this bottleneck remains a key focus in improving the precision and reliability of lattice QCD studies.

Recently, a novel procedure was introduced \cite{Wagman:2024rid,Hackett:2024xnx} for extracting energy levels by computing eigenvalues of the transfer matrix. Given the infinite-dimensional nature of this operator, a recursive Lanczos method was employed, making use of the Kaniel-Paige-Saad (KPS) bound for energy estimation \cite{2ce3c337-18f7-3f29-bc34-fcad79263ab3,be49459e-702a-3abe-83be-a616c86c5038}. The efficacy of the above method was shown using toy model calculations as well as for pion and nucleon. However, as noted there,  a major challenge with this method is filtering out spurious eigenvalues, which can introduce subjectivity and systematic errors into the estimation of energy levels. 

Following this developments, we propose a similar approach to extract energy levels from the eigenvalues of the transfer matrix in lattice QCD. Specifically, we introduce a Krylov subspace-based method for eigenvalue extraction from two-point correlation functions. We show that the oblique Lanczos method introduced in Ref.~\cite{Wagman:2024rid} can be reformulated as a generalized eigenvalue problem (GEVP), as a variant of the Prony GEVP method~\cite{Prony1795,Fleming:2004hs,Kunis2015AMG,Fischer2020,PhysRevD.79.114502,PhysRevD.81.054505}. We henceforth refer to this framework as the Transfer Matrix GEVP (TGEVP). Both analytically and numerically, we demonstrate the equivalence of this formulation with the oblique Lanczos approach. The same conclusion has also been reached in Refs.~\cite{Ostmeyer:2024qgu,abbott2025filteredrayleighritzneed}\footnote{Ref.~\cite{Ostmeyer:2024qgu} appeared simultaneously with this work, while Ref.~\cite{abbott2025filteredrayleighritzneed} came subsequently.}.

A major challenge in these TGEVP/Lanczos-type methods lie in identifying and removing spurious eigenvalues, which—if not address properly—can bias the extracted energy levels and increase systematic uncertainty. Ref.~\cite{Wagman:2024rid} advocated a bootstrap generalization of the Cullum--Willoughby (CW) test for filtering these spurious modes. However, as noted there, the CW test can be imperfect in the presence of noise. Additionally, we find its performance depends critically on tuning three hyperparameters that vary across datasets. This makes it difficult to apply uniformly without dataset-specific calibration. As we demonstrate in Subsec.~\ref{Lanczos_TGEVP}, even when tuned, the CW method may fail to produce stable estimates.

To address this, we propose a data-driven, non-parametric filtering method based on estimating the statistical distribution of eigenvalues using Gaussian-smoothed histograms and adaptive kernel density estimators (KDEs). Peaks in the resulting distribution and their full-width at half-maxima (FWHM) are utilized to identify and exclude outlier eigenvalues that deviate significantly from the dominant physical modes. While histogram- and KDE-based peak estimation and distribution estimations are well-established in other fields (see e.g.,~\cite{article,refId0,McCarthy2014PeaKDEck,Sadiq:2024xsz,Poluektov_2015}), and also utilized in lattice QCD in different contexts \cite{particles3010007,Bruno:2023bue}, here we employ such techniques in spectral filtering of spurious eigenvalues of TGEVP matrix. Our histogram/KDE-based filtration method for extracting physical eigenvalues from noisy TGEVP spectra provides a data-driven, hyperparameter-light alternative to traditional filtering techniques (such as the CW test), and achieves comparable or superior stability across a variety of datasets without the need for manual tuning, as demonstrated in Sec.~\ref{sec:results} with the numerical results.

The remainder of the draft is structured as follows. In Sec.~\ref{sec:methodology}, we present the details of the proposed TGEVP method. Sec.~\ref{sec:statistics_background} outlines how histogram and kernel density estimation (KDE) can be employed to estimate the distribution of eigenvalues, following established techniques from the statistics literature~\cite{Hardle2004}. In Sec.~\ref{eigs_filtration}, we provided a flowchart to describe the filtration workflow used to identify and remove spurious eigenvalues. Results obtained using TGEVP with the proposed filtering procedure are presented in Sec.~\ref{sec:results}. In Subsec.~\ref{results:comp1}, we compare the outcomes with those from the standard effective mass approach and Prony’s method. A detailed comparison between the Lanczos and TGEVP approaches is provided in Subsec.~\ref{Lanczos_TGEVP}. Subsecs.~\ref{results:charm}--\ref{results:nuclei} present results for a wide range of hadrons, including light and heavy baryons and nuclei, for a variety of lattice ensembles, illustrating the broad applicability and effectiveness of our method for extracting energy levels. In Subsec.~\ref{corltn}, we examine correlations across TGEVP iterations for testing the reliability of the method. In Subsec.~\ref{subsec:overlap}, we discuss the extraction of overlap factors, along with supporting numerical results, for studying decay properties of hadrons. Subsec.~\ref{subsec:TGEVP_GEVP} demonstrates how the TGEVP framework can be incorporated into multi-operator GEVP methods to access excited states and finite-volume amplitude analysis. Finally, Sec.~\ref{sec:summary} provides a summary of our findings along with concluding remarks and future directions. The mathematical equivalence between the TGEVP and oblique Lanczos approaches is shown in Appendix~\ref{app:equiv}. The asymptotic convergence properties of histogram and KDE-based distribution estimation are discussed in Appendix~\ref{app:convergence}. Additional results and supporting figures are presented in Appendix~\ref{app:additional_results}. In Appendix \ref{app:complx_eigs} we discuss the impact of complex eigenvalues in estimating the physical energy levels.

\section{\label{sec:methodology}Methodology} 
In lattice QCD calculations, the two-point correlation function for an interpolating field \( \mathcal{O} \), between a source at \( (\mathbf{x}_i, t_i) \) and a sink at \( (\mathbf{x}_f, t_f) \), is given by
\begin{equation}
C(t = t_f - t_i) = \sum_{\mathbf{x}} \langle \Omega \vert \mathcal{O}(\mathbf{x}_f, t_f)\, \overline{\mathcal{O}}(\mathbf{x}_i, t_i) \vert \Omega \rangle,
\end{equation}
where \( \vert \Omega \rangle \) denotes the vacuum state of the theory. In Euclidean space-time, this can also be expressed in terms of the Hamiltonian \( \mathcal{H} \) as
\begin{equation}
C(t) = \langle \Omega \vert \mathcal{O}(0)\, e^{- \hat{\mathcal{H}} t}\, \overline{\mathcal{O}}(0) \vert \Omega \rangle.
\end{equation}
Defining the Euclidean transfer matrix as \( \hat{\mathcal{T}} = e^{-a \hat{\mathcal{H}}} \), and the state \( \vert \chi \rangle \equiv \overline{\mathcal{O}}(0) \vert \Omega \rangle \), we obtain
\begin{equation}
C(t) = \langle \chi \vert \hat{\mathcal{T}}^{t/a} \vert \chi \rangle.
\end{equation}
In lattice simulations, time is discretized as \( t = na \), where \( n \in \{1, 2, \ldots, n_T\} \), and the above simplifies to
\begin{equation}
C(n) = \langle \chi \vert \hat{\mathcal{T}}^{n} \vert \chi \rangle.
\end{equation}

The eigenvalues of the transfer matrix correspond to the eigenstates of the Hamiltonian $\mathcal{H}$. In order to study the transfer matrix explicitly, one must work with a truncated version defined on a finite-dimensional basis. This is in contrast to path integral Monte Carlo methods, which sample configurations and can implicitly encode contributions from the full, effectively infinite-dimensional Hilbert space. In our implementation, we choose the Krylov subspace, $\mathcal{K}_{m}$, as a set of linearly independent basis vectors for this purpose of truncating $\hat{\mathcal{T}}$. These basis vectors are outlined below. 
\begin{eqnarray}
    \mathcal{K}_{m} &\equiv& \{\vert \chi \rangle, \hat{\mathcal{T}}\vert \chi \rangle,\hat{\mathcal{T}}^2\vert \chi \rangle,\cdots, \hat{\mathcal{T}}^m\vert \chi \rangle \}, \label{KrylovSpace}\\
    &\hspace{0.05cm}&\text{for\hspace{0.05cm} $2m-1  \leq n_T$}. \nonumber
\end{eqnarray}
Since this set of vectors is neither orthogonal nor normalized,  we normalize those as, $\vert v_i\rangle = \hat{\mathcal{T}}^i\vert \chi \rangle/\sqrt{\langle \chi \vert \hat{\mathcal{T}
}^{2i}\vert \chi \rangle}$. Next, in terms of these basis vectors $\vert v_i\rangle$s we define following two-matrices,
\begin{eqnarray}
    T^m_{ij} &=& \langle v_i \vert \hat{\mathcal{T}} \vert v_j \rangle  = \frac{C(i+j+1)}{\sqrt{C(2i)C(2j)}}\hspace{0.15in}\text{for $0\leq i,j \leq m$} \label{eq:eq7},\\
    V^m_{ij} &=& \langle v_i \vert v_j \rangle  = \frac{C(i+j)}{\sqrt{C(2i)C(2j)}}\hspace{0.30in}\text{for $0\leq i,j \leq m$} \label{eq:eq8}.
\end{eqnarray}
It can be shown that the eigenvalues of  $\hat{\mathcal{T}}$ matrix, truncated in the Krylov-subspace $\mathcal{K}_m$, can be obtained by solving the following generalized eigenvalue problem (GEVP),
\begin{eqnarray}
T^mx^m_{n}=\lambda^m_nV^mx^{m}_n, \hspace{0.2in}\text{for $0\leq n \leq m$}\;,\label{GEVPeq1}
\end{eqnarray}
where $\lambda_n^m$'s and $x^m_n$'s are eigenvalues and eigenvectors of the transfer matrix truncated on the $(m+1)$ dimensional Krylov-subspace. We provide the proof in the Appendix~\ref{app:equiv}. 
Note that, $V^m$ and $T^m$ both are positive-definite matrices for a given source-sink setup. 
Hence the eigenvalues, $\lambda_n^m$, in Eq.~(\ref{GEVPeq1}) are real and positive. The energies of the eigenstates of the truncated transfer matrix are given by, $E^m_n = -\log\lambda_n^m$. Since the  eigenenergies are positive, we expect $0 \le \lambda_n^m \le 1 $
\footnote{As, described in \cite{Wagman:2024rid}, thermal states appear in the spectrum of $\mathcal{T}^m$ with $\lambda_n^m>1$. However we are not interested in those states here, and so are removed.}. Note in this formalism, changing the problem as a standard GEVP enables the use of well-established linear algebra routines, avoiding the need for a recursive oblique Lanczos implementation. Moreover, as we will show later that overlap factors and matrix elements can directly be extracted from the GEVP eigenvectors, unlike in the oblique Lanczos method, which requires an additional basis transformation. This makes TGEVP conceptually simpler and easier to implement.

Note that for the case of staggered quarks, the two-point correlation functions will have contribution from opposite parity states with an oscillating phase in time $(-1)^t$. Consequently, along with the positive eigenvalues, there are negative real eigenvalues, $\lambda_n^o$, corresponding to the oscillating term of the correlation functions. They will corresponds to energy, $E_n^o=-\log \lambda_n^o+i\pi=-\log(\vert\lambda_n^o \vert)$, 
with a  phase $(-1)^t$. We will elaborate this while presenting results with staggered quarks. Similarly, for a thermal correlator the eigenvalues will be complex and a separate method is needed to gain information on thermal modes. However, that is beyond the scope of this work.

In the absence of statistical noise in a correlation function, $C(n)$, selecting eigenvalues across different iterations of the matrix $T^m$ is straightforward. The eigenvalues computed in the $m$-th iteration will be arranged in increasing order,
$E^m_0 < E^m_1 < E^m_2 < \cdots < E^m_m\;$,
where $E^m_0$ represents the ground-state energy estimate for the $m$-th iteration, while $E^m_i$ for $1 \leq i \leq m$ corresponds to the subsequent excited states.  In lattice QCD simulations, however, the correlation functions are evaluated stochastically using the Monte Carlo path integral approach with finite statistics. This introduces statistical noise, causing fluctuations in the computed matrices $C^m$ and $T^m$ which can then deviate from their positive-definite structure, resulting 
spurious eigenvalues in the spectrum of the $T^m$. This issue is particularly problematic in systems with a dense spectrum, as spurious eigenvalues can disrupt the proper ordering of energy levels, making the extraction of reliable results challenging. The problem was also highlighted in Ref.~\cite{Wagman:2024rid}, where a filtering procedure based on the Cullum–Willoughby (CW) method~\cite{CULLUM1981329} was employed to remove such spurious modes. However, as noted in that work, the CW test can be unreliable in the presence of statistical noise. Moreover, we find that its effectiveness depends sensitively on the tuning of three hyperparameters, which vary significantly across different datasets. This sensitivity makes it difficult to apply the method in a consistent and fully automated way.

To overcome these limitations, we introduce a filtration strategy based on Gaussian-convoluted histograms and adaptive kernel density estimators (KDEs). These methods are  data-driven, and while the former requires minimal manual tuning, KDE is fully automated  offering a more robust and broadly applicable alternative.  Using these methods, from a lattice two-point function, one can extract the energy levels reliably without subjective tuning of hyperparameters.
A full technical description of our filtration methodology is presented in the following section, and subsequently we demonstrate the efficacy and usefulness of these methods over any other methods.

\section{\label{sec:statistics_background}Filtering Spurious Eigenvalues in GEVP}

Our strategy for removing spurious eigenvalues arising in the solution of the generalized eigenvalue problem (GEVP) is based on the observation that such modes typically manifest as outliers relative to the central bulk of the eigenvalue distribution of the transfer matrix under bootstrap resampling. If left unfiltered, these spurious modes can significantly bias both the extracted central values and the associated uncertainties in an statistical analysis.

To address this issue, we identify the dominant peaks in the eigenvalue distribution and define peak-specific thresholds to eliminate outliers lying beyond the main support of each peak. A key ingredient in this procedure is a reliable estimation of the eigenvalue density obtained from the bootstrap-resampled correlators. To this end, we utilize two complementary non-parametric techniques: Gaussian-convoluted histograms and adaptive kernel density estimators (KDEs). These provide smooth and robust estimates of the underlying eigenvalue distribution without assuming a specific parametric form.

From the estimated eigenvalue density, we identify the prominent peaks corresponding to the dominant eigenvalue modes. The FWHM of each peak is measured, yielding an estimate of the spread of the eigenvalue distribution around the central modes. This FWHM serves as a quantitative measure of the uncertainty or fluctuation of the eigenvalue modes in the correlation function. Using the FWHM estimates, we then apply a cut (to be discussed later) to the data by discarding any eigenvalues that fall outside the expected range, defined as a factor of the FWHM from the center of the peak. This cut removes outliers, eigenvalues that lie far from the main distribution and could distort further analysis, and hence yields a cleaner eigenvalue dataset, enhancing the reliability of subsequent computations.

The use of histograms and kernel density estimators as non-parametric tools for distribution estimation from stochastic samples is well-established in statistics. For detailed treatments, we refer  to Refs.~\cite{Silverman1998,10.1214/aos/1176348768,Hardle2004,Hastie2009}. These techniques are also utilized in various other scientific areas for peak detection and peak-width estimation in noisy distributions~\cite{article,refId0,McCarthy2014PeaKDEck,Sadiq:2024xsz,Poluektov_2015}. To the best of our knowledge, however, this work constitutes the first systematic application of such methods to outlier removal in the spectral analysis of lattice QCD correlators.

\subsection{Gaussian-Convoluted Histogram}
\label{Gaussian_Histogram}
As a first step toward estimating the eigenvalue density from bootstrap-resampled data, we begin with the construction of a histogram—one of the simplest non-parametric estimators of a probability density function (PDF). Given a set of eigenvalues \(\{x_1, x_2, \cdots, x_n\}\), the histogram groups the data into bins and normalizes the counts to approximate the underlying distribution. The resulting estimator \(\hat{f}_h(x)\), with bin width \(h\) and origin \(x_0\), is defined by:
\begin{eqnarray}
    \hat{f}_h(x) &=& \frac{1}{nh} \sum_{i=1}^n \mathcal{I}(x_i \in B_j), \nonumber \\
    && \text{where } B_j = [x_0 + (j-1)h, x_0 + jh), \label{eq:gauss_conv}
\end{eqnarray}
where \(\mathcal{I}(\cdot)\) denotes the indicator function. Under suitable conditions, this estimator converges to the true density as \(n \to \infty\) and \(h \to 0\)~\cite{Hardle2004,Hastie2009}. We summarize its statistical properties, including bias, variance, and convergence rate, in Appendix~\ref{app:convergence}. Note that, in this case the error in \(\hat{f}_h(x)\) scales as \(n^{-2/3}\) \cite{Hardle2004}, which is slower than that expected from a more refined estimator.

While intuitive and easy to implement, the histogram suffers from several known drawbacks. It is piecewise constant by construction, sensitive to the choice of bin origin, and prone to introducing artificial discontinuities. These limitations become especially problematic in our context, where statistical fluctuations are present, and fine spectral structure must be preserved without over-interpreting noise.

To address these issues, we smooth the histogram by convolving it with a Gaussian kernel, producing a Gaussian-convoluted histogram. This approach retains the non-parametric nature of the histogram while mitigating its roughness. The smoothed estimator \(\tilde{f}_h(x)\) is given by,
\begin{equation}
    \tilde{f}_h(x) = \int_{-\infty}^{\infty} \hat{f}_h(u) \cdot K_\sigma(x - u) \, du,
\end{equation}
where the Gaussian kernel takes the form,
\begin{equation}
    K_\sigma(x) = \frac{1}{\sqrt{2\pi}\sigma} \exp\left(-\frac{x^2}{2\sigma^2}\right),\label{eq:gauss}
\end{equation}
and \(\sigma\) is the kernel width that controls the degree of smoothing. This convolution serves to suppress high-frequency noise and resolve genuine features of the distribution by replacing abrupt bin edges with a continuous smooth curve. In particular, it improves the reliability of peak detection, which is central to our filtration strategy.

However, as with all smoothing techniques, an appropriate care must be taken in choosing the parameter \(\sigma\). If \(\sigma\) is too small, the estimator remains noisy and prone to false peaks; if too large, it can over-smooth the distribution, merging nearby peaks or shifting modal locations. We therefore explore a range of \(\sigma\) values and assess the stability of key features, such as peak positions and full width at half maximum (FWHM), to guide the selection of both \(\sigma\) and the histogram bin width \(h\).

Importantly, since our application to correlation functions uses these peaks only to define outlier-rejection thresholds, the robustness of the final mean and percentile estimates remains intact, and this is valid even if the smoothed FWHM marginally overestimates the true peak width. This makes the Gaussian-convoluted histogram a practical and stable tool in the first stage of our density estimation pipeline.

\subsection{Adaptive Kernel density Estimator}\label{KDE}
In addition to the Gaussian-convoluted histogram method, we employ kernel density estimation (KDE) to obtain a smooth, non-parametric estimate of the eigenvalue density. As we show below this method is found to be completely data-driven and more robust for the purpose of extracting energy levels from two-point correlation functions.  By constructing the density as a superposition of smooth kernels centered at each data point, 
KDE avoids manual tuning of $\sigma$ associated with histogramming, such as bin widths and edge placements. We explain the method below. Given a bootstrap-resampled set of eigenvalues, $\{x_1, x_2, \cdots, x_n\}$, the kernel density estimate $\hat{f}_h(x)$ is given by,
\begin{equation}
    \hat{f}_h(x) = \frac{1}{n h} \sum_{i=1}^n K\left(\frac{x - x_i}{h}\right),
\end{equation}
where \( K(\cdot) \) is a continuous kernel function and \( h \) is the bandwidth that governs the degree of smoothing. For our purposes, we adopt the Gaussian kernel,
\begin{equation}
    K(u) = \frac{1}{\sqrt{2\pi}} e^{-u^2/2},
\end{equation}
yielding the Gaussian KDE,
\begin{equation}
    \hat{f}_h(x) = \frac{1}{n h \sqrt{2\pi}} \sum_{i=1}^n \exp\left(-\frac{(x - x_i)^2}{2h^2}\right). \label{eq:kde}
\end{equation}
This approach has several advantages. It produces a smooth, differentiable estimate of the density that is less sensitive to local noise and discretization effects than histograms. Moreover, it enables accurate identification of distinct peaks and their widths, both of which are essential for our automated outlier rejection procedure. Importantly, KDE does not require the manual specification of bin edges or origins, thereby removing a key source of arbitrariness in traditional histogram-based approaches. We discuss its statistical properties, including bias, variance, and convergence rate, in Appendix~\ref{app:convergence}. As discussed there, another advantage of KDE over histogram is its faster asymptotic convergence as $n^{-4/5}$ \cite{Silverman1998, Hardle2004, Hastie2009} with sample size $n$. It can be shown
that, under weak assumptions, there is no non-parametric estimator that converges at a faster rate than that of 
the kernel estimator \cite{10.1214/aos/1176342997}.

Just as with histograms, however, the effectiveness of KDE depends crucially on the choice of smoothing parameter \( h \). Too small a bandwidth results in noisy estimates dominated by statistical fluctuations, while too large a bandwidth merges neighboring modes and washes out genuine features. We therefore adopt a data-driven strategy to determine the optimal smoothing scale, especially suited to the potentially multimodal structure of bootstrap-resampled eigenvalue spectra. We explain the implementation procedure below in details:

\subsubsection*{Bandwidth Selection and Local Adaptivity}
To balance resolution with stability, we employ an adaptive KDE scheme based on the balloon estimator~\cite{10.1214/aos/1176348768,Mills10112011}. Unlike fixed-bandwidth KDE, which imposes a uniform smoothing scale across the entire spectrum, this approach modulates the bandwidth locally, guided by the density of the data around each evaluation point. Such adaptivity is crucial in our context, where the spectral landscape may feature narrow peaks associated with low-lying energy states alongside broader, and also flatter structures characteristic of excited levels. By allowing the degree of smoothing to vary with local data density, the estimator captures fine structure without over-fitting statistical noise in the tails. Below we elaborate the procedure in a few sequential steps.
\begin{enumerate}[wide, labelwidth=!, labelindent=0pt]
    \item \textbf{Pilot Estimation:} We begin by constructing a pilot estimate using standard KDE with a globally optimized bandwidth \( h_{\text{global}} \). The global bandwidth, \( h_{\text{global}} \), is  chosen using Silverman's rule of thumb \cite{Silverman1998, Hardle2004, Hastie2009}, which is given by $h = 0.9 \times \min\left( \sigma, \frac{\text{IQR}}{1.34} \right) \times n^{-1/5}$, where IQR is the inter-quartile range, $\sigma$ is the standard deviation and $n$ is the sample size. For the adaptive KDE procedure, we focus on using this to generate the pilot estimate for each data point.
    \item \textbf{Bandwidth Modulation:} After obtaining the pilot estimate, the local bandwidth, \( h_i \), for each point, \( x_i \), is calculated adaptively based on the density of points in the vicinity of \( x_i \) as,
    \begin{eqnarray}
        h_i = h_{\text{global}} \left( \frac{f_{\text{pilot}}(x_i)}{g} \right)^{-\alpha},
    \end{eqnarray}
    where \( f_{\text{pilot}}(x_i) \) is the pilot density at the point \( x_i \), and \( g \) is the geometric mean of the pilot densities across all data points:
    \begin{eqnarray}
        g = \exp\left( \frac{1}{n} \sum_{i=1}^n \log f_{\text{pilot}}(x_i) \right).
    \end{eqnarray}
    The parameter \( \alpha \) adjusts the strength of the adaptivity. 
    \item \textbf{Adaptive KDE Estimation:} The final adaptive KDE is then computed using the locally adjusted bandwidths. The density estimate at each point \( x \) is:
    \begin{eqnarray}
    \hat{f}(x) = \frac{1}{n} \sum_{i=1}^n \frac{1}{h_i} K\left( \frac{x - x_i}{h_i} \right),\label{eq:aKDE}
    \end{eqnarray}
    where \( K \) is the kernel function, which we choose to be Gaussian for this work. This step incorporates the locally adaptive bandwidths, allowing for finer resolution near denser regions and smoother estimates in sparser regions.
   \item \textbf{Choosing the Optimal \texorpdfstring{$\alpha$}{alpha}:} To determine the optimal $\alpha$ in our adaptive KDE framework, we employ the method of $K$-fold cross-validation~\cite{Silverman1998, Hardle2004, Hastie2009}, a widely utilized technique 
to assess the performance of a model and to mitigate over-fitting. In $K$-fold cross-validation, the entire dataset is partitioned into $K$ almost equal-sized, mutually-exclusive datasets. For each datasets $k \in \{1, \cdots,K\}$, we perform the following procedure:
\begin{enumerate}
\item Treat the $k$-th dataset as the validation set and rest of the $(K-1)$ dataset as the training set.
\item Using the training set, we estimate the density $\hat{f}(x;\alpha)$ with a given $\alpha$ via our adaptive KDE procedure which is laid out in the first three steps.
\item Then, the robustness of the density estimated by the training set is evaluated by computing negative log-likelihood of the validation set as below,
\begin{eqnarray}
\mathcal{L}^k(\alpha)=-\sum_{x_i\in \text{validation set}}\log \hat{f}(x_i;\alpha).
\end{eqnarray}
\item This procedure is repeated for each of the $K$ data partitions, and the average cross-validated log-likelihood is estimated as,
\begin{eqnarray}
CV(\alpha) =\frac{1}{K}\sum_{k=1}^K\mathcal{L}^k(\alpha).
\end{eqnarray}
\end{enumerate}
We repeat the above procedure for a large number of  $\alpha$ values on a grid. The optimal $\alpha^\ast$ is then chosen 
as the one that minimizes the average $CV(\alpha)$. Finally this optimal $\alpha^\ast$ is plugged into Eq.(\ref{eq:aKDE}) to perform final density estimation using adaptive KDE. Note that the entire procedure, including the choice of $\alpha$, as outlined above is data-driven and does not need any manual tuning, making this method robust and the preferred choice. 
\end{enumerate}
This adaptive construction provides finer resolution in regions where the eigenvalue distribution is sharply peaked, thereby capturing closely spaced modes, while applying greater smoothing in the sparse tails, where statistical noise dominates. In effect, it allows the KDE to adapt to the spectral structure revealed by the bootstrap ensemble, offering a statistically robust and automated tool for the identification and rejection of spurious modes.

As with histogram-based approaches, we emphasize that the role of KDE in our workflow is not merely visualization, but rather to enable quantitative detection of peak locations and widths. These features directly inform the thresholds used in our filtering strategy, ensuring that our treatment of spurious modes is grounded in the underlying statistical structure of the spectrum.

\subsection{Analysis of eigenvalue distributions with the Histogram and KDE methods}
In this subsection, we apply the previously described histogram-based techniques and the KDE method to analyze the eigenvalue distributions and assess their effectiveness in filtering out spurious eigenvalues produced by the TGEVP method. We choose two examples: two-point correlation functions of nucleon and $J/\psi$ meson. The goal is to demonstrate how these methods can be utilized to estimate the FWHM of the peaks in the distributions and remove outliers based on these estimates.

\begin{figure}[h!]
    \centering
    \includegraphics[width=0.45\textwidth]{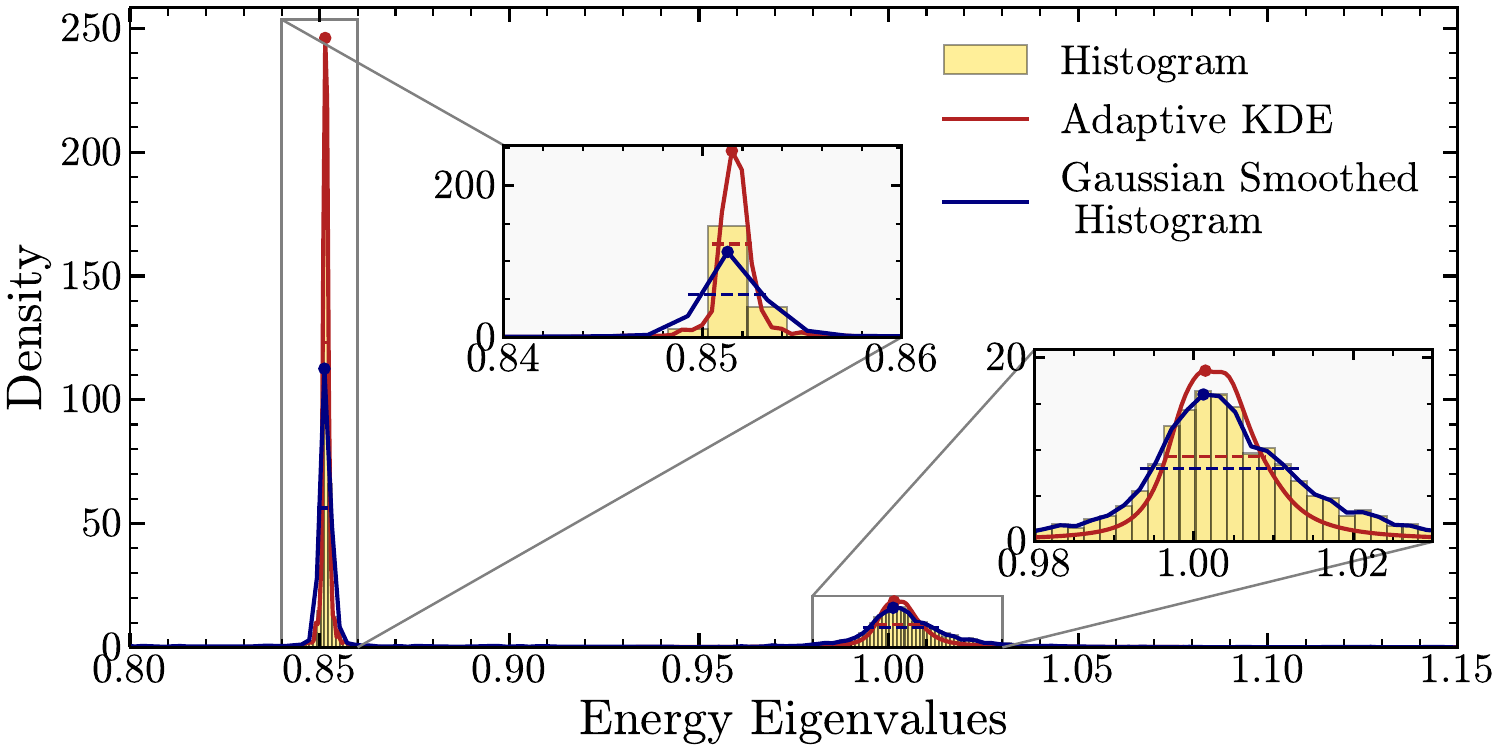}
    \includegraphics[width=0.45\textwidth]{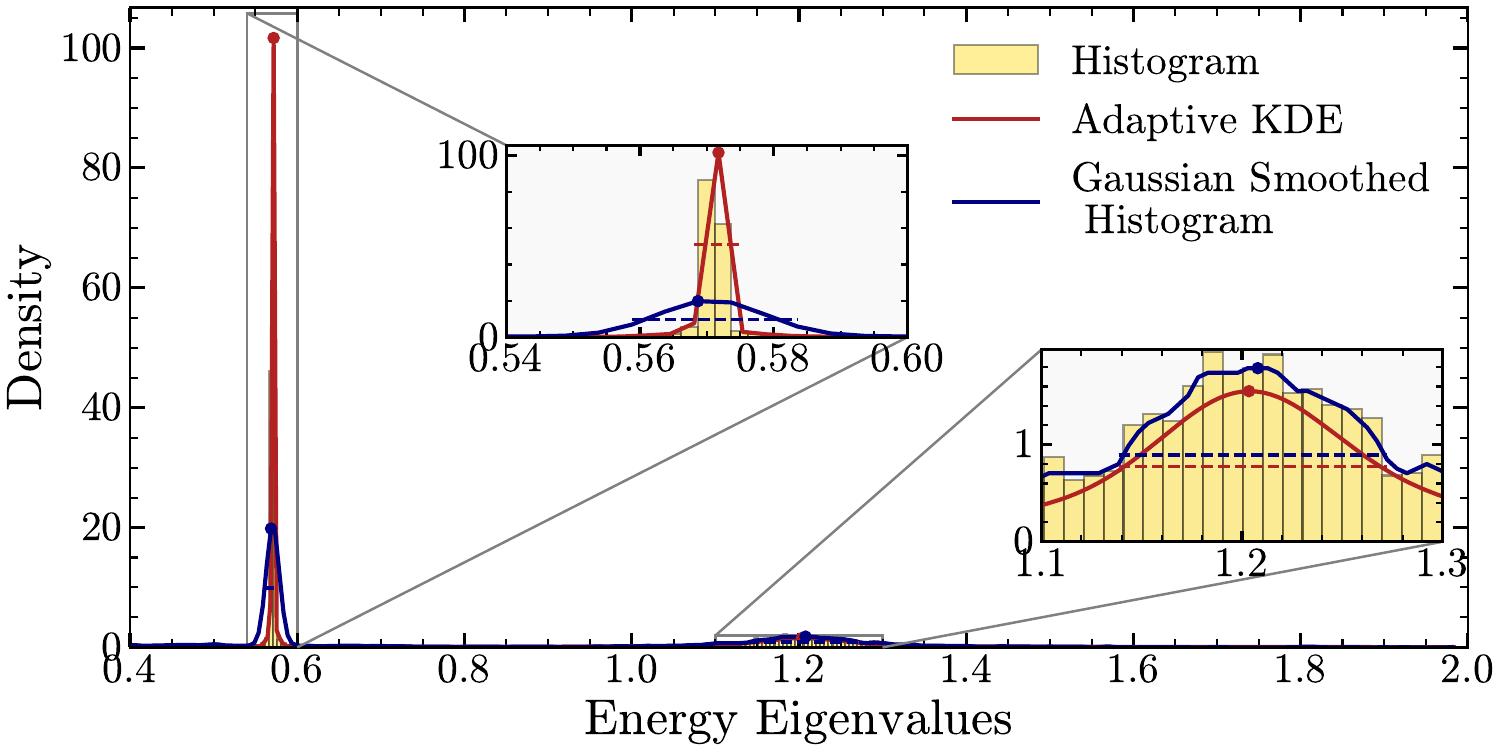}
    \caption{Histogram, Gaussian-smoothed histogram, and kernel density estimate (KDE) for the eigenvalue distribution of the  $1^-$ charmonium (\textit{top panel}) and nucleon (\textit{top panel}) 2-point correlation function. The histogram shows the bootstrapped eigenvalue distribution, the smoothed histogram suppresses noise, and the KDE provides a continuous estimate of the probability density.}
    \label{fig:etac_hist}
\end{figure}

We begin by constructing a histogram of the eigenvalues, which 
are shown in Fig. \ref{fig:etac_hist}. The histograms  reveal the distribution of eigenvalues, with some statistical noise due to the discrete nature of the data. Following the procedure described in \ref{Gaussian_Histogram}, we then apply a Gaussian smoothing filtration to the histograms. The smoothed distribution is shown by the blue line, highlighting the central features of the distribution while suppressing fluctuations.

\begin{figure}[h!]
    \centering
    \includegraphics[width=0.42\textwidth]{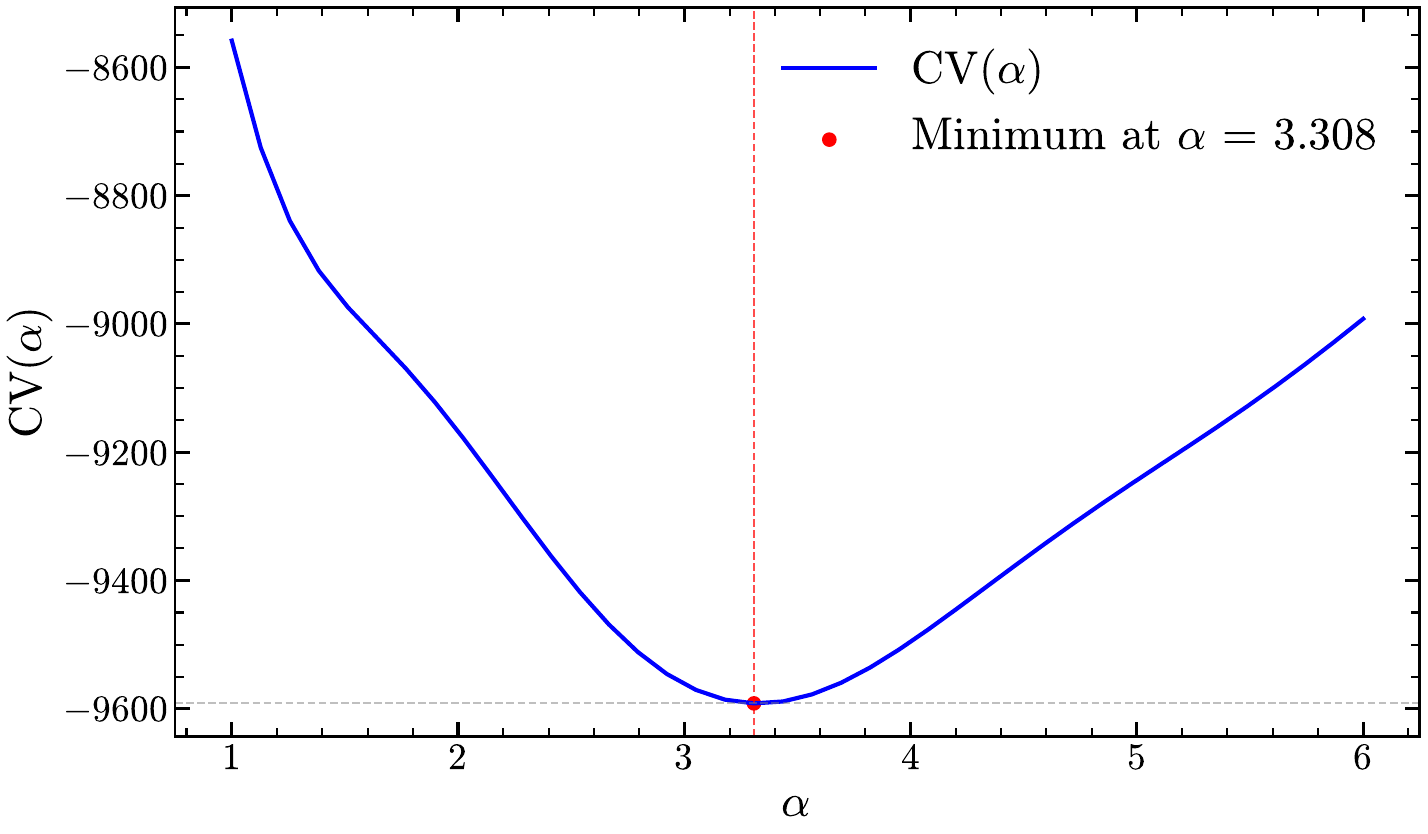}
    \includegraphics[width=0.42\textwidth]{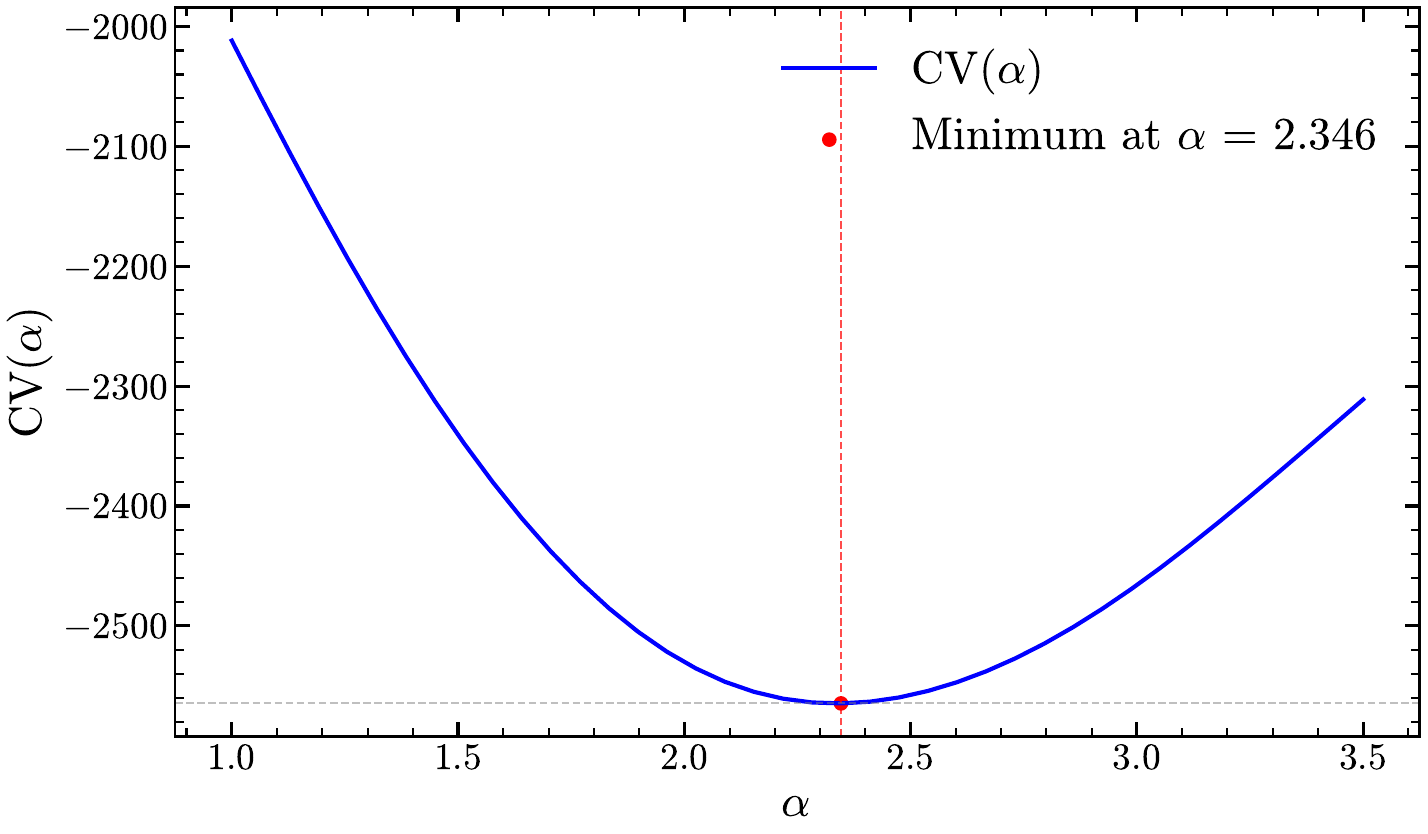}
    \caption{Cross-validation score $\text{CV}(\alpha)$ as a function of the regularization parameter $\alpha$ for $1^-$ charmonium (\textit{top panel}) and single nucleon (\textit{bottom panel}). The minimum of the curve indicates the optimal value $\alpha^\star$. The red marker denotes the location of the minimum cross-validation score.}
    \label{fig:cv_score_vs_alpha}
\end{figure}

Next we compute the kernel density estimate, as described in \ref{KDE}, and the resulting distribution is depicted by the red line alongside the smoothed histograms in Fig. \ref{fig:etac_hist}. We find the adaptive KDE result provides a continuous estimate of the probability density function of the eigenvalues and captures the shape of the distribution more accurately than the discrete histogram. Note the adaptive nature of the KDE adjusts the bandwidth depending on the local density of the data, yielding a smoother and more reliable density estimate. To ensure an optimal balance between smoothness and fidelity, we determine the regularization parameter, $\alpha$, using the 
$K$-fold cross-validation method, as described in the previous subsection. In Figure~\ref{fig:cv_score_vs_alpha}, we present the results of this optimization procedure, with optimal values of $\alpha$, for both the cases of nucleon and the $J/\psi$ meson correlators. 
For all other correlators investigated (as presented in the Results section), we observe similar advantages of the adaptive KDE method, achieved in a fully data-driven manner through hyperparameter optimization. Based on these findings, we recommend the adaptive KDE method as the most robust and reliable approach for eigenvalue filtration. The Gaussian-convoluted method serves as a complementary option, useful for cross-verification.

\section{\label{eigs_filtration}Flowchart for filtration of spurious eigenvalues}
\label{flowchart}

Having broadly discussed the methodologies for eigenvalue filtration in the previous section, we now present a complete flowchart outlining the implementation of the above-mentioned procedures for extracting energy levels within our proposed TGEVP framework.

\begin{enumerate}[wide, labelwidth=!, labelindent=0pt]
    \item First, perform a bootstrap resampling on the original dataset to generate $N_b$ bootstrap samples. For each of these samples, one needs to carry out GEVP calculations, as defined in Eq.~(\ref{GEVPeq1}), computing the energy eigenvalues $ \lambda_n^{m,b}$. Here, the index $ n $ enumerates the eigenvalues obtained for the $ m $-th iteration and the $ b $-th bootstrap sample. The ranges are defined as, $ 0 \leq n \leq m $, $ 1 \leq m \leq \lfloor (n_T+1)/2 \rfloor $, and $ b \in \{1, 2, \dots, N_b\} $.
    \item Next, we eliminate all eigenvalues for which \( \left\vert \arg \lambda_n^{(m,b)} \right\vert > \epsilon \), retaining only those that are real within numerical precision (the value of \( \epsilon \) is set based on the required floating-point tolerance). Eigenvalues outside the range
    $0 < \lambda \le 1$ are discarded, as the transfer matrix is a positive operator. However, if the correlation functions are affected by poor statistics, the removal of negative and complex eigenvalues may introduce statistical bias in the final energy estimates. To guard against this, we provide diagnostic tools for evaluating the impact of such removals, with further details presented in Appendix~D.

    \item 
For each iteration, estimate
the density of eigenvalues obtained across all bootstrap samples using adaptive KDE \ref{KDE} and/or Gaussian-convoluted histogram \ref{Gaussian_Histogram} method as described in the previous section (note KDE is our recommended method for its robustness). 
The peaks in the estimated eigenvalue density are then identified by determining their centers and calculating their full width at half maximum. These  peak-centers and their corresponding widths are denoted by $C_{r}^{m}$ and $\Delta_r^{m}$, 
respectively, where 
$m$ refers to the iteration number, and the subscript $r$ indexes the peaks in descending order of their center values. 

\item 
Each peak thus identified in the estimated eigenvalue density represents an eigenstate of the transfer matrix. The peak-centers and their widths are then utilized to guide the procedure for selecting eigenvalues, ensuring that only physically meaningful eigenstates are retained. This procedure enhances the reliability of the extracted energy levels by systematically filtering out spurious eigenvalues and maintaining a consistent ordering of eigenvalues.
    
    \item 
    For each peak in the $m$-th iteration, then select the eigenvalue $\lambda_n^{m,b}$ that is closest to $C_r^m$ and lies within the range $\left[C_r^m - 3\Delta_r^m, C_r^m + 3\Delta_r^m\right]$ for each bootstrap sample. This procedure results in a set of filtered eigenvalues, denoted as $\tilde{\lambda}_r^{m,b}$, where the index $r$ represents the peaks in the histogram, or equivalently, the well-resolved states in the spectrum of $T^m$. For example, $-\log \tilde{\lambda}_{r=0}^{m,b}$ represents the ground state energy eigenvalue for the $m$-th iteration and the $b$-th bootstrap sample.
\end{enumerate}
Naturally, a concern here is on the accurate estimation of errors in the eigenvalues obtained using the above procedure, particularly when comparing them with traditional method such as, effective masses. One needs to
account for the potential influence of outliers in the
bootstrap eigenvalue estimates. 
In Refs. \cite{Wagman:2024rid, Ostmeyer:2024qgu} a double-bootstrap method was utilized to address that. Instead, we adopt the 68th percentile as the bootstrap confidence interval. Despite the filtration procedure applied to the eigenvalues \( \tilde{\lambda}_r^{m,b} \), some outliers may still remain. These outliers can skew the bootstrap mean and inflate the bootstrap standard error, particularly in certain iterations. To mitigate this, we define the bootstrap confidence interval by using the 68th percentile, which corresponds to the interval between the 16th and 84th percentiles of the bootstrap distribution. The mean of the eigenvalues within this 68th percentile interval is then quoted as the bootstrap mean. 
However, to be conservative with errors, we estimate those using the 68th percentile of the {\it full samples} without removing any outlier.  
We find that this approach effectively reduces the influence of extreme values while providing reliable error bars, resulting in more stable and robust estimates. We automate the whole procedure described above employing the adaptive KDE framework for density estimation of bootstrap resampled eigenvalues. In the case of Gaussian-convoluted histogram, the smearing parameter ($\sigma$) for Gaussian kernel in (Eq.~\ref{eq:gauss} in \ref{Gaussian_Histogram}) needs to be provided manually.

\section{\label{sec:results}Results}
Here, we present results obtained using our proposed method, applied to a wide range of two-point correlation functions, with the aim of extracting the corresponding energy levels of the associated states.

Before presenting the results obtained using the TGEVP method, it is worthwhile to first compare its effectiveness in extracting energy levels with that of other existing approaches. We begin by comparing results obtained from a given set of two-point correlation functions using different analysis methods: the standard effective mass method, Prony’s method \cite{Prony1795,Kunis2015AMG}, Lanczos method, as described in Ref. \cite{Wagman:2024rid}, and the TGEVP method incorporating the proposed eigenvalue filtration. The effective mass is defined as $m_{eff}(t) = {\mathrm{log}}[C(t)/C(t+1)]$, where
$C(t)$ denotes a two-point correlation function. In the TGEVP method, eigenvalues are computed using two independent approaches—\texttt{scipy.linalg.eig} \cite{python} and the FEAST eigensolver \cite{PhysRevB.79.115112}—to cross-validate results and ensure high-precision evaluation. Subsequently, we discuss results from different correlation functions across a variety of hadrons, including nuclei.

\subsection{Comparison between Prony's method and TGEVP method}\label{results:comp1}

\label{subsec:prony_vs_tgevp}

In this subsection, we present a comparative analysis of commonly used methods for extracting energy levels from lattice QCD two-point correlation functions: the traditional Prony method \cite{Prony1795,Kunis2015AMG} and the TGEVP framework introduced in this work. Our aim is to assess the relative reliability and stability of these approaches in determining ground-state energies.

The Prony method models the correlator as a sum of exponentials, reconstructing the parameters (masses and amplitudes) by solving a system of non-linear equations derived from the exponential time dependence of the two point correlation function. It is particularly effective at short time separations and allows for straightforward multi-state fits, but it is sensitive to statistical noise and the choice of fit window as is shown below.

\begin{figure}[htbp]
\centering
\includegraphics[scale=0.35]{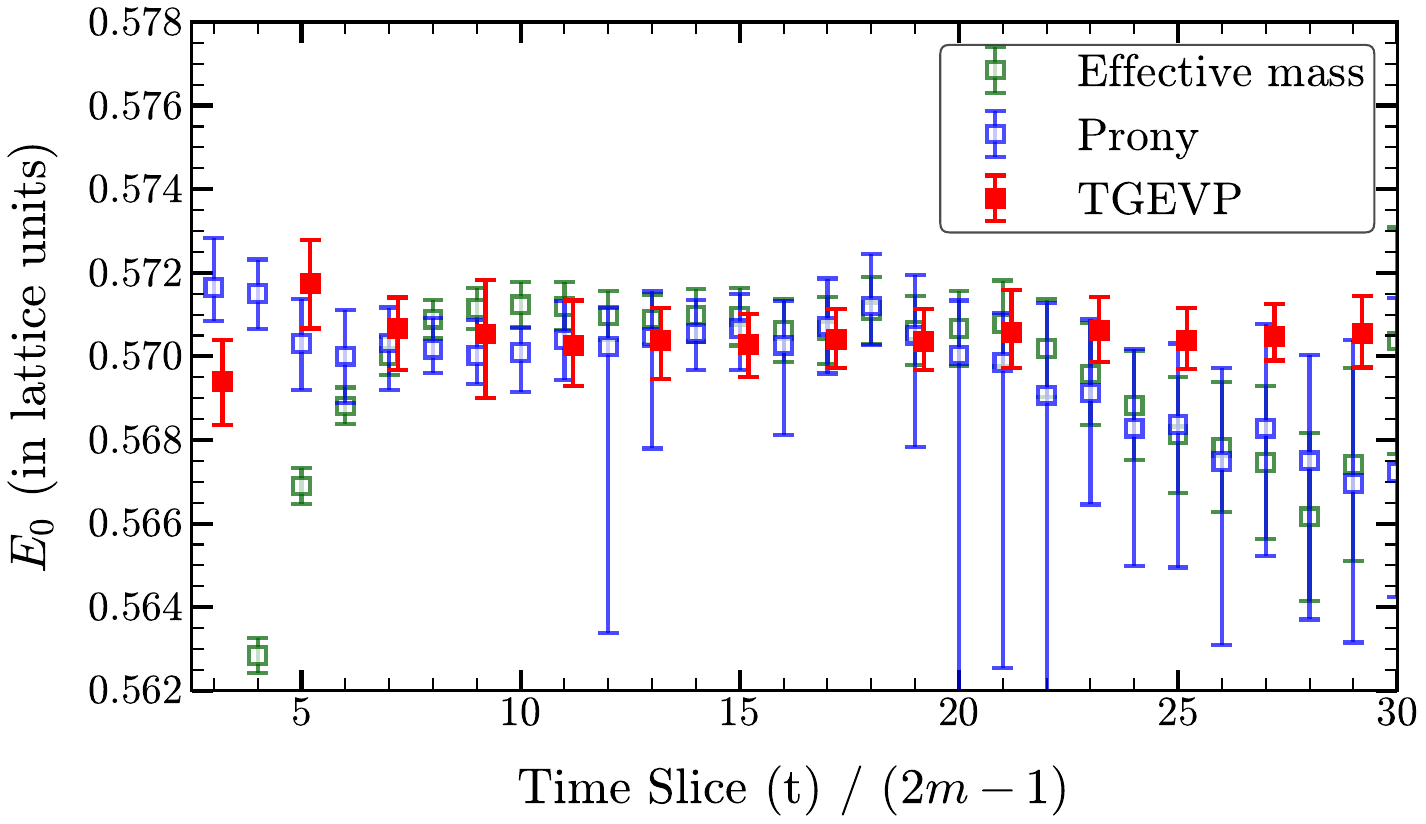}
\caption{Comparison of ground-state energy extraction using TGEVP (red), the two-state Prony method (blue) and effective mass (green) for a given set nucleon two-point functions. The $x$-axis corresponds to the minimum time slice $t_{\text{min}}$ for the Prony fits and the iteration index $m$ for the TGEVP estimate.}
\label{fig:prony_vs_tgevp}
\end{figure}

To illustrate the comparison, we apply both methods to a same set of nucleon two-point correlation functions computed using clover valence quarks on HISQ gauge action. 
Figure~\ref{fig:prony_vs_tgevp} presents the results: the red point marks the TGEVP estimate, while the light green points indicate the Prony method’s estimates as a function of the minimum time slice $t_{\text{min}}$ included in the fit.
 At small $t_{\text{min}}$, the TGEVP estimates exhibit fluctuations due to excited-state contamination but quickly stabilize. It also agrees with the effective mass plateau within uncertainties, but is more stabilized. For both the TGEVP and Prony methods, the 16th and 84th percentiles are shown as error bars.  Both the Prony and TGEVP estimates converge faster than the effective mass. However, the Prony method is found to be less stable than the TGEVP method, particularly at intermediate time slices. This instability becomes more pronounced at larger time separations, where the Prony results are consistently noisier.

This comparison demonstrates that both the TGEVP and Prony methods can yield consistent estimates of the ground-state energy when applied appropriately. However, the TGEVP method exhibits greater robustness against excited-state contamination at early times and provides a cleaner signal at intermediate and large time separations compared to the Prony method.

\subsection{Numerical comparison between Lanczos and TGEVP methods}\label{Lanczos_TGEVP}
The analytical equivalence between the Lanczos and TGEVP methods are shown in the Appendix \ref{app:equiv}.  Here we compare the numerical results obtained with the Lanczos method, as described in Ref. \cite{Wagman:2024rid}, and the TGEVP method, as proposed here.
 We follow a similar implementation of Oblique Lanczos method (Lanczos A) as in Ref. \cite{Wagman:2024rid}. We also use another method (Lanczos B) which is similar to Ref. \cite{Wagman:2024rid},  except differences in the procedure of filtering and sorting the eigenvalues. 
 We outline the implementation below.
\begin{enumerate}[wide, labelwidth=!, labelindent=0pt]
    \item We first resample our data to obtain $N_b$ bootstrap samples and perform $M$ iterations of Oblique Lanczos for each, yielding $\lambda^{m,b}_n$. The index $1\leq M$ and $1\leq b\leq N_b$ correspond to Lanczos iterations and the bootstrap sample being used respectively, while $n$ enumerates the computed eigenvalues. 
    \item Of these $\lambda^{m,b}_n$, we discard all results satisfying $\arg\lambda_n^{m,b}>\epsilon$ with $\epsilon>10^{-12}$ as spurious. We also discard $\lambda^{m,b}_n>1$, for reasons described earlier.
    \item For each Lanczos iteration, we plot a histogram of the bootstrap eigenvalues and use a Gaussian filter to smoothen the data. We then identify the peaks $P^m_n$, and their corresponding widths $\Delta^m_n$, in this histogram and sort the $\lambda^{m,b}_n$ on the basis of the peak they correspond to.  
Iterations where no eigenvalue aligns with a particular peak are treated as though no result is obtained, and so any apparent outliers are discarded. This ensures that only eigenvalues corresponding to the same state are compared or binned together in subsequent analysis, reducing the risk of contamination from spurious or unrelated states.
    \item Lanczos A: To these eigenvalues, we apply the Cullum-Willoughby procedure with a threshold $\epsilon_{CW}$. This is utilized to filter out spurious eigenvalues, using the criteria and parameters described in Ref. \cite{Wagman:2024rid}. This leads to the eigenvalues $\lambda^m_n$ . 
    \item Lanczos B: For the mean and error estimation, as in the case of TGEVP, we further use the $68$th-percentile bootstrap confidence interval of the eigenvalues obtained in the previous step. 
\end{enumerate}

\begin{figure}[htb!]
 \includegraphics[width=0.45\textwidth, height=0.28\textwidth]{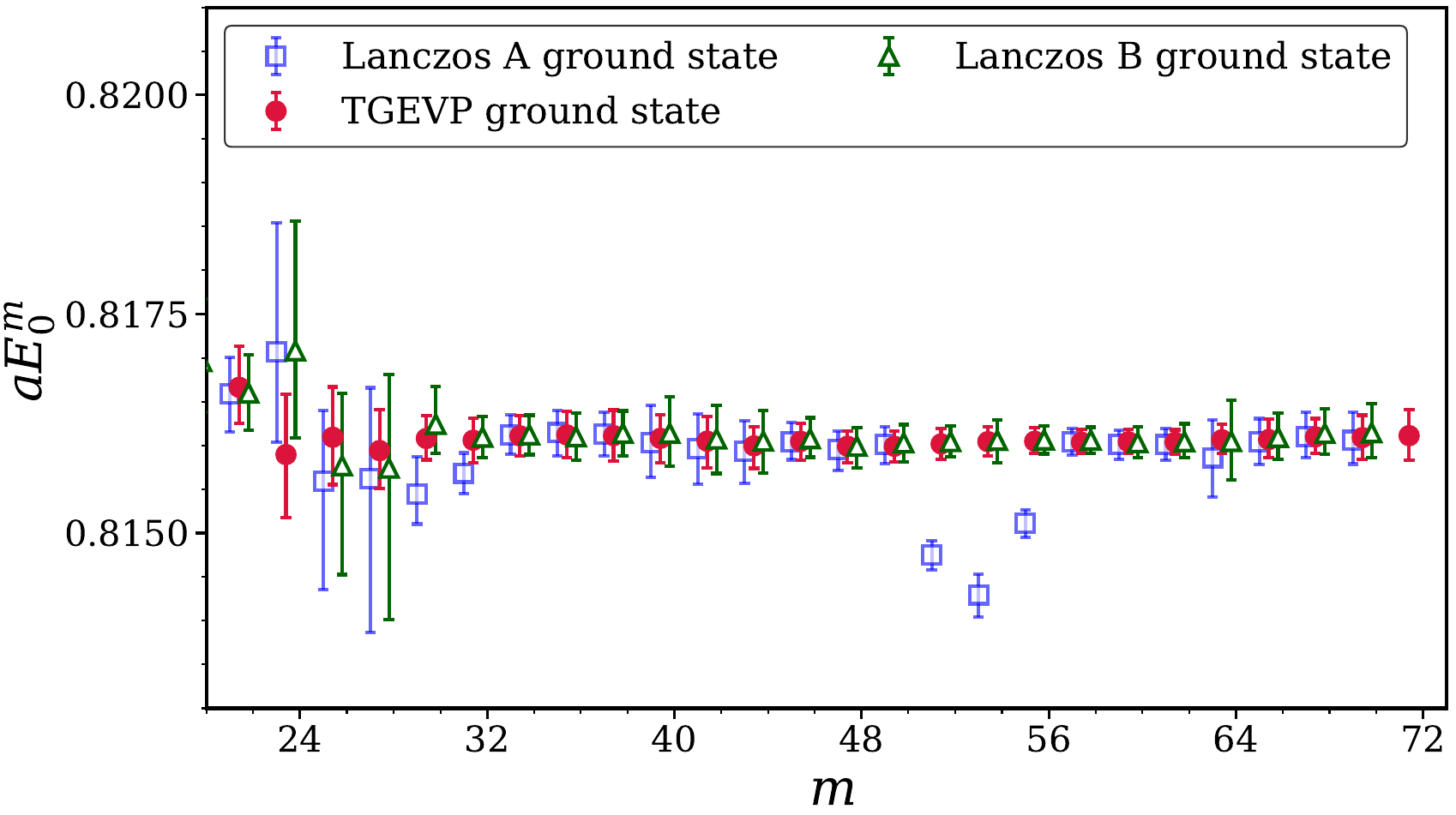}
\includegraphics[width=0.43\textwidth, height=0.28\textwidth]{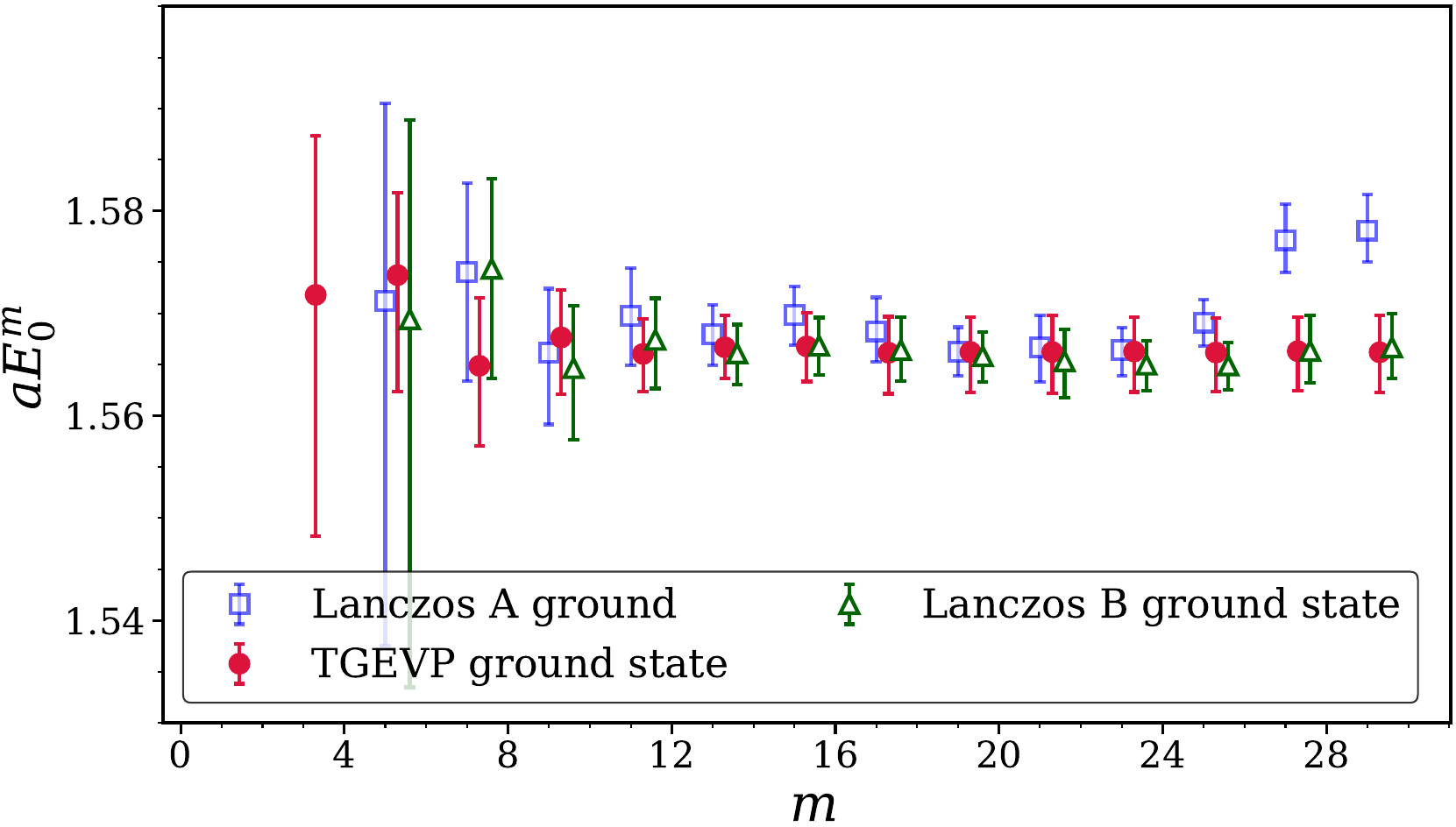}
        \caption{Comparison of eigenvalues determined by Lanczos and TGEVP methods. Lanczos A is the method as in Ref. \cite{Wagman:2024rid}, Lanczos B utilizes the original Lanczos method \cite{Wagman:2024rid} but with mean and errors estimated using 68th-percentile, as in the TGEVP method.}
        \label{fg:comp1}
  \end{figure}

Following the above procedure for Lanczos method, in Fig. \ref{fg:comp1}, we compare the eigenvalues obtained using the Lanczos and TGEVP methods for the same set of correlation functions corresponding to a meson: $\eta_c$ (top),  and a baryon: $\Xi_{cc}$ (bottom). Results for Lanczos A and B are both shown. One can clearly observe the numerical equivalence of the two methods, though Lanczos A, where our filtration procedure has not been not employed show fluctuations on some iterations for both the cases.

Having compared the TGEVP method with other existing approaches—and demonstrated the robustness and efficiency of our approach in extracting energy levels—we now present additional results obtained using our proposed method.
In Figs. (\ref{fg:charm1}, \ref{fg:cont_extp1},
\ref{fg:cont_extp2}, \ref{fg:cont_extp3},
\ref{fg:heavy_baryon1}, \ref{fg:heavy_baryon2},
\ref{fg:nu1}, \ref{fg:mult_nu1}) we show our results for a variety of hadrons, both for mesons and baryons.  In each of these figures we present the extracted
lowest energy levels as estimated by the above method, along with the standard effective mass, defined as $m_{eff}(t) = {\mathrm{log}}[C(t)/C(t+1)]$. In each case, throughout this article from here, blue circles represent the regular effective mass $m_{eff}(t)$, and the red circles represent the lowest eigenvalues as computed over the iterations. The magenta band represents the correlated fit of the two-point function using a single exponential, illustrating the fit range and the associated $1\sigma$ error. The faded-green band corresponds to the eigenvalues, fitted with a constant term over multiple iterations, also using a correlated fit. This comparison highlights the stability and error behavior of the TGEVP eigenvalues over regular effective mass approach. Below we discussed each of the cases. A large set of other results, including the first excited states and nuclei, are shown in Appendix \ref{app:additional_results}.

\subsection{Case I: Charmonia}\label{results:charm}
\begin{figure}[htb!]
  \includegraphics[width=0.46\textwidth, height=0.26\textwidth]{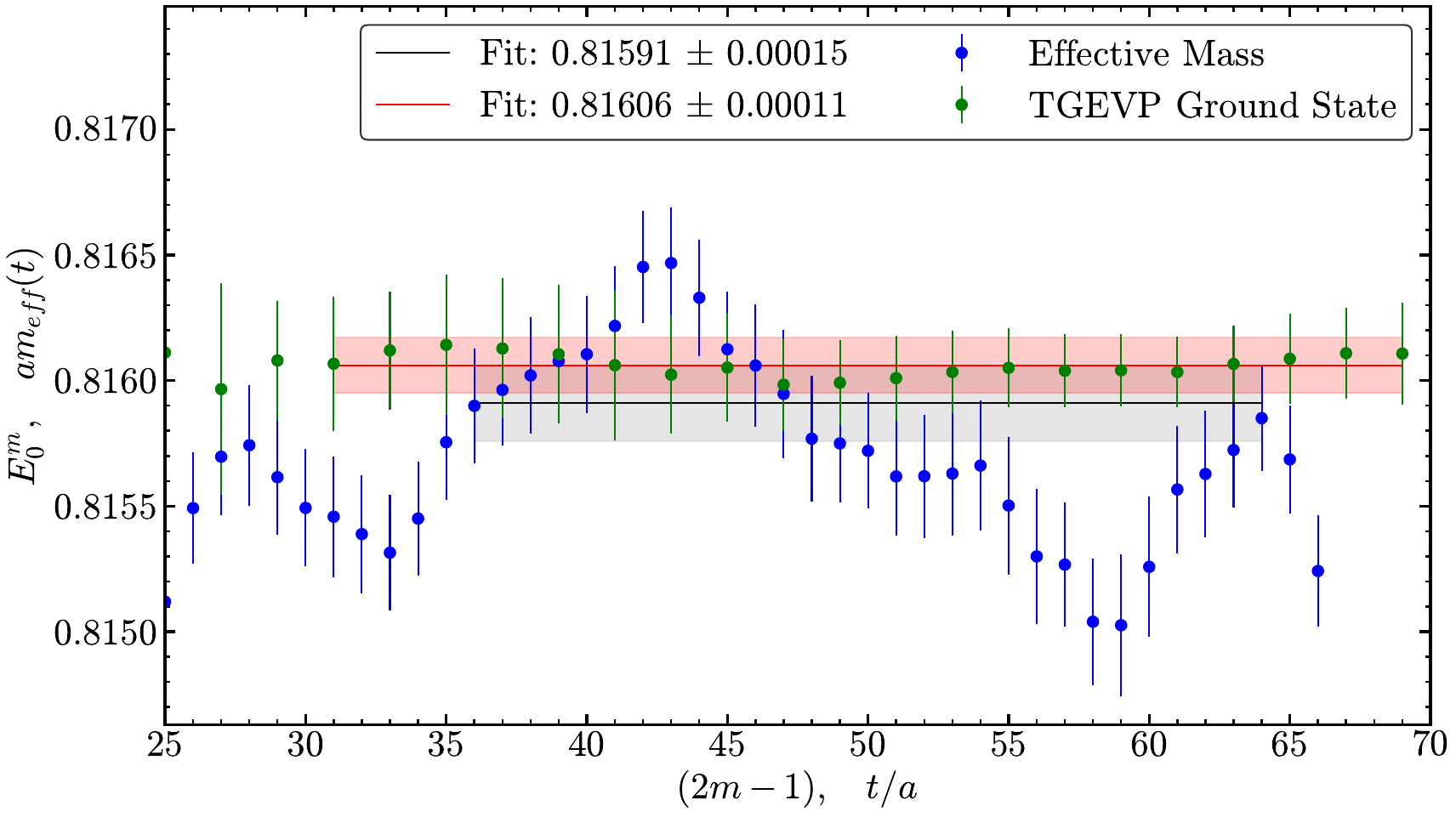}
  \includegraphics[width=0.46\textwidth, height=0.26\textwidth]{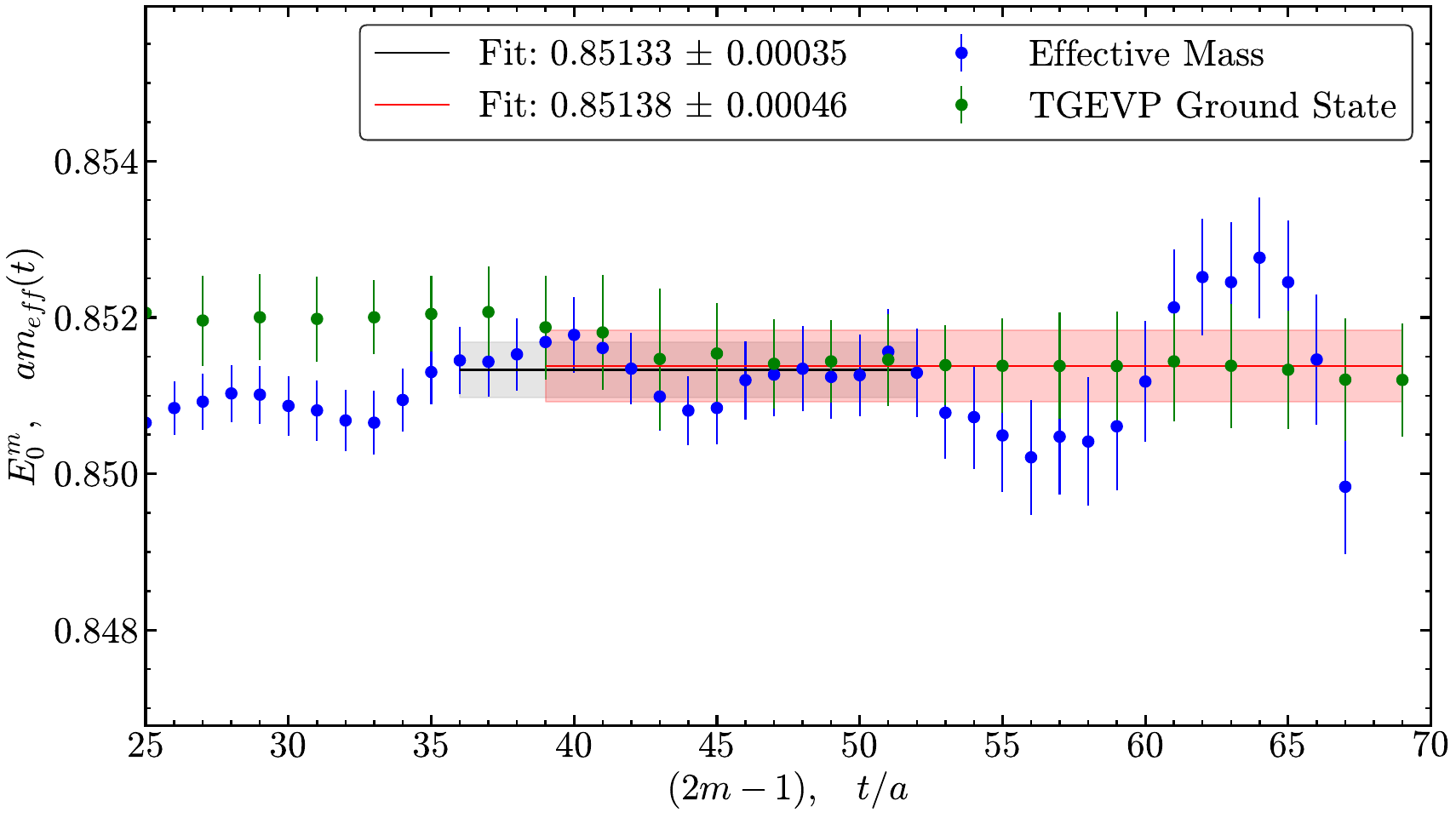}
        \caption{Comparison of the results between the standard effective mass of the ground state (blue circles)
        and the lowest eigenvalue (red circles) corresponding to the two-point correlation functions of 
        the pseudoscalar ($J^P \equiv 0^-$) (top) and vector ($1^-$) (bottom) states of charmonia. Number of measurements ($n_{meas}$) in this case is $180$.}
        \label{fg:charm1}
\end{figure}

We first present the results for the 1S-charmonia in Fig. \ref{fg:charm1}.  The results corresponds to the two-point correlation functions of  $0^-$ (top) and $1^-$ (bottom) states of charmonia, as computed using overlap valence quarks with a HISQ lattice ensemble ($48^3\times 144$, $a = 0.0582$ fm, $n_{meas} = 180$). As can be observed, while the effective mass is fluctuating over time-slices, the lowest eigenvalue estimated over different iterations is very stable. In Fig \ref{fg:charm2} in Appendix \ref{app:additional_results}  we show additional similar results for other lattice ensembles along with the first excited states. 

\begin{figure}[htb!]
\includegraphics[width=0.47\textwidth, height=0.31\textwidth]{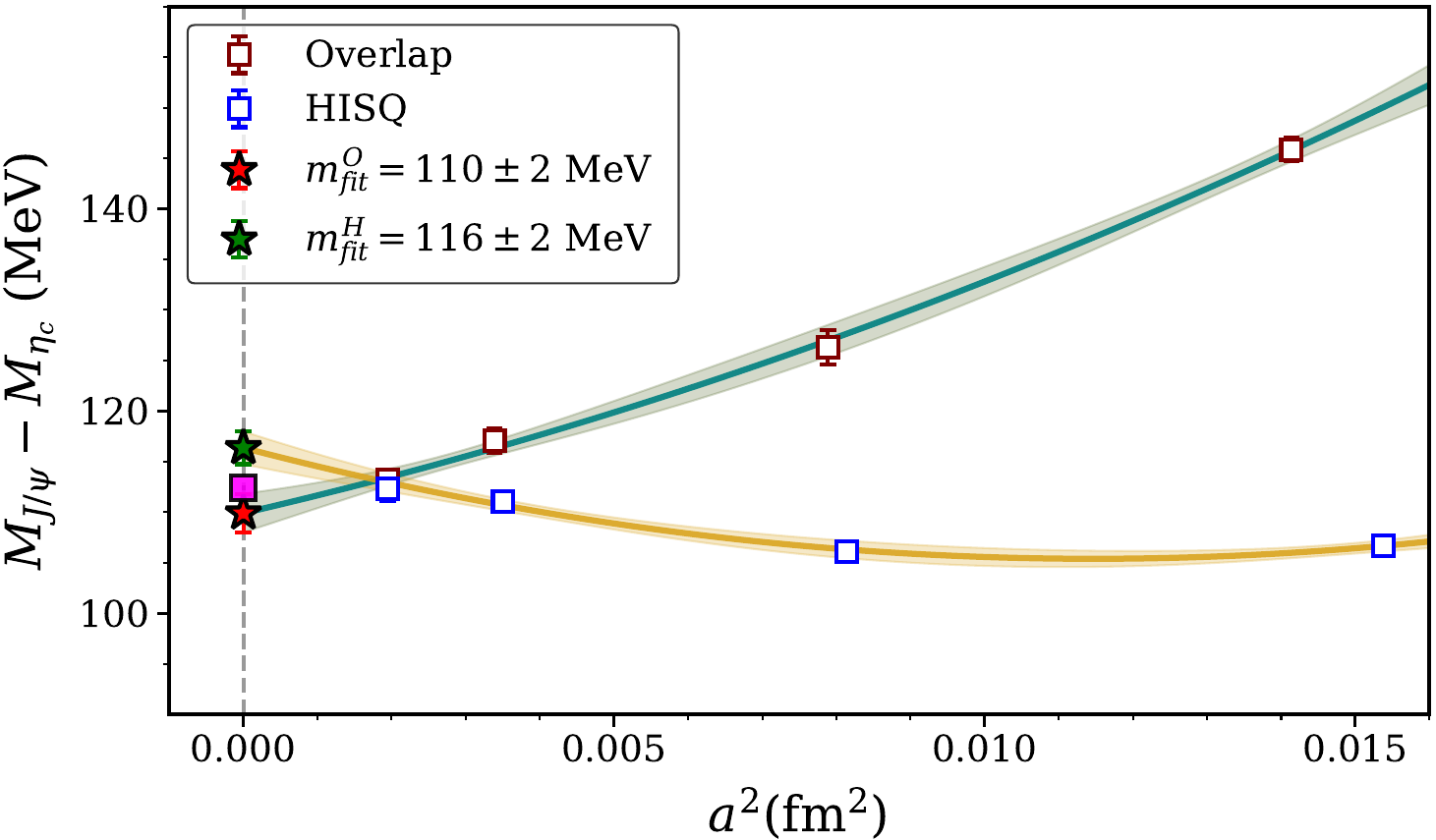}
        \caption{Hyperfine splitting of 1S-charmonia as extracted from the lowest eigenvalues of $1^-$ and $0^-$ correlators, on different lattice ensembles. The data point with red-square (coming from above) are obtained with overlap valence quarks and those with blue-square are due to HISQ valence quarks. Continuum extrapolated results are shown by star-symbols along its physical value (solid magenta square).}
        \label{fg:cont_extp1}
  \end{figure}

\begin{figure}[htb!]
\includegraphics[width=0.47\textwidth, height=0.31\textwidth]{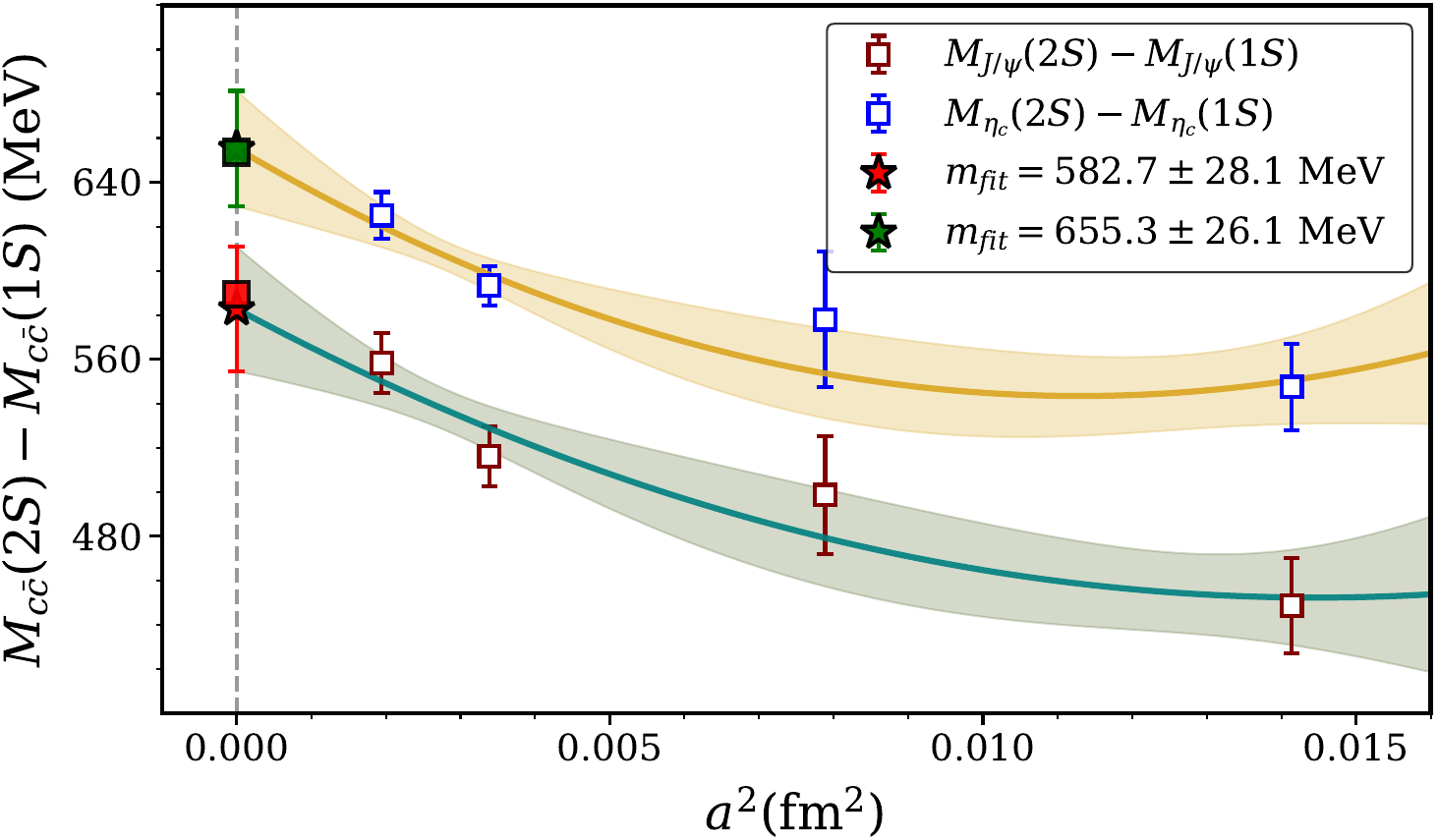}
\includegraphics[width=0.48\textwidth, height=0.31\textwidth]{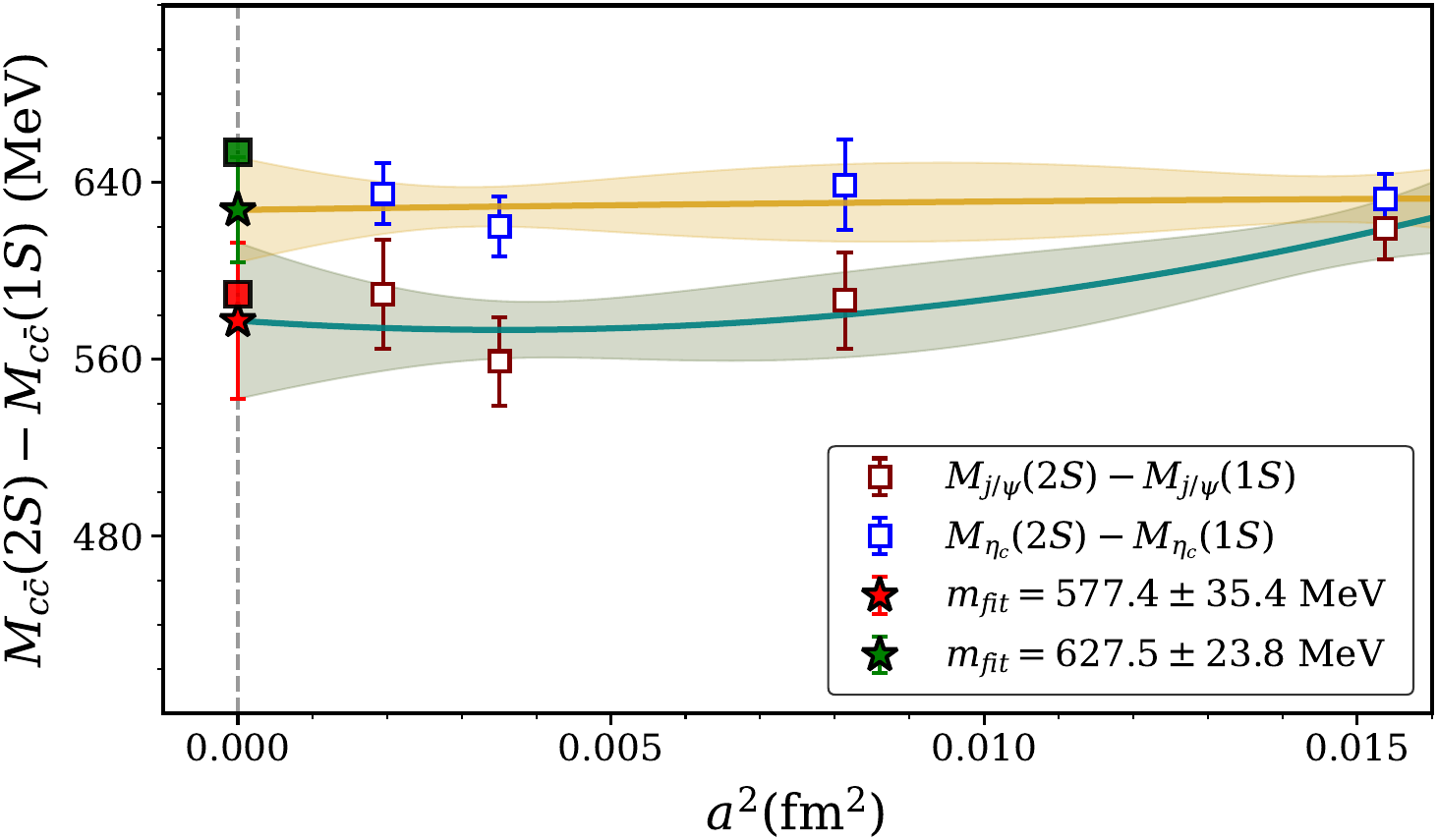}
        \caption{Continuum extrapolation of the charmonia energy splittings, using the lowest two eigenvalues, obtained by TGEVP method as shown in \ref{fg:charm2}. Top:   results with overlap valence quarks, Bottom: same with HISQ valence quarks. The star symbols represent the continuum extrapolated values while solid squares are their experimental values.}
        \label{fg:cont_extp2}
  \end{figure}

\begin{figure}[htb!]
\includegraphics[width=0.47\textwidth, height=0.31\textwidth]{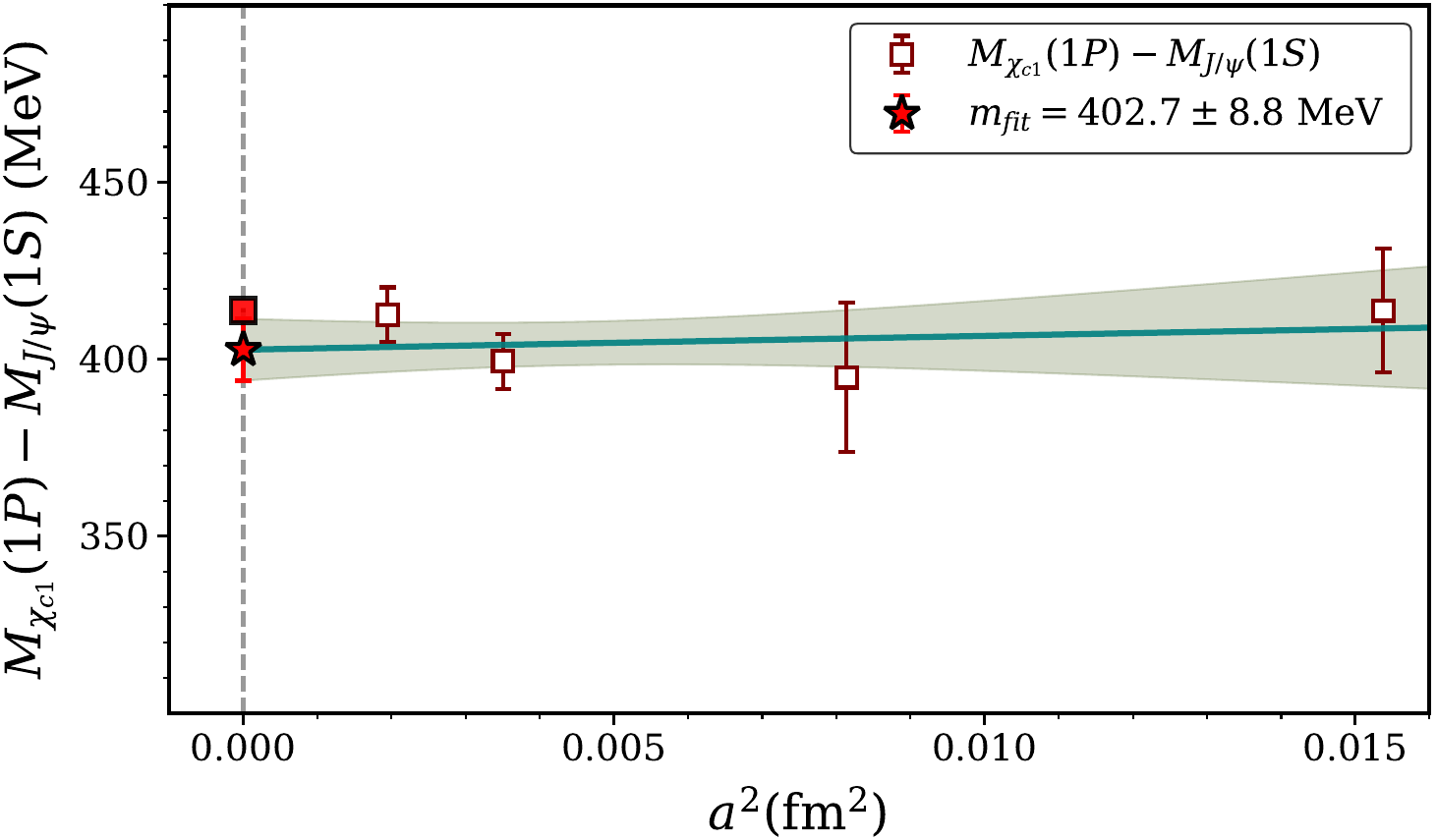}
        \caption{Continuum extrapolation of the energy splittings, 1P$-$1S, in charmonia, using the 1st and 3rd eigenvalues as shown in the right pane of Fig. \ref{fg:charm3}.}
        \label{fg:cont_extp3}
  \end{figure}

Based on these results on the lowest two/three eigenstates we then calculate various energy splittings between them, for example, hyperfine splitting in 1S, along with 2S$-$1S and 1P$-$1S splittings in charmonia.  Since we could extract these values at multiple lattice spacings we then extrapolate them to the continuum limit. Since both the overlap and HISQ actions do not have  $\mathcal{O}(n \times  ma)$ errors ($n$ odd integers, $ma$ being quark mass in lattice unit), we extrapolate these energy splittings with a fit ansatz, $f(a) = c_1+c_2a^2+c_3 a^4$. We show the continuum extrapolated results in Figs. [\ref{fg:cont_extp1},\ref{fg:cont_extp2}, 
\ref{fg:cont_extp3}].
Fig. \ref{fg:cont_extp1} represents the hyperfine splitting for 1S-charmonia obtained with both overlap (red square) and HISQ (blue square) actions. The  continuum extrapolated results are shown by the red and green star-symbols respectively, while the experimental value is shown by the magenta square. 
Note that the HISQ result agrees with that obtained by HPQCD collaboration using the same lattice ensembles \cite{PhysRevD.102.054511}. 
In Fig. \ref{fg:cont_extp2}, we present the 2S $-$1S splitting with overlap results at the top and the HISQ results at the bottom pane. Here again, we use the same fit form and extrapolations are shown with the same symbols as above. In Fig. \ref{fg:cont_extp3}, we show the 1P$-$1S splitting.


These results collectively demonstrate that the TGEVP method can reliably extract not only the ground state but also the excited state spectra from a single correlator. With increased statistics, the precision of excited state measurements can further be improved. This highlights the potential of TGEVP, particularly when combined with multiple operators, to provide highly reliable results for both ground and excited states in lattice QCD calculations.

\subsection{Case II: Heavy Baryons}\label{results:heavy_baryons}
\begin{figure}[htb!]
\includegraphics[width=0.46\textwidth, height=0.26\textwidth]{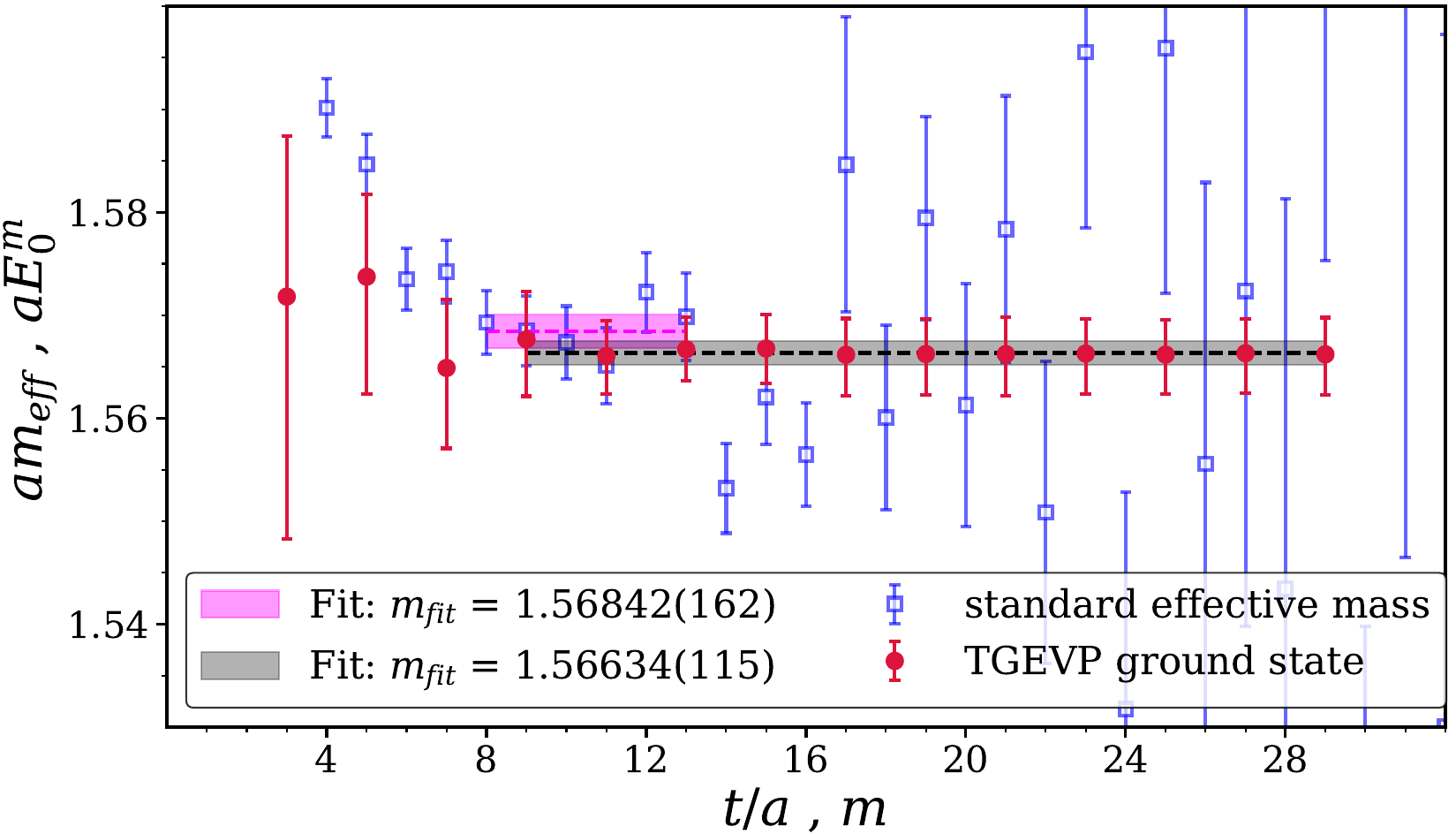}
\includegraphics[width=0.46\textwidth, height=0.26\textwidth]{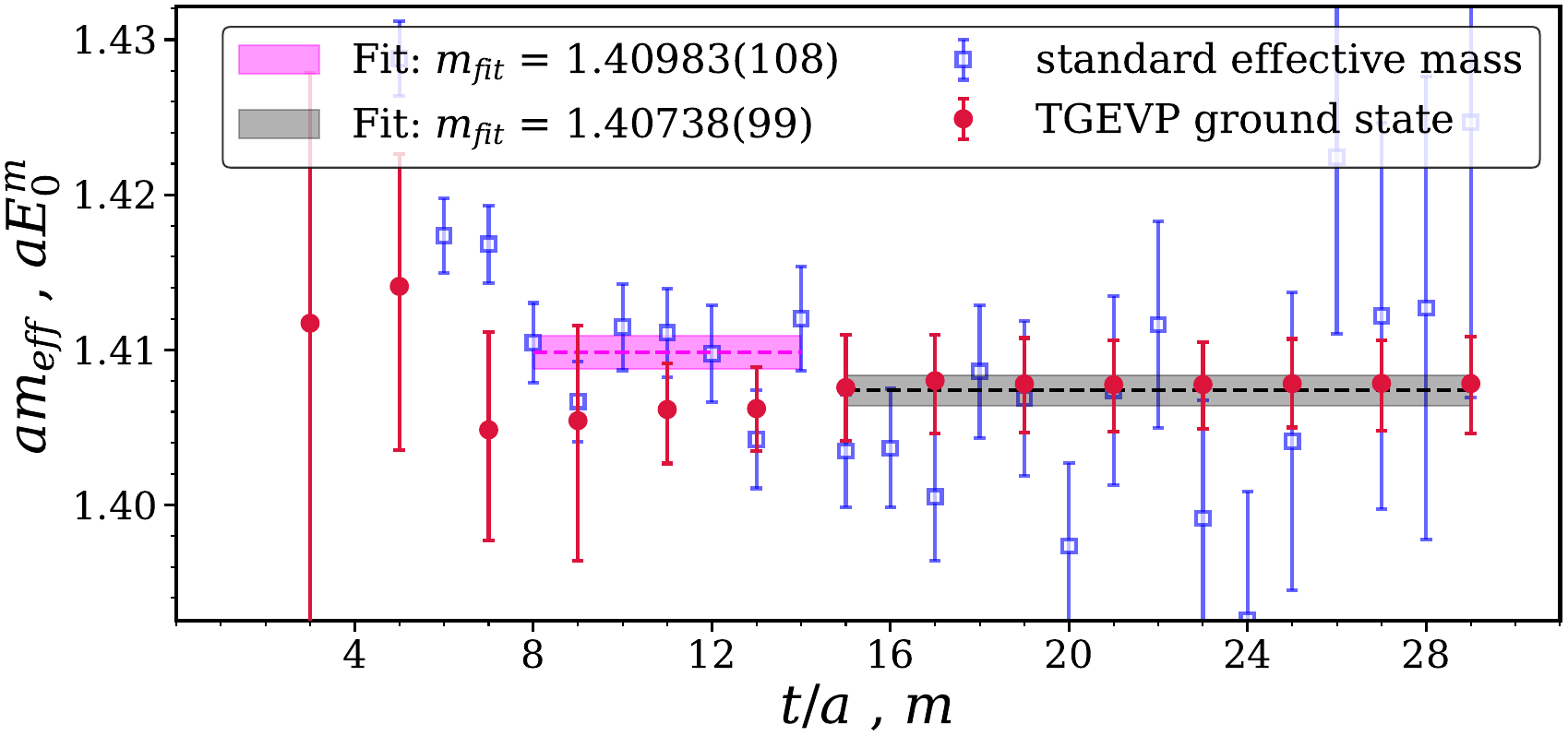}
        \caption{Similar to Fig. \ref{fg:charm1}, corresponding to smeared-smeared two-point functions for the doubly charmed baryon $\Xi_{cc}$, computed at two different lattice spacings with number of measurements ($n_{meas}$) $1612$ and $2000$, respectively.}
        \label{fg:heavy_baryon1}
  \end{figure}

\begin{figure}[htb!]
\includegraphics[width=0.46\textwidth, height=0.26\textwidth]{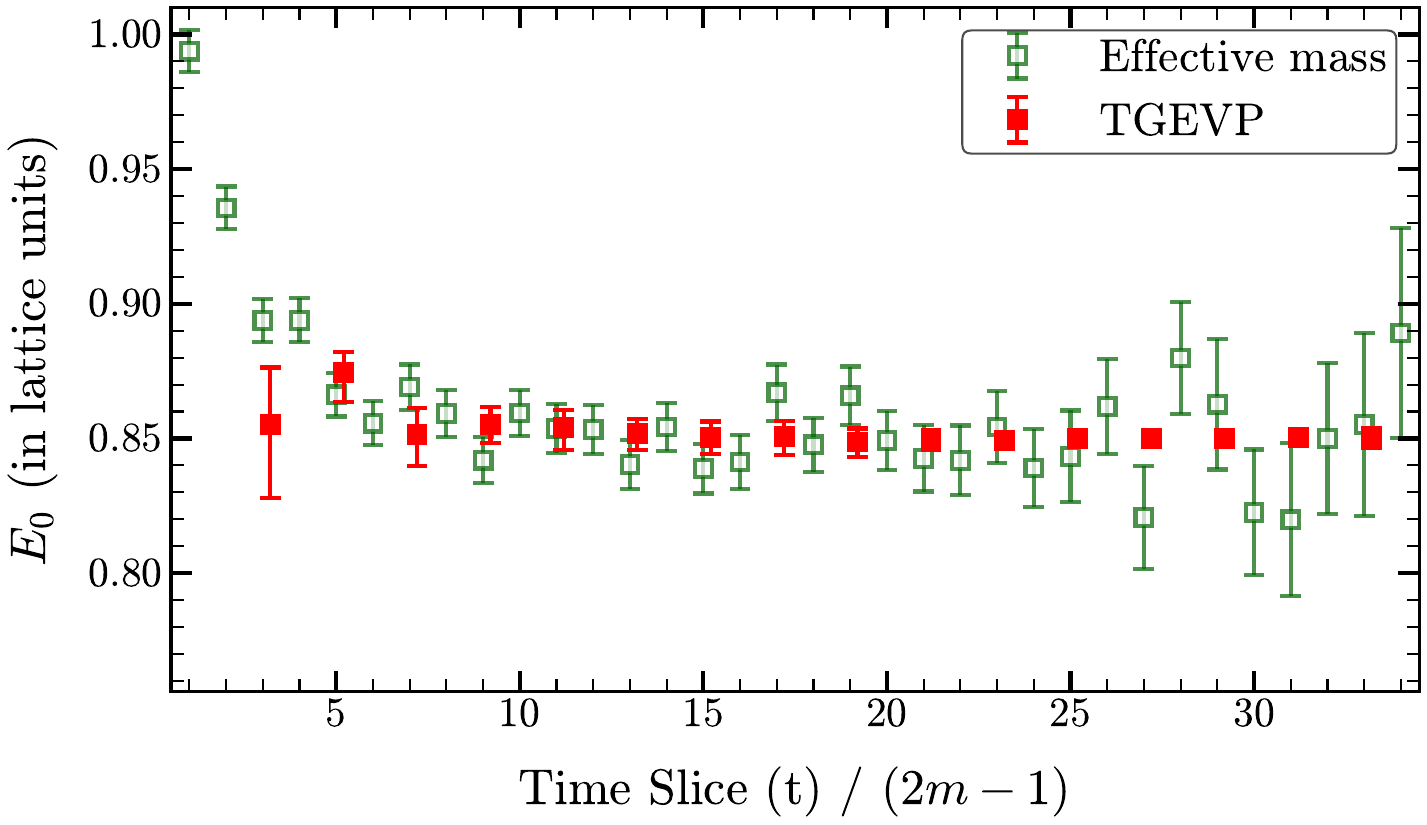}
\includegraphics[width=0.46\textwidth, height=0.26\textwidth]{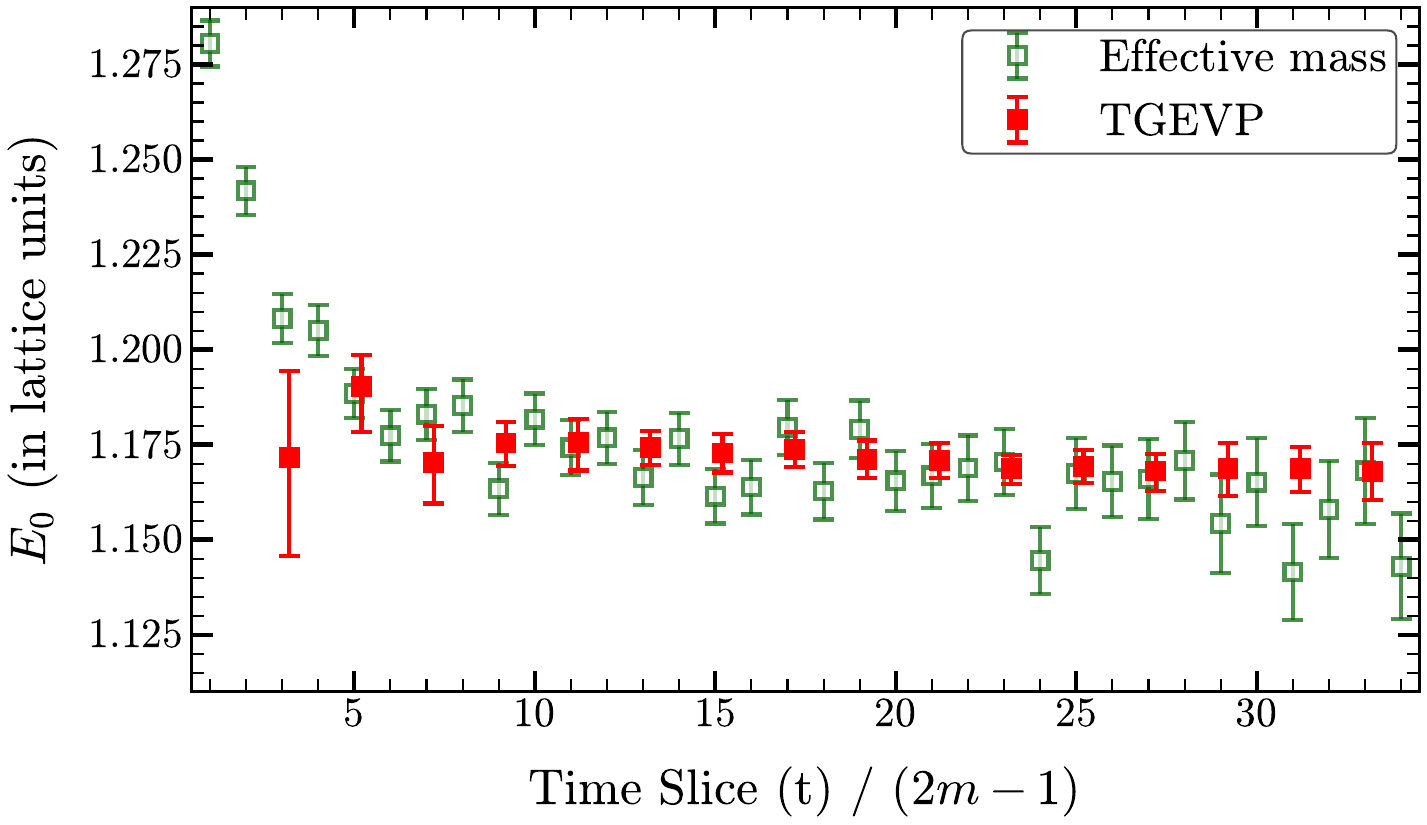}
        \caption{Similar to Fig. \ref{fg:charm1}, corresponding to smeared-smeared two-point functions for the singly and doubly charmed baryon $\Omega_c$ (top) and $\Omega_{cc}$ (bottom), with number of measurements ($n_{meas}$) $1009$.}
        \label{fg:heavy_baryon2}
  \end{figure}

After discussing mesons, we now move on to application of TGEVP method in extracting energy levels for baryons, which are generally noisier than those of mesons. This makes it particularly important to assess the efficiency of the TGEVP method in such cases. In Figs. \ref{fg:heavy_baryon1} and \ref{fg:heavy_baryon2}, we show lowest energy levels extracted by TGEVP method for the charmed-baryons, $\Xi_{cc}$ (Fig. \ref{fg:heavy_baryon1}), $\Omega_{c}$ and $\Omega_{cc}$ (Fig. \ref{fg:heavy_baryon2}), and compare those with their respective effective masses. The results are obtained from their respective two point functions computed on CLS ensembles generated on $N_f = 2+1$  nonperturbatively improved Wilson fermions \cite{PhysRevD.108.034512}.  It again very clearly shows the stability of the lowest eigenvalue over the iteration numbers, as opposed to unstable and fluctuating effective mass data. 
In such cases, the TGEVP results clearly demonstrate superiority over the effective mass method and related single-exponential fits, showing that it can reliably extract the ground state while avoiding the systematic uncertainties associated with the choice of fit ranges in traditional fitting procedures.

\subsection{Case III: Nucleon and Nuclei}\label{results:nuclei}
\begin{figure}[htb!]
  \includegraphics[width=0.4\textwidth, height=0.32\textwidth]{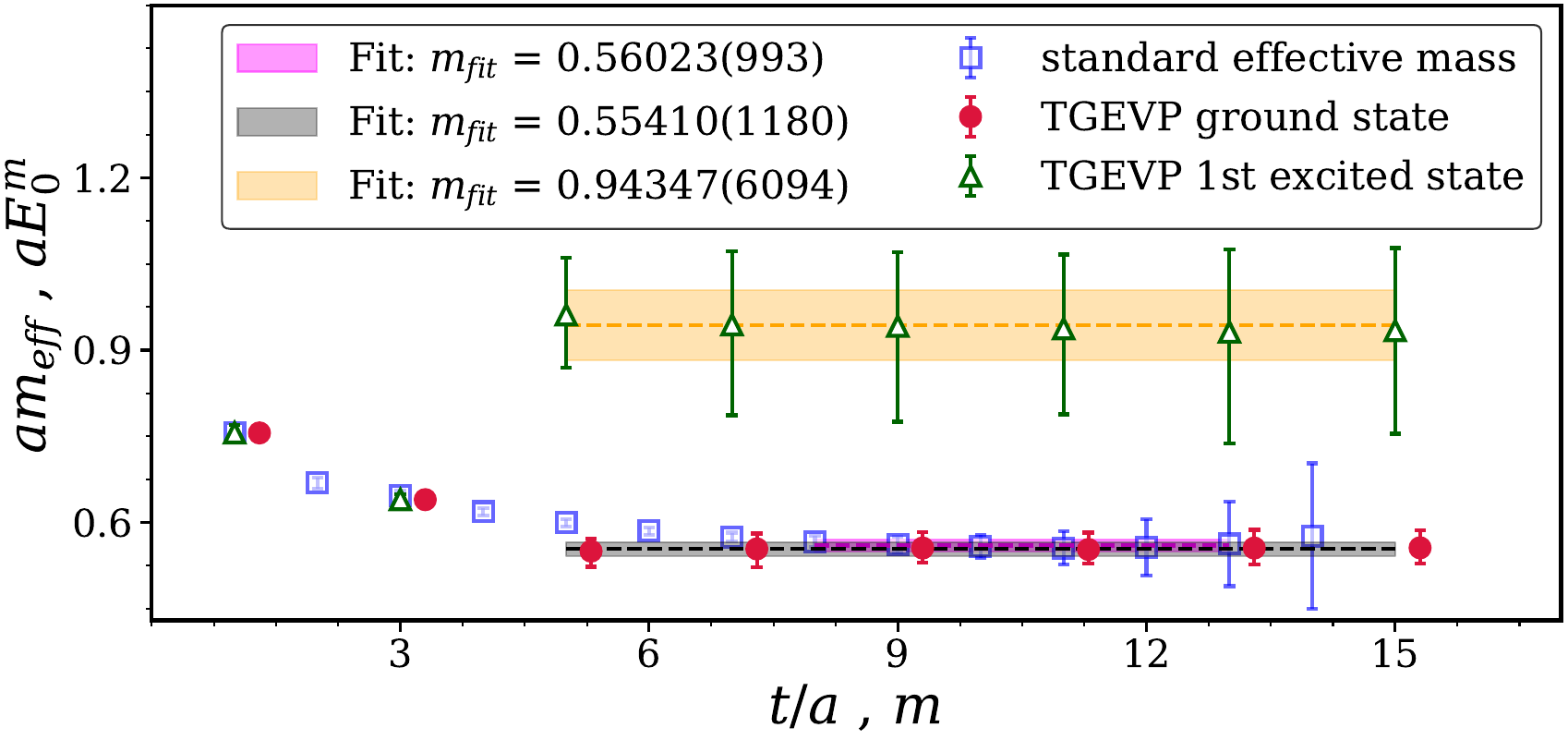}
        \caption{First two energy levels, corresponding to a smeared-point two-point function of nucleon $(n_{meas} = 300)$.
       Note the second level may correspond to the Roper resonance.}
        \label{fg:nu1}
  \end{figure}

In general, the signal-to-noise ratio (SNR) decreases at lighter pion masses, making it especially important to extract energy levels reliably at those quark masses in order to obtain physically meaningful results. 
In Fig. \ref{fg:nu1}, we show the lowest energy level obtained from a smeared-point two-point function of the nucleon computed at a near physical pion mass. 
Note that while the results from the two estimations are consistent, the eigenvalues at the larger iterations have smaller errors compared to errors in the effective mass at the larger times.

\begin{figure}[htb!]
\includegraphics[width=0.48\textwidth, height=0.26\textwidth]{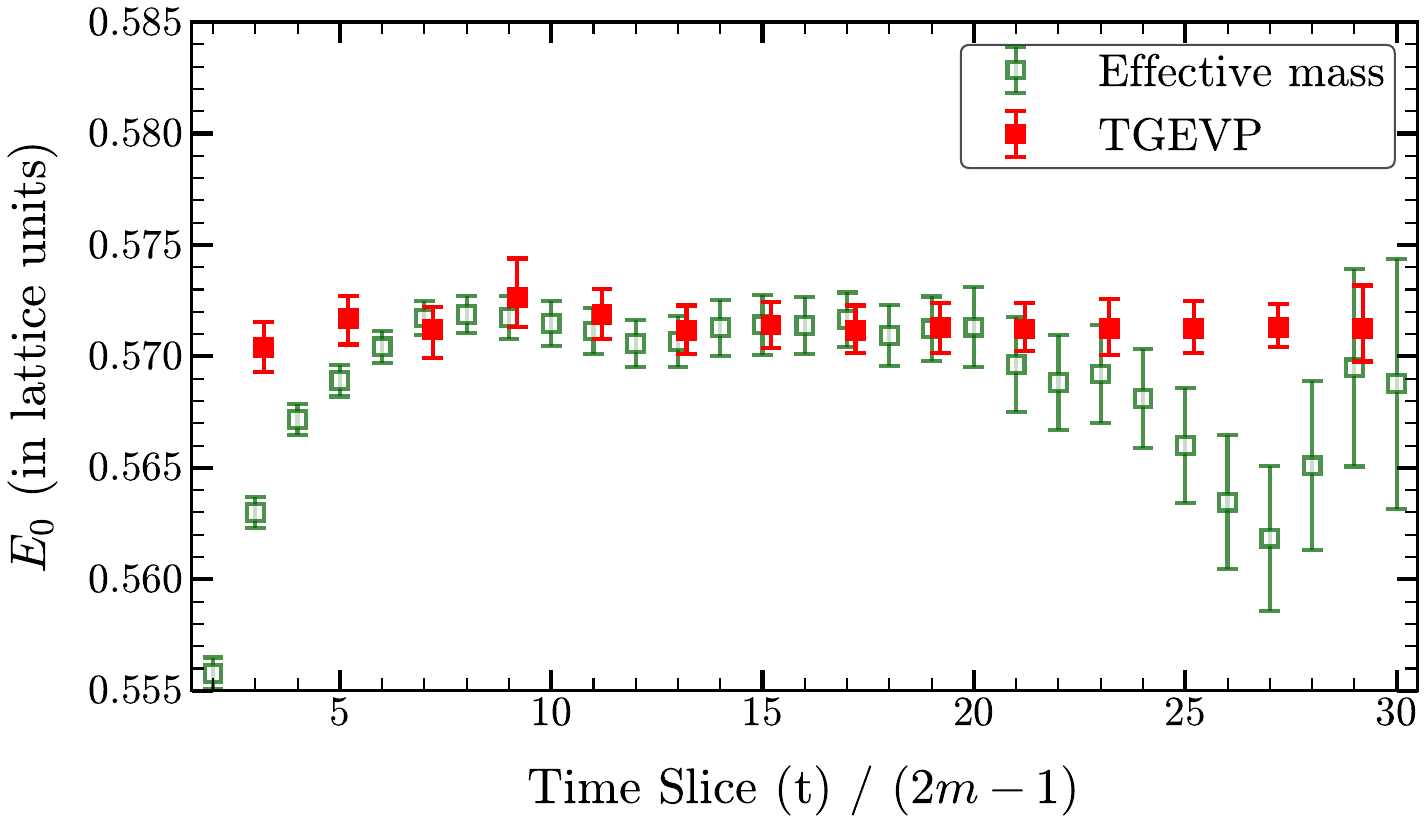}
\includegraphics[width=0.48\textwidth, height=0.26\textwidth]{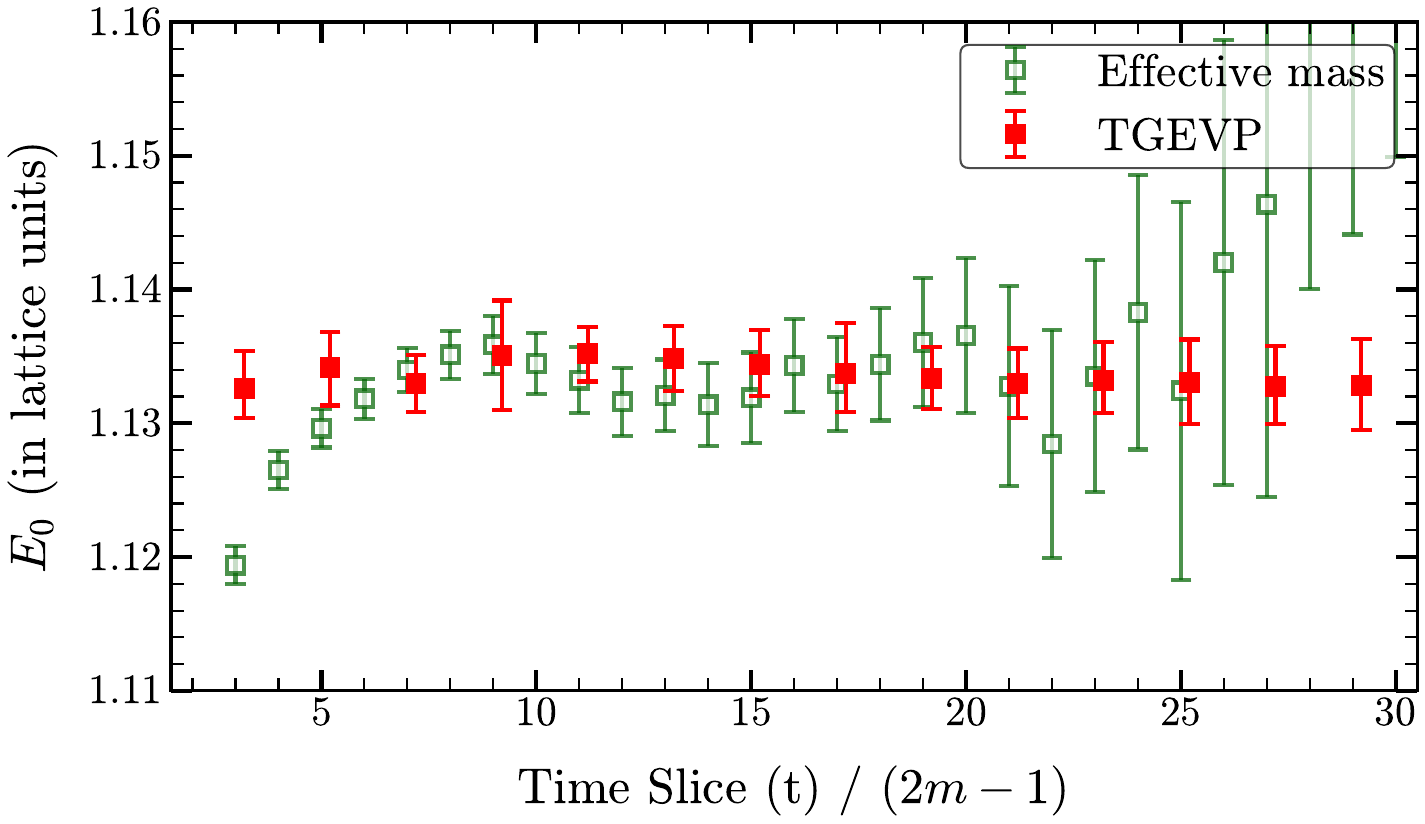}
        \caption{Comparison of the results between the regular effective mass (green squares) and TGEVP ground state (red circles) corresponding to two-point functions of a Nucleon (top pane) and Deuteron (bottom pane) with the quark masses, $m_u = m_d = m_s$ $(n_{meas} = 9600)$.}
        \label{fg:mult_nu1}
  \end{figure}

In contrast to regular hadrons, the study of nuclei using lattice QCD remains at a relatively early stage of development. One of the main bottlenecks in that is the lack of reliability in extracting the energy levels, even for the ground states of low-lying nuclei \cite{PhysRevD.87.114512,PhysRevD.90.034503,PhysRevD.93.094507,Chakraborty:2024oym}. This is primarily due to poor SNR  in the two-point correlation functions at lighter quark masses, resulting very noisy and small plateau at a large distance in the effective mass plots. As a result, statistical error together with fit-uncertainties, becomes large making the whole process unreliable. 
It is thus quite important to check if TGEVP method can be helpful in this. In Fig. \ref{fg:mult_nu1}, we show the effective mass of the ground state and lowest eigenvalue of TGEVP spectrum for 
nucleon and $^2$H.
The associated two-point functions are obtained with clover valence quarks on the $N_f = 2+1$ HISQ gauge ensembles ($64^3 \times 64, ~ a = 0.076$ fm) at the quark masses, $m_u = m_d = m_s$. 
Additional results for $^2$H, $^3$He, $^4$He and $^7$Li are provided in Fig. \ref{fg:mult_nu2} of Appendix \ref{app:additional_results}, where overlap valence quarks on HISQ gauge ensembles are utilized to compute the two point functions ($N_f = 2+1+1$ HISQ gauge ensembles ($48^3 \times 144, ~ a = 0.0582$ fm) at the quark masses, $m_u = m_d = m_s$: left panel, and $m_u = m_d = m_c$: right panel).
The method for computing the correlation functions for these nuclei has been provided in Ref. \cite{Chakraborty:2024oym}. As can be observed that the effective masses of these nuclei exhibit significant noise at large times, and sometime also from the beginning (Fig. \ref{fg:mult_nu2}), leading to substantial fit-window error in traditional fitting methods. In contrast, the TGEVP eigenvalues remain stable across numerous iterations. With a large number of measurements, it is expected that errors in eigenvalues in each iteration can be reduced. 
 However, it is also necessary to assess the method's effectiveness, with the availability of higher statistics, for the cases with closely spaced states carrying finite momentum, particularly when dealing with varied source-sink setups in nuclear systems. Another point of concern is correlation over iterations which we will discuss next. Nevertheless, as the results suggest, this method holds excellent promise for extracting reliable energy levels for nuclei in the future.

\subsection{Correlation in TGEVP eigenvalues}\label{corltn}
The eigenvalues obtained at different iterations of the TGEVP method are inherently correlated, as each iteration builds upon the results of the previous one through random sampling.
Addressing this correlation mathematically, where data in each step in a matrix are generated randomly from the previous step, is non-trivial.
In Refs. \cite{Wagman:2024rid, Ostmeyer:2024qgu} a double-bootstraping is done to address that.
In this work, we define the bootstrap confidence interval by using the $68$th percentile. 

\begin{figure}[htb!]
\includegraphics[scale=0.33]{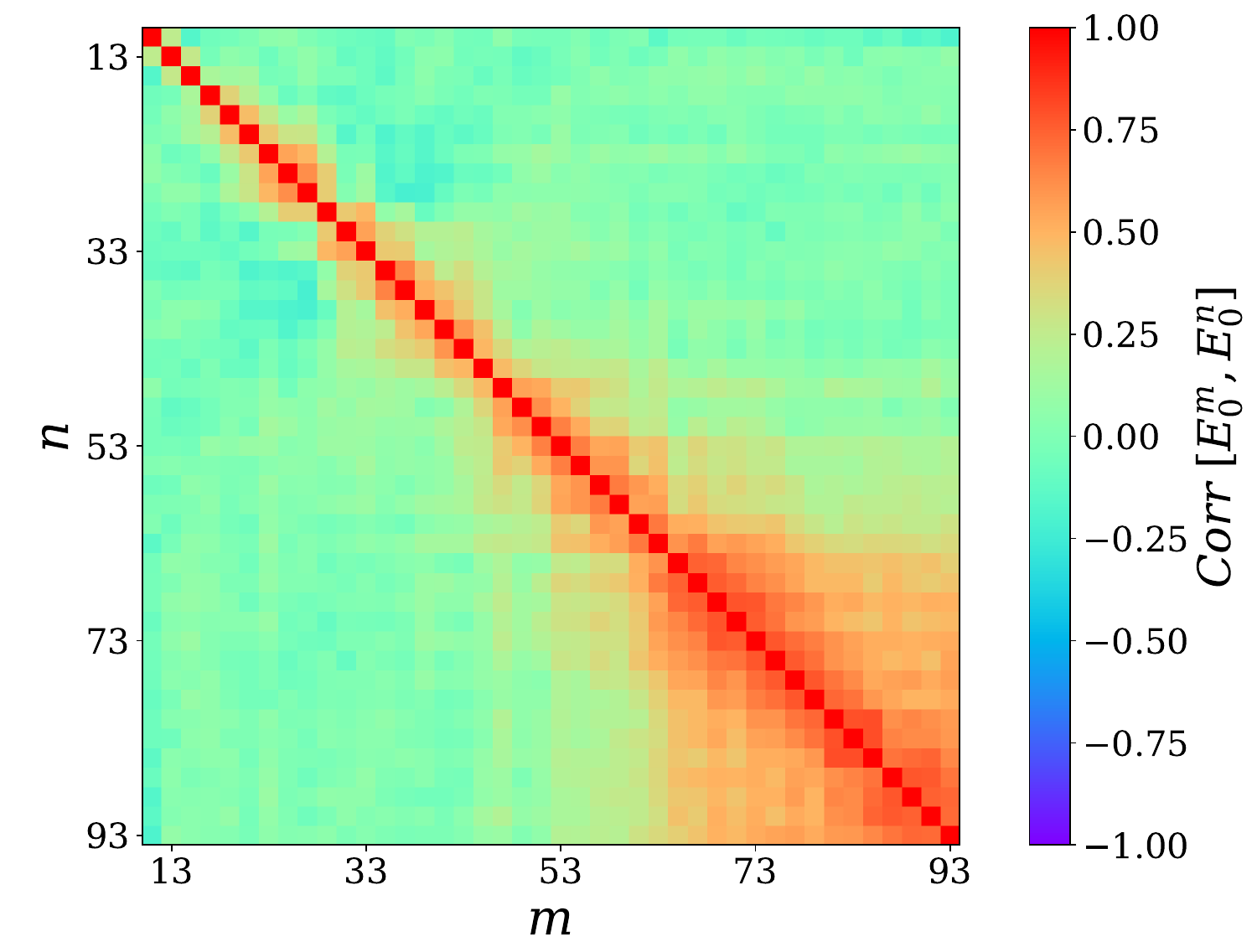}
        \caption{Color-coded correlation matrix showing the correlations between the eigenvalues across iterations. This dataset corresponds to the two-point correlation functions of the lowest $0^{-}$ state ($\eta_c$ meson), showed in the top pane of Fig. \ref{fg:charm1}.}
        \label{fg:cor_mat1}
  \end{figure}

To assess the extent of this correlation, we compute the associated correlation matrix using the bootstrap method from the eigenvalues that fall within the 68th percentile interval.  The first construct the covariance matrix as,
\begin{equation}
    \text{Cov}(E_n^{i}, E_n^{j}) = \frac{1}{\tilde{N} - 1} \sum_{b=1}^{\tilde{N}} \left(E_n^{(i,b)} - \bar{E}_n^{i}\right)\left(E_n^{(j,b)} - \bar{E}_n^{j}\right), \label{covariance}
\end{equation}
where $E_n^{(i,b)}$ and $E_n^{(j,b)}$ are the \(b\)-th bootstrap samples for $n$-th energy eigenvalues at the $i$ and $j$-th TGEVP iterations, and $\bar{E}_n^{i}$ and $\bar{E}_n^{j}$ are their respective bootstrap means. In Eq.~(\ref{covariance}), $\tilde{N}$ is the number of bootstrap samples for which the energy eigenvalues, $E_n^{(i,b)}$ and $E_n^{(j,b)}$, fall within the 68th percentile. The normalized correlation matrix is then constructed as,
\begin{equation}
    \text{Corr}\left[E_n^{i}, E_n^{j}\right] = \frac{\text{Cov}(E_n^i, E_n^j)}{\sqrt{\text{Cov}\left(E_n^{i}, E_n^{i}\right)\text{Cov}\left(E_n^{j}, E_n^{j}\right)}}\,.
\end{equation}
We present here this correlation matrix in a color-coded format for some of the above correlators that we analyzed. In Fig. \ref{fg:cor_mat1}, we show such a color-coded correlation-plot for the two-point correlation functions for $\eta_c$ that was shown in the top pane of Fig. \ref{fg:charm1}. As observed, the correlation of eigenvalues across iterations is insignificant at the beginning and increases towards the end. Particularly in the fit region (Fig. \ref{fg:charm1}), it remains small enough to justify a correlated fitting with a constant term. 

\begin{figure}[htb!]
\includegraphics[scale=0.33]{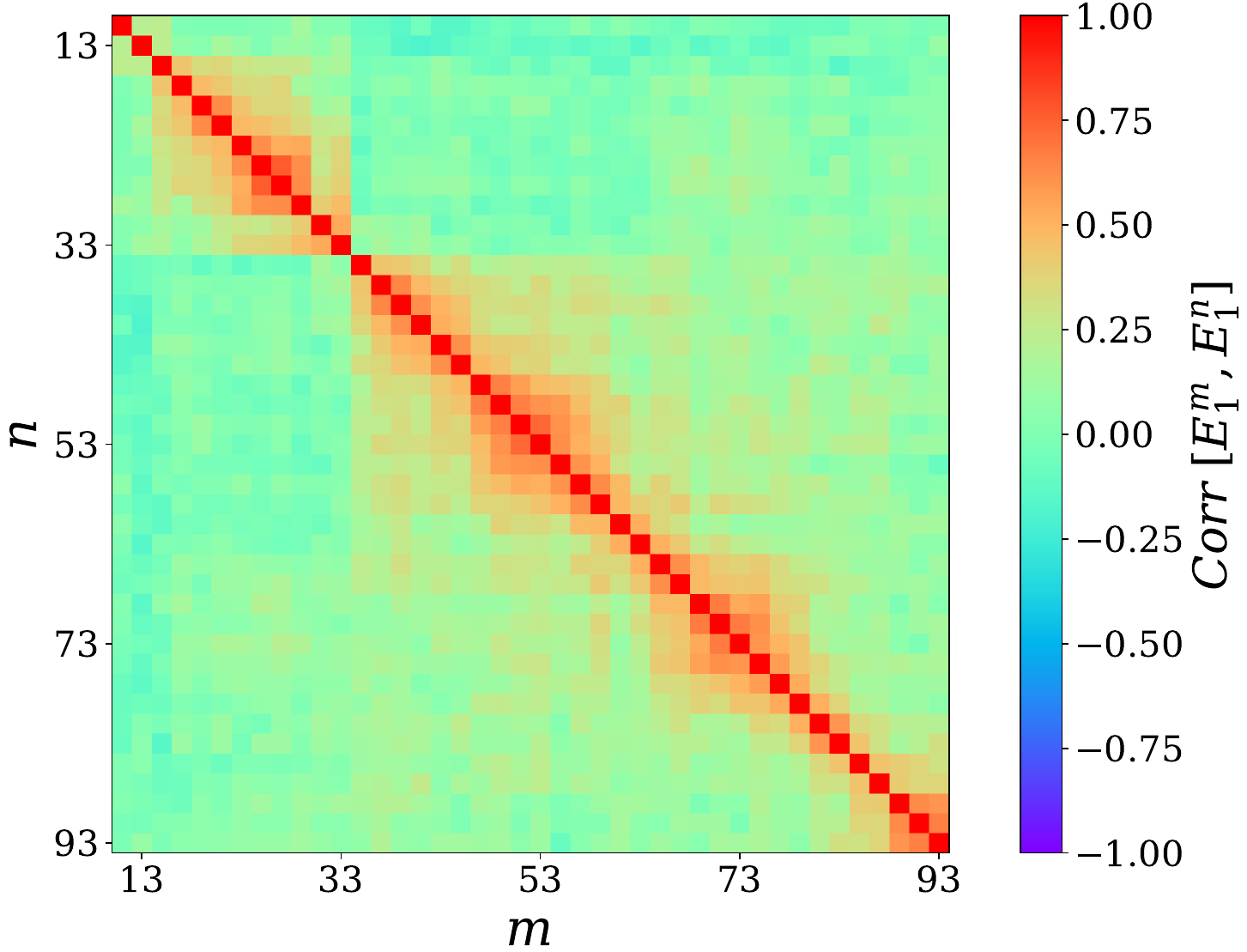}
        \caption{Correlation matrix plot corresponding to the second eigenvalue of the two-point correlation function of $1^-$ state in charmonia  ($J/\psi(2S)$ state, corresponding to the first excited state of the top write plot of Fig. \ref{fg:charm2}}
        \label{fg:cor_mat2}
  \end{figure}

\begin{figure}[htb!]
\includegraphics[scale=0.33]{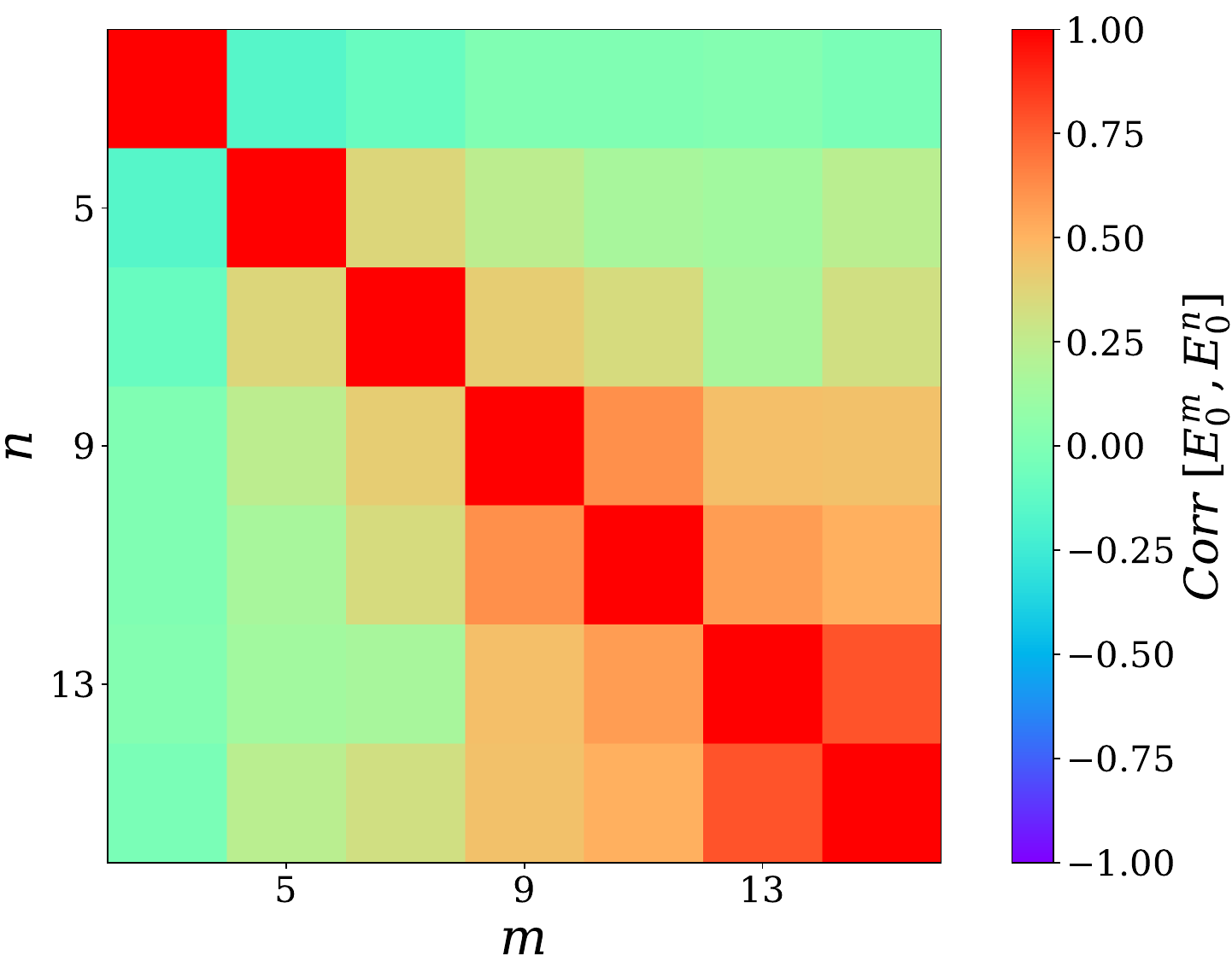}
\includegraphics[scale=0.33]{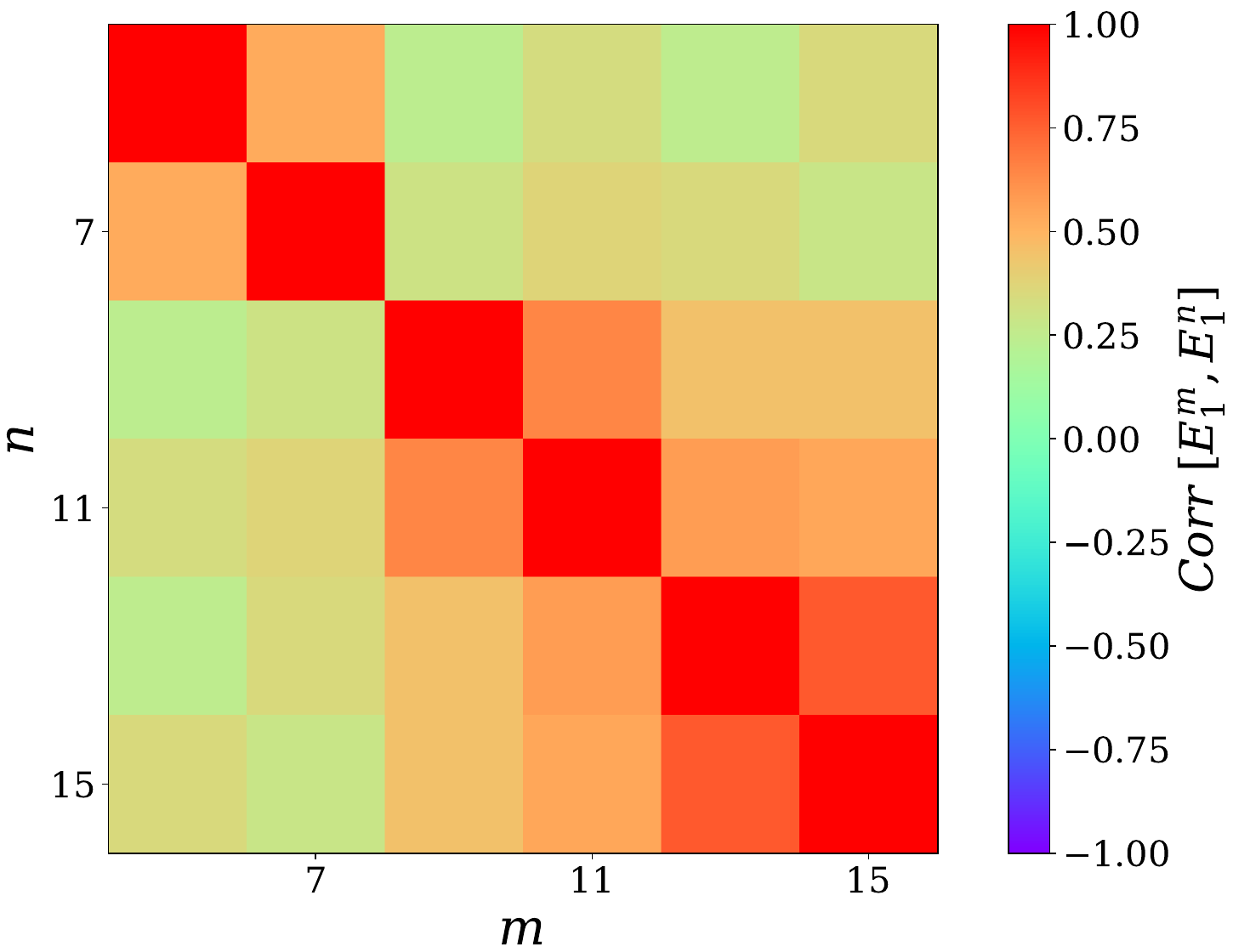}
        \caption{Correlation matrix plot for the first two eigenvalues of the nucleon correlators corresponding to the two states shown in Fig. \ref{fg:nu1}.}
        \label{fg:cor_mat3}
  \end{figure}

In general, extracting higher energy levels is more challenging than extracting the ground state. Therefore, it is crucial to examine the correlations for the excited states to ensure they can be reliably obtained using the TGEVP method.
 In Fig. \ref{fg:cor_mat2}, we show correlation matrix corresponding to the excited state of $1^{-}$ state in charmonia, This corresponds to the second eigenvalue of the top right plot of Fig. \ref{fg:charm2}. As can be seen that correlation is somewhat larger compared at the fit window to that of the lowest eigenvalues shown in Fig. \ref{fg:charm1}. The higher eigenvalues corresponding to other correlation functions also show the same pattern of higher correlation than that present in the lowest eigenvalue.  In Fig. \ref{fg:cor_mat3} we depict the correlation matrix plot of the first two eigenvalues of the nucleon two-point functions corresponding to the two states shown in Fig. \ref{fg:nu1}.
Here again, the second state, which may correspond to the Roper resonance, shows more correlation than the ground state.

In Fig. \ref{fg:cor_mat4} of Appendix \ref{app:additional_results} we show additional correlation matrix for the light nuclei. The top pane corresponds to Deuteron correlators at $m_u = m_d = m_s$ (left) and $m_u = m_d = m_c$ (right), respectively. The bottom pane represents the same for $^4$He. It clearly shows that correlation increases at the lighter quark mass and also increases with the atomic number. The eigenvalues corresponding to $^4$He correlators at the strange quark mass are correlated to a large extent, and increases further as the quark mass become lighter. 

This analysis leads us to the following observations:
\begin{enumerate}[wide, labelwidth=!, labelindent=0pt]
\item If the statistics is high and the effective mass is reasonably good we do not find much correlation of eigenvalues across iterations. In that case fitting these eigenvalues with a constant term will be reliable and provide better estimate compared to exponential fitting. This could be a good choice in multi-operator GEVP setup, that is merging of GEVP and TGEVP together. We will explore that in future
\item For the higher eigenvalues, correlations are larger than that for the case of lowest eigenvalue. With higher statistics we find the correlation in the higher eigenvalues also decreases.
\item For multi-nucleons we find correlations to be higher than the single nucleon. This is a concern and needs to be addressed properly in a multinucleon calculation, particularly considering the presence of dense spectrum. It remains to be seen how this correlation evolves with higher statistics in these systems.
\end{enumerate}

\begin{figure*}[htb!]
\includegraphics[scale=0.33]{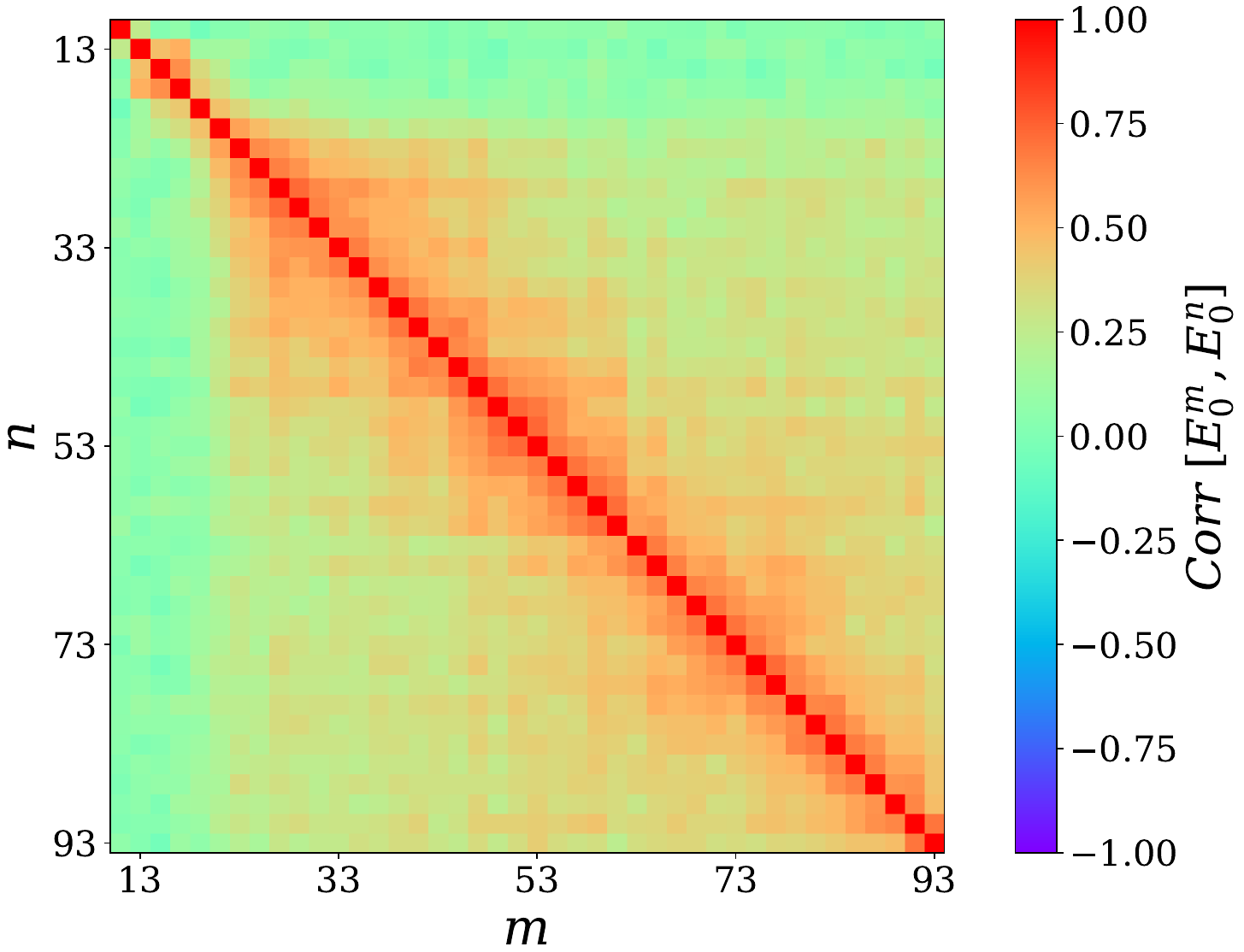}
\includegraphics[scale=0.33]{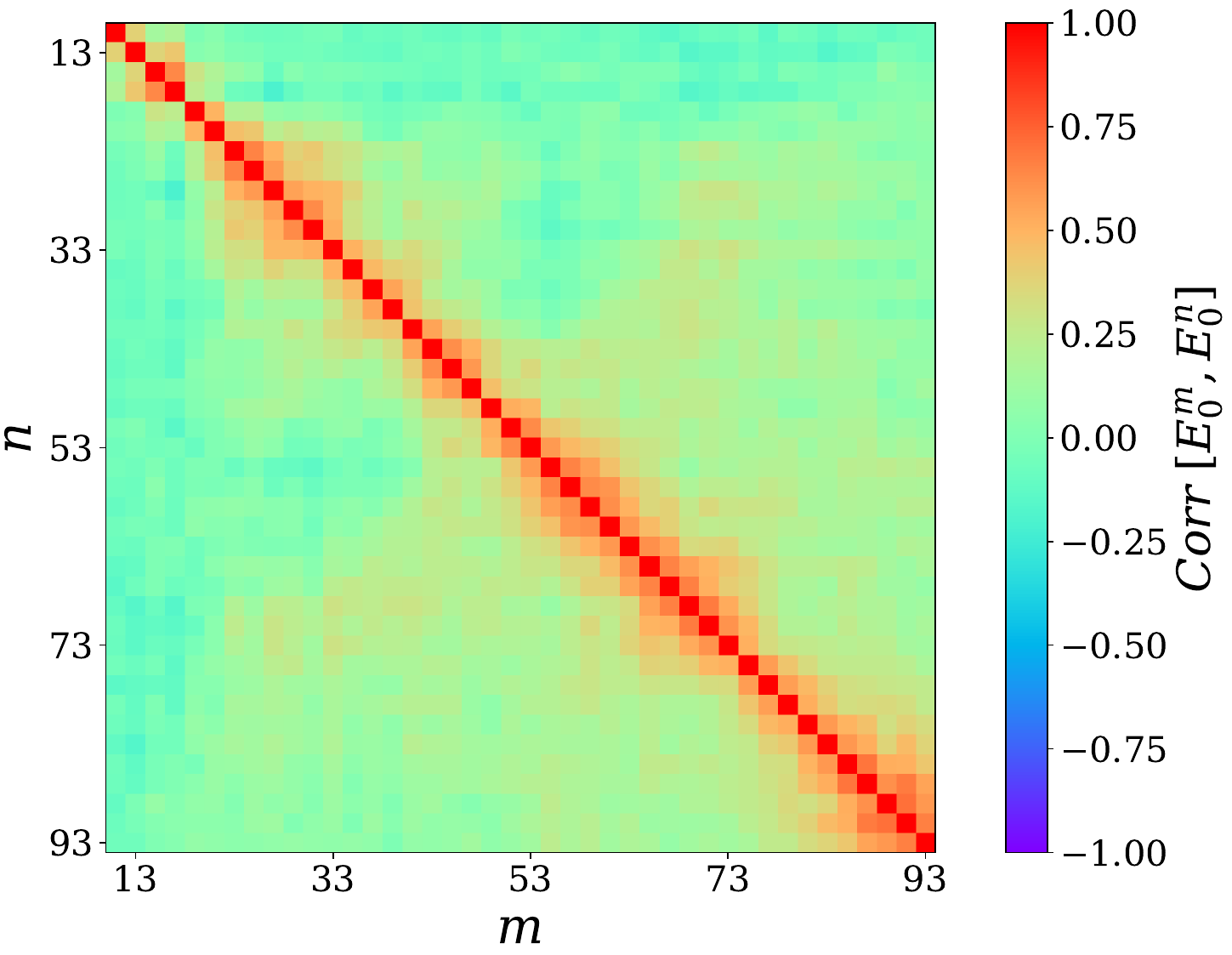}
        \caption{Correlation matrix plot corresponding to the lowest eigenvalues of  $\Omega_c$ (left) and $\Omega_{cc}$ (right), corresponding to Fig.\ref{fg:heavy_baryon2}}
        \label{fg:cor_mat4}
  \end{figure*}

\begin{figure*}[htb!]
\includegraphics[scale=0.45]{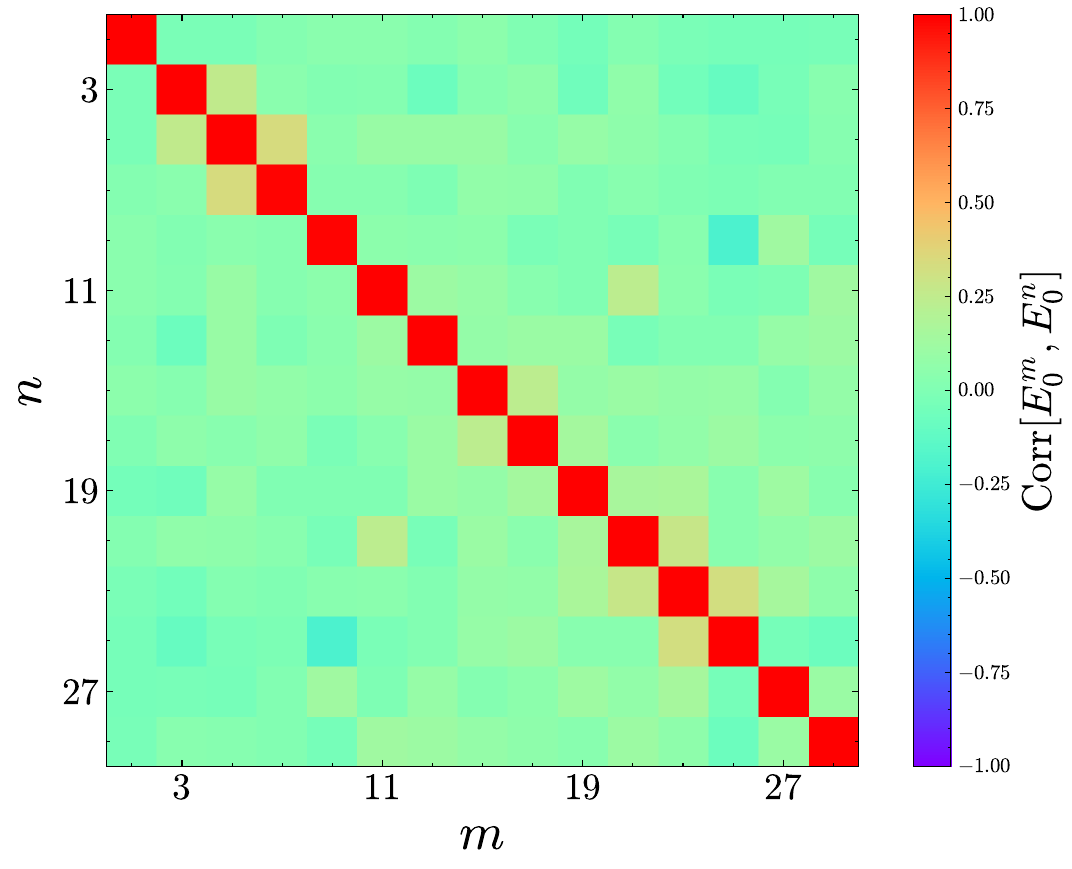}
\includegraphics[scale=0.45]{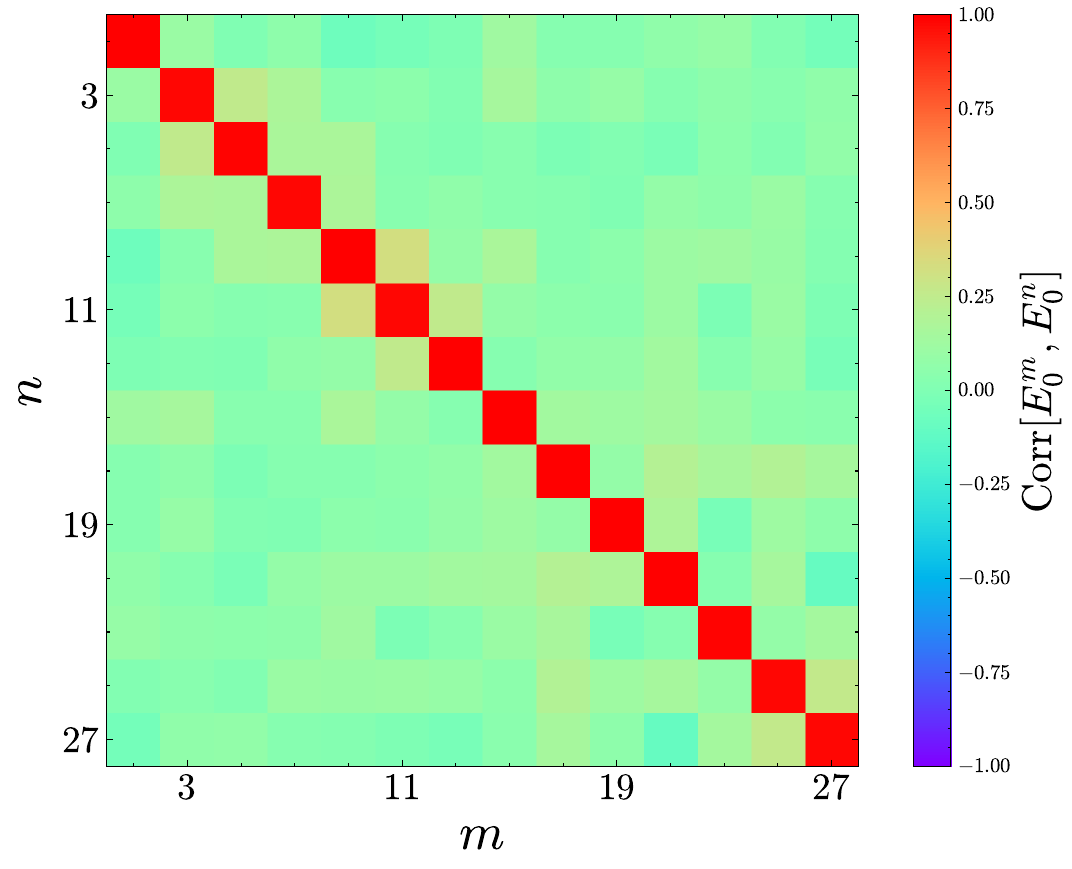}
        \caption{Correlation matrix plot corresponding to the lowest eigenvalues of  Nucleon and $^2$H, corresponding to Fig. \ref{fg:mult_nu1}.}
        \label{fg:cor_mat5}
  \end{figure*} 


\subsection{\label{subsec:overlap}Overlap Factors}
Along with the extraction of energy levels, determining the overlap factors of an operator with a specific state is crucial, as such factors are related to the decay properties and internal structure of the states corresponding to the energy levels.
The two-point correlation function can be defined in terms of the transfer matrix $\hat{\mathcal{T}}$ as, 
\begin{eqnarray} C^{\psi}_{2\text{pt}}(t) &=& \langle\psi\vert \hat{\mathcal{T}}^t\vert \psi\rangle \nonumber\\
&=& \sum_{n,m}\langle \psi\vert n\rangle \langle n\vert\hat{\mathcal{T}}^t\vert m\rangle\langle m\vert\psi\rangle\nonumber\\
&=& \sum_n\vert \langle \psi \vert n \rangle \vert^2 e^{-E_n t} \nonumber\\\
&=& \sum_n\vert \mathcal{Z}^{\psi}_n \vert^2 e^{-E_n t}, \label{2pt_spec_decomp}
\end{eqnarray} 
where in Eq.~(\ref{2pt_spec_decomp}), we have inserted a complete set of states $\mathrm{1}=\sum_n\vert n\rangle \langle n\vert = \mathrm{1}$, with $\vert n \rangle$ denoting the orthonormal eigenstates of the lattice Hamiltonian (and thus of the transfer matrix). The overlap factors $\mathcal{Z}^{\psi}_n$ are defined as, $\mathcal{Z}^{\psi}_n = \langle \psi\vert n\rangle$.

Similarly, the three-point function used in lattice QCD calculations to investigate the matrix elements of a current between states, with a current insertion at time  $\tau$ between source (at $t = 0$) and sink (at $t = t$), is given by,
\begin{eqnarray} 
C_{3\text{pt}}(t,\tau) &=& \langle \Omega\vert \mathcal{O}{\psi^\prime}(t) J(\tau) \mathcal{O}{\psi}(0) \vert\Omega\rangle\nonumber\\ 
&=& \langle\psi^\prime\vert \hat{\mathcal{T}}^{t-\tau} J(0) \hat{\mathcal{T}}^{\tau} \vert\psi\rangle\nonumber\\
&=& \sum_{m,n}\langle\psi^{\prime}\vert n \rangle \langle n \vert J(0) \vert m\rangle \langle m \vert\psi\rangle e^{-E_n(t-\tau)} e^{-E_m\tau}\nonumber\\
&=& \sum_{m,n} \mathcal{Z}^{\psi^\prime}_n \left( \mathcal{Z}^{\psi}_m \right)^\ast J_{nm}~ e^{-E_n t - \Delta E_{nm} \tau}, \label{3pt_spec_decomp} 
\end{eqnarray} 
where $J_{nm}$ is the matrix element of the current $J$ between the states $\vert n \rangle$ and $\vert m \rangle$, whereas $\Delta E_{nm}=E_m-E_n$.

By solving the generalized eigenvalue problem (GEVP) associated with the two-point function $C^{\psi}_{2\text{pt}}(t)$, as described in section~(\ref{sec:methodology}), one can express the energy eigenstates $\vert n\rangle$ in terms of normalized vectors within the Krylov space. 
Specifically, by solving the GEVP for a matrix of size $m \times m$, the eigenvector $\vec{v}_m^i$, corresponding to the $i$-th eigenvalue defines the associated state vector as,
\begin{eqnarray} 
\vert i \rangle_m = \sum_{a=1}^m (v_m^i)_a \frac{\hat{\mathcal{T}}^a \vert\psi\rangle}{\sqrt{C_{2\text{pt}}^{\psi}(2a)}}. \label{eig_vec_KS_decomp} 
\end{eqnarray} 
The state vectors, $\vert i \rangle_m$, as defined above satisfy the expected orthonormality conditions. The corresponding overlap factors, $\mathcal{Z}i^m$, can then be determined as,
\begin{equation}
\mathcal{Z}_i^m  =  \langle \psi \vert i \rangle_m 
 = \sum_{a=1}^m (v_m^i)_a \frac{C^{\psi}_{2\text{pt}}(a)}{\sqrt{C^{\psi}_{2\text{pt}}(2a)}}. \label{overlap}
\end{equation}
This overlap quantifies the projection of the interpolating operator onto the ground state, as reconstructed within a truncated Krylov subspace of dimension \( m \). 

Similarly, the matrix elements, $J_{ij} = \langle i \vert J \vert j \rangle$, can be computed using the state vectors, $\vert i\rangle_m$ and $\vert j \rangle_m$ as defined in Eq. (\ref{eig_vec_KS_decomp}), and then plugging that in Eq. (\ref{3pt_spec_decomp}), as below: 
\begin{eqnarray} 
J_{ij}^m &\equiv& _m\langle i \vert J \vert j \rangle_m \nonumber\\
&=& \sum_{a=1}^m \sum_{b=1}^m \frac{(v^j_m)_a (w^i_m)_b^\ast C_{3\text{pt}}(a+b,b)}{\sqrt{C_{2\text{pt}}^{\psi^\prime}(2b) C_{2\text{pt}}^{\psi}(2a)}}, 
\end{eqnarray} 
where $w^i_m$ and $v^j_m$ are the eigenvectors corresponding to the states $\vert i \rangle_m$ and $\vert j \rangle_m$, respectively. Note in the case where the transfer matrix is not Hermitian (for example, due to an asymmetric source-sink setup or due to finite-statistics data), it becomes necessary to use the right and left eigenvectors of the GEVP to construct $\vert i \rangle_m$ and $_m\langle i \vert$, respectively. A similar methodology for extracting overlap factors and matrix elements from two-point and three-point functions using the oblique Lanczos algorithm is discussed in Ref. \cite{Hackett:2024xnx}. Aside from differences in the treatment of spurious eigenvalues, the two setups are mathematically equivalent, as both utilize the same Krylov subspace basis vectors to approximate the energy eigenstates. However, the TGEVP setup is comparatively simpler in implementation: in the Lanczos approach, an additional conversion step is required to relate the eigenvectors to the original Krylov subspace basis vectors.

\begin{figure}[h!]
\vspace*{0.2in}
    \centering
    \includegraphics[scale=0.35]{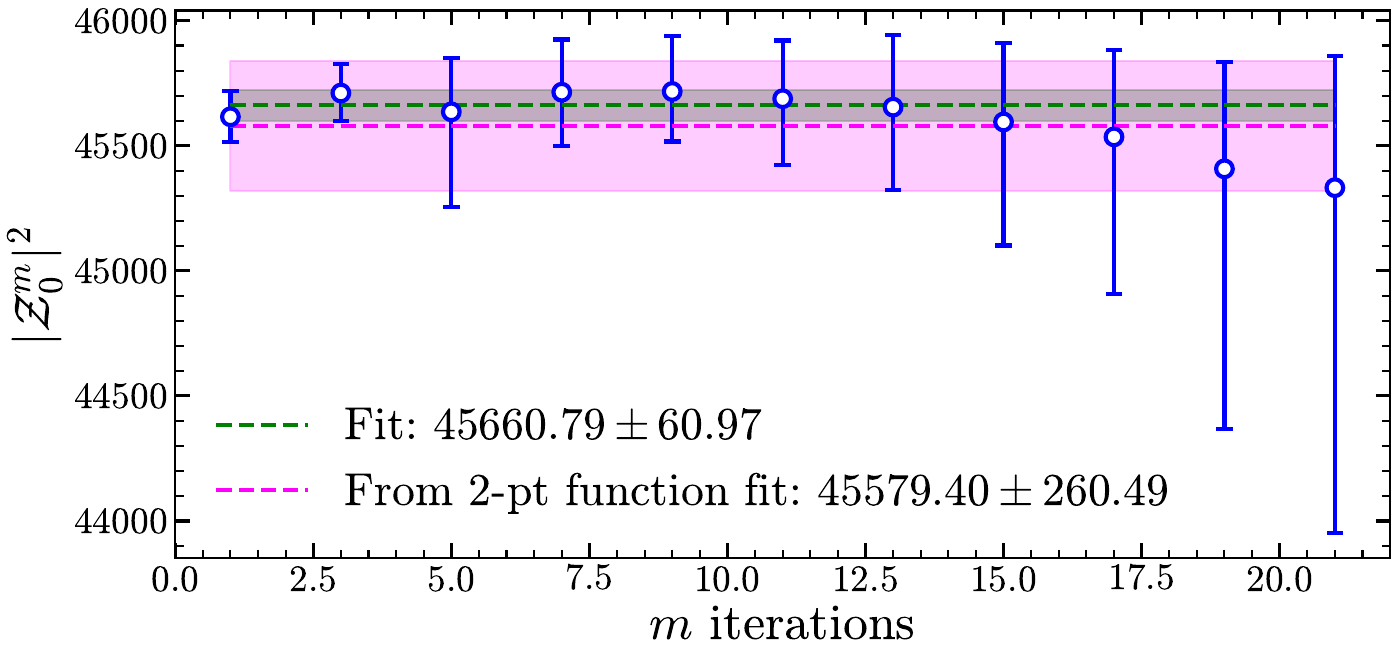}
    \caption{
    Ground-state overlap factor \( |\mathcal{Z}_0^m|^2 \) as a function of the GEVP matrix size \( m \), extracted from the nucleon two-point correlation function. The light green band represents a linear fit to the GEVP-based estimates, while the magenta band shows the corresponding value obtained from a single-exponential fit to the same correlator. The agreement validates the effectiveness of the GEVP approach in isolating the ground-state contribution.
    }
    \label{fig:overlap_gevp}
\end{figure}

Below we show how the procedure described above can be employed  
in estimating the overlap factor \( |\mathcal{Z}_0^m|^2 \) for various values of the GEVP matrix size \( m \). We choose the example of the ground state of nucleon, corresponding to the Fig. (\ref{fig:prony_vs_tgevp}). Using Eq.~(\ref{overlap})  we extract the values of \( |\mathcal{Z}_0^m|^2 \) as a function of \( m \) and show the results in Fig.~\ref{fig:overlap_gevp}.
  A clear trend is observed, indicating a mild but consistent linear behavior with increasing \( m \), reflecting the improvement in isolating the ground-state contribution through the GEVP procedure.
To guide the eye, we perform a correlated constant fit to the data points under bootstrap, depicted as a light green band in the figure. For reference, the overlap factor obtained from a standard single-exponential fit to the same nucleon correlator is shown as a magenta band. The close agreement between the GEVP-based estimates and the direct fit result at moderate values of \( m \) demonstrates the reliability and consistency of the GEVP method in extracting ground-state overlap factors, even with a modest basis size. Further, note that the error in TGEVP method is comparatively smaller than the standard method.

\subsection{\label{subsec:TGEVP_GEVP} Block TGEVP}
The use of the generalized eigenvalue problem (GEVP) \cite{MICHAEL198558, Luscher:1990ck} with a large operator basis has become a standard method for extracting energy levels in contemporary lattice QCD calculations. The inclusion of TGEVP for correlation matrix with a large set of operators can bring added advantage in such process.  Below we discuss how the block TGEVP can be formulated and applied to extract energy levels. Similar framework using oblique Lanczos algorithm can found in Ref. \cite{Hackett:2024nbe}.
For a given basis of $n$ interpolating operators, $\{ \mathcal{O}_\alpha \}$, the  $n \times n$ correlation matrix, $\mathbf{C}(t)$, at time $t$,  constructed out of  two point correlation functions is given by,
\begin{equation}
    C_{\alpha\beta}(t) = \langle \Omega \vert \mathcal{O}_\alpha(t)\, \overline{\mathcal{O}}_\beta(0) \vert \Omega \rangle, \qquad \alpha, \beta = 1, \dots, n,
\end{equation}
The TGEVP framework can easily be extended to this matrix-valued case by promoting each scalar element of the transfer and overlap matrices to an $n \times n$ matrix block. As a result, the transfer matrix $\mathbf{T}^m$ and overlap matrix $\mathbf{V}^m$, at the Krylov order $m$, become block matrices of size $(nm) \times (nm)$, where the block elements of these matrices are given by,
\begin{align}
    \mathbf{T}^m_{ij} &= \mathbf{C}(2i)^{-1/2} \, \mathbf{C}(i + j + 1) \, \mathbf{C}(2j)^{-1/2}, \label{eq:blockT}\\
    \mathbf{V}^m_{ij} &= \mathbf{C}(2i)^{-1/2} \, \mathbf{C}(i + j) \, \mathbf{C}(2j)^{-1/2}, \label{eq:blockV}
\end{align}
for $i, j = 0, \dots, m-1$. Here, the square roots of the matrices and inverse operations are defined via spectral decomposition. The corresponding generalized eigenvalue problem takes the form,
\begin{equation}
    \sum_j \mathbf{T}^m_{ij} \, \mathbf{v}^{(n)}_j = \lambda_n \sum_j \mathbf{V}^m_{ij} \, \mathbf{v}^{(n)}_j,
\end{equation}
where the eigenvalues, $\lambda_n$, are related to the energy levels through
\begin{equation}
    E_n = -\ln \lambda_n.
\end{equation}
Compared to the scalar case (one element in $\mathbf{C}(t)$), the effective dimension of the Krylov space increases from $\lfloor n_T/2 \rfloor$ to $n \times \lfloor n_T/2 \rfloor$, where $n_T$ is the temporal extent of the lattice. This expansion enhances the resolution of excited states and improves ground-state isolation. Spurious solutions in this case can also be removed using the same methodology described in Sec.~(\ref{eigs_filtration}).

In the following we provide the efficacy of 
 the block TGEVP method, described above, to extract the lowest and the first excited state energy spectra. We choose the example of $\Omega_{ccc}$ baryon in the $\frac{3}{2}^+$ channel. Results from this method are compared with those obtained using the standard GEVP method. The analysis employs a $2 \times 2$ correlation matrix constructed from two variationally optimized interpolating operators, as described in Ref.~\cite{Dhindsa:2024erk}. The calculation uses overlap valence quarks on $2+1+1$ flavor HISQ sea configurations, generated by the MILC collaboration~\cite{MILC:2012znn}. Details of the operator basis construction can be found in Ref.~\cite{Dhindsa:2024erk}.

\begin{figure}[htbp]
\centering
\includegraphics[scale=0.30]{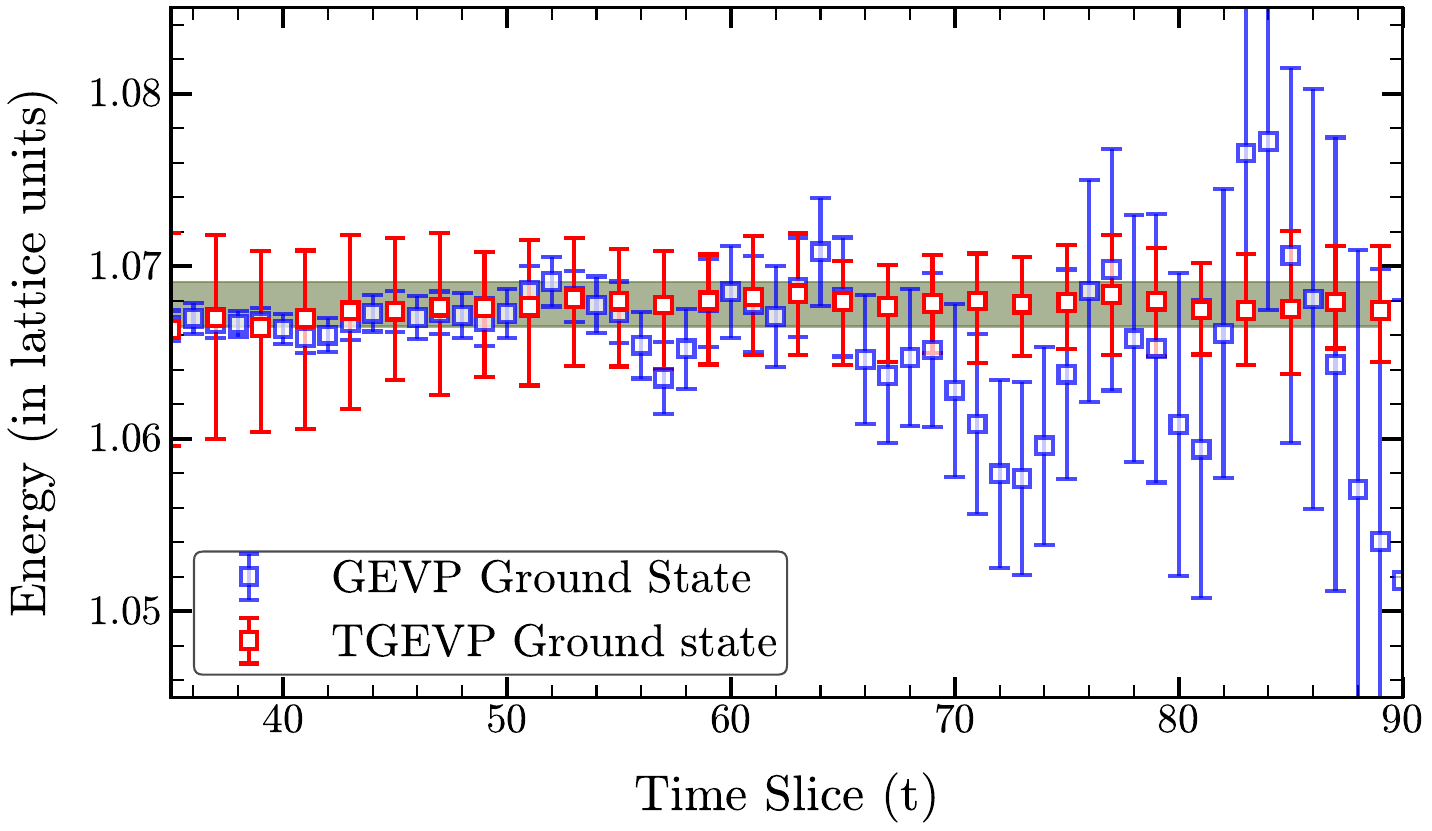}
\includegraphics[scale=0.30]{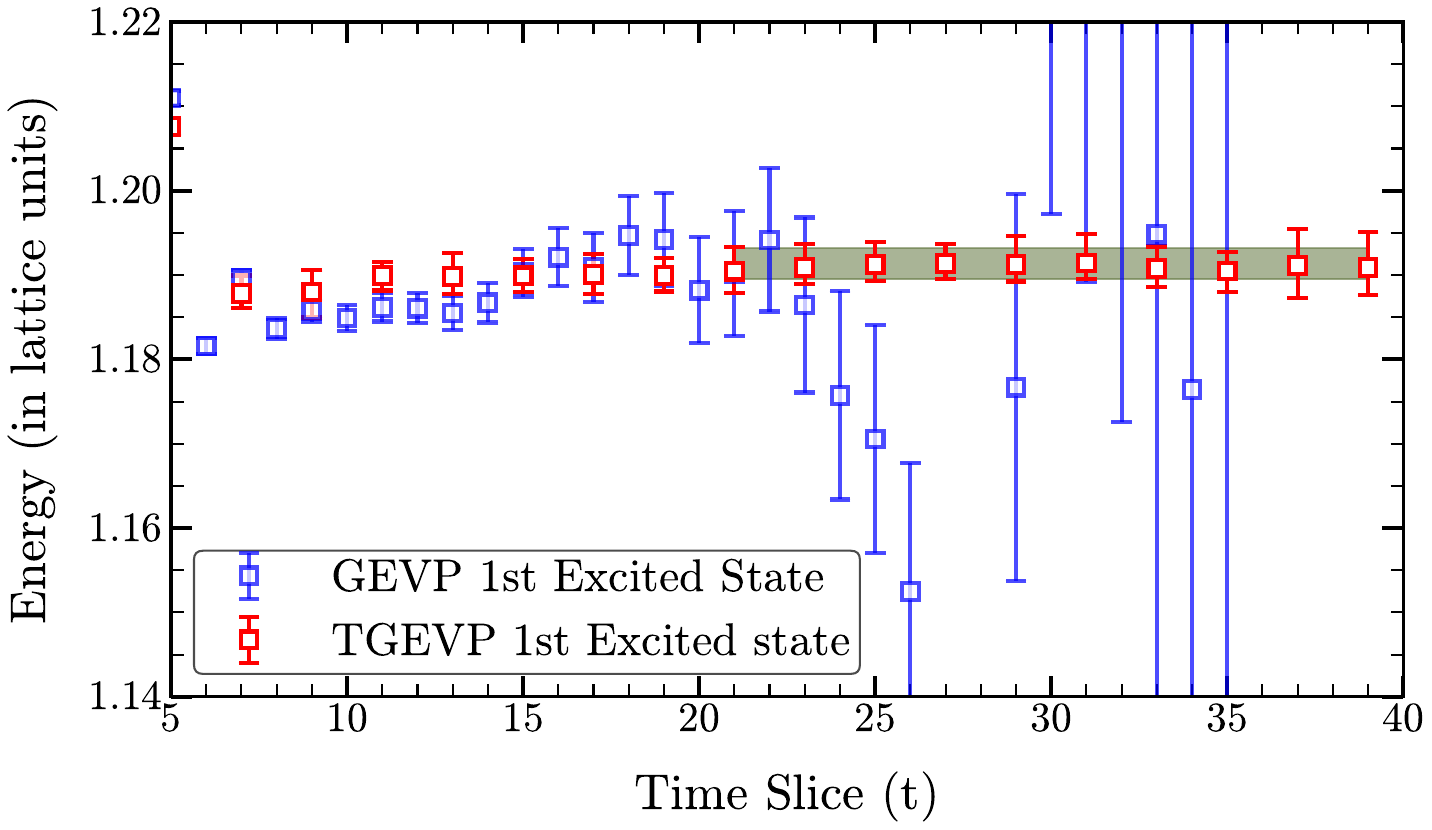}
\caption{Comparison of ground state (top panel) and first excited state (bottom panel) energy estimates obtained from the standard GEVP and block TGEVP methods. The standard GEVP results are shown with light blue data points, while the block TGEVP results are shown in red. A fit to the block TGEVP data is overlaid as a light olive green band, illustrating improved resolution and reduced uncertainty in the excited-state energy extraction.}
\label{fig:excited_state}
\end{figure}

Figure~\ref{fig:excited_state} presents the results for the ground and the first excited state energies of $\Omega_{ccc}$ baryon. In the top panel, we show the ground state energies extracted using both the standard GEVP (blue points) and the block TGEVP method (red points). A fit to the block TGEVP data is overlaid as a light olive green band, illustrating improved statistical precision and stability, especially at larger time separations.

The bottom panel displays the corresponding comparison for the first excited state. As in the ground-state case, the block TGEVP framework yields more precise and stable estimates, with reduced statistical uncertainties. The improved resolution of the energy gap between the ground and excited states highlights the effectiveness of the block TGEVP approach, particularly in suppressing excited-state contamination and mitigating noise. In summary, these results underscore the effectiveness of the block TGEVP formalism in improving the efficiency of extracting both ground and excited state energies, particularly in scenarios where statistical fluctuations are significant.

\section{Conclusion and outlook}\label{sec:summary}
We present an efficient method, TGEVP, for extracting energy levels by applying the generalized eigenvalue framework to the transfer matrix constructed from time-shifted two-point correlation functions. Reliable extraction of energy levels and matrix elements from lattice QCD correlation functions, for a given finite number of statistical samples, is an interesting and  long-standing challenge in lattice QCD computations. The conventional approach, based on exponential fitting guided by an effective mass plateau, is widely used. More recently, the Lanczos algorithm has been applied to the transfer matrix constructed from two-point functions to extract energy levels~\cite{Wagman:2024rid,Ostmeyer:2024qgu,abbott2025filteredrayleighritzneed} and matrix elements~\cite{Hackett:2024nbe,Hackett:2024xnx}. These approaches fall within the broader class of Prony methods. We find that our proposed TGEVP method is equivalent to the Oblique Lanczos method.
In our TGEVP formalism, as the problem is cast as a standard generalized eigenvalue problem (GEVP), it allows the use of standard time-tested linear algebra routines and thereby bypassing the need to implement the recursive oblique Lanczos method. Moreover, when extracting overlap factors and matrix elements, the state vectors can be directly obtained from the eigenvectors of the GEVP. In contrast, the oblique Lanczos method requires an additional change of basis. These aspects make the TGEVP framework conceptually simpler and easier to implement.

Moreover, a central issue for all methods in this class is the filtration of spurious eigenvalues, which arise due to finite statistics in the correlation functions. Failing to address them properly leads to unreliable estimates of energy levels or matrix elements. The standard Prony method exhibits significant instability when extended to three or more states, limiting its practical utility. The Cullum–Willoughby approach \cite{CULLUM1981329}, which was utilized in Ref. \cite{Wagman:2024rid}, requires non-trivial manual tuning of three hyperparameters—$(\Delta, K_{CW}\; \text{and}\;F_{CW})$-to effectively discard outliers. Improper tuning of these hyperparameters can lead to unstable energy estimates and introduces subjectivity into error estimation. 

In contrast, we propose a novel, fully data-driven, automatic, non-parametric filtration method, based on estimating the statistical distribution of eigenvalues using adaptive kernel density estimators (KDEs). Peaks in the resulting distribution and their full-width at half-maximum (FWHM) are used to identify and exclude outlier eigenvalues that significantly deviate from the dominant physical modes. We employ K-fold cross-validation \cite{Silverman1998, Hardle2004, Hastie2009} to optimize the regularization parameter $\alpha$, as detailed in Section \ref{KDE}. This framework eliminates the need for manual hyperparameter tuning, enabling consistent application across a wide range of datasets. When compared with the Cullum–Willoughby test (as used in Ref.\cite{Wagman:2024rid}), our method demonstrates at least equivalent or superior stability across diverse datasets without requiring manual tuning. As a complementary technique, we also use a Gaussian-smoothed histogram method, which is nearly fully data-driven and requires only a single manual input, the kernel width $\sigma$, which can itself be tuned over a reasonable range. While KDE and histogram-based peak detection techniques have been applied in other scientific domains~\cite{article,refId0,McCarthy2014PeaKDEck,Sadiq:2024xsz,Poluektov_2015}, and also in lattice QCD in different contexts~\cite{particles3010007,Bruno:2023bue}, this work marks the first application of such techniques to spectral filtering of spurious eigenvalues in the TGEVP framework.

We demonstrate the efficacy of the method by calculating the first two eigenvalues from two-point correlation functions for a variety of hadrons, including nuclei, across multiple lattice ensembles. For charmonia, along with the individual energy levels we also compute energy splittings between different states and show that, when extrapolated to the continuum limit, they align very well with experimental values, even with limited statistics. In most cases with higher statistics, we find the correlation between eigenvalues across iterations is sufficiently low, allowing for a constant fit, which yields more reliable results than traditional exponential fitting methods. For nuclei, we find the correlation between eigenvalues across iterations to be higher than that for single nucleons, particularly when statistics are low.
The increased correlation poses a challenge for solving the SNR problem, and a thorough evaluation is necessary to account this correlation for accurate estimation of energy levels in such cases. However, within the limit of a given statistical dataset, the TGEVP eigenvalues remain more stable across iterations compared to the noisy effective masses obtained through traditional methods. 
We also demonstrate how the proposed method can be used to extract the overlap factor of a given interpolating operator to a particular state, an essential quantity for studying decay properties. In addition, we show that the TGEVP method can be seamlessly integrated into the traditional multi-operator GEVP framework that is commonly employed in modern lattice QCD calculations for extracting excited energy levels and performing finite-volume amplitude analyses \textit{\`a la}  L\"uscher formalism~\cite{Luscher:1990ck}.
In conclusion, we find that TGEVP method with the data-driven filtration of spurious eigenvalues, through adaptive kernel density estimator,  provides the most robust, simple and reliable method for extracting energy levels and matrix elements from lattice QCD correlation functions. With increased statistics, the use of a multi-operator setup, and proper accounting of correlations, this procedure holds promise for accurately determining energy levels even in challenging multi-nucleon systems.


\section{Acknowledgment}
This work is supported by the Department of Atomic Energy, Government of India, under Project Identification Number RTI 4002. 
Computations related to some of the correlators were done on the Cray-XC30 of ILGTI, TIFR (which has recently been closed), and the computing clusters at the Department of Theoretical Physics, TIFR, Mumbai. NM is thankful to Keh-Fei Liu, Gunnar Bali, Wolfgang Soeldner, Swagato Mukherjee and Xiang Gao for providing us with the nucleon, H$^2$ and heavy baryon data. 
NM and DC also acknowledge multiple discussions with Piyush Srivastava, and Vipul Arora at various stages of this work.
We would also like to thank  Ajay Salve, Kapil Ghadiali, and T. Chandramohan for computational support. 

\appendix
\section{Equivalence between Lanczos and TGEVP methods}\label{app:equiv}
Assuming $\vert u_i \rangle$, $i=0,1,\cdots,m$, form a complete set of orthonormal basis in the Krylov-subspace $\mathcal{K}_m$, the vectors $\vert v_i \rangle$, that enter in Eqs. (\ref{eq:eq7}) 
and (\ref{eq:eq8}) in the main text, can be expanded as,
\begin{eqnarray}
    \vert v_i \rangle = \sum_{j = 0}^m\alpha_{ij}\vert u_j \rangle .\label{lin_exp}
\end{eqnarray}
The eigenvalue equation for the transfer matrix $\mathcal{T}$ using this orthonormal basis $\vert u_i \rangle$'s is given by,
\begin{eqnarray}
    \tilde{T}_{ij}(\tilde{x}_n)_j &=& \lambda_n(\tilde{x}_n)_i\hspace{0.5cm}\text{for $0\leq n < m$},\label{EVeq1}\\
    &&\text{where  }\tilde{T}_{ij} = \langle u_i \vert \mathcal{T} \vert u_i \rangle\nonumber.
\end{eqnarray}
We need to show that the above  generalized eigenvalue problem is exactly equivalent to the eigenvalue problem in Eq.~(\ref{EVeq1}). Starting with Eq. (\ref{GEVPeq1}) 
of the main text, this equivalence can be demonstrated through the following steps,
\begin{eqnarray}
    T_{ij}(x_n)_j &=& \lambda_nV_{ij}(x_n)_j\,,\\
\alpha^{\ast}_{ik}\tilde{T}_{kl}\alpha_{jl}(x_n)_j &=&\lambda_n \alpha_{ik}^{\ast}\alpha_{jk}(x_n)_j\,,\\
    (\alpha^{-1})^{\ast}_{pi} \alpha^{\ast}_{ik}\tilde{T}_{kl}\alpha_{jl}(x_n)_j &=&\lambda_n  (\alpha^{-1})^{\ast}_{pi}\alpha_{ik}^{\ast}\alpha_{jk}(x_n)_j\,,\\
    \tilde{T}_{il}\alpha_{jl}(x_n)_j &=&\lambda_n  \alpha_{ji}(x_n)_j\,,\\
    \tilde{T}_{ij}(\tilde{x}_n)_j &=&\lambda_n (\tilde{x}_n)_i \quad {\mathrm{proved.}}
\end{eqnarray}
Here, in  the second step, we have substituted $\vert v_i\rangle$ utilizing Eq.~(\ref{lin_exp}). In the third step, we have multiplied the both sides of the equation using inverse of $\alpha_{ij}$ (given that $\vert v_i\rangle~$ are linearly-independent, inverse of the matrix $\alpha_{ij}$ exists). In the final step, we have redefined $(\tilde{x}_n)_i = \alpha_{ji}(x_n)_j$ to arrive at the desired form.


\section{Convergence of Histogram Estimator and Kernel Density Estimator}\label{app:convergence}
In this appendix, we describe the statistical properties—including bias and variance—of the two non-parametric methods used in this work for estimating probability density functions from a given dataset: the histogram and the kernel density estimator (KDE). The optimal choices of bandwidth for both estimators are derived, and their respective convergence rates are compared. The discussion in this appendix follows the framework outlined by \cite{Hardle2004}, providing a detail understanding of these methods in non-parametric density estimation. 

\subsection{Histogram as a Non-parametric Estimator}
We begin with the discussion of histogram estimator as discussed in subsection \ref{Gaussian_Histogram}. For ease of reading we reproduce Eq. \ref{eq:gauss_conv} below. Mathematically, the histogram estimator \(\hat{f}_h(x)\) for a dataset, \(\{x_1, x_2, \cdots, x_n\}\), with bin width \(h\) and origin \(x_0\), evaluated at a point \(x \in B_j\), is given by:
\begin{eqnarray*}
    \hat{f}_h(x) &=& \frac{1}{nh} \sum_{i=1}^n \mathcal{I}(x_i \in B_j), \nonumber \\
    && \text{where } B_j = [x_0 + (j-1)h, x_0 + jh),
\end{eqnarray*}
where \(\mathcal{I}(\cdot)\) is the indicator function. The histogram groups the data into discrete bins and counts the frequency of values falling within each bin, normalized appropriately to approximate the density. For sufficiently small bin widths and large sample sizes, the histogram can be shown to converge in probability to the true density under appropriate regularity conditions. We investigate the statistical properties of \(\hat{f}_h(x)\), including its bias and variance:
\begin{align}
    \text{Bias}\{\hat{f}_h(x)\} &= \mathbb{E}\{\hat{f}_h(x)\} - f(x), \\
    \text{Var}\{\hat{f}_h(x)\} &= \mathbb{E}\{(\hat{f}_h(x) - \mathbb{E}\{\hat{f}_h(x)\})^2\},
\end{align}
where \(f(x)\) is the true density. It can be shown that, under mild assumptions and for sufficiently small bin width \(h\), the bias and variance are approximately given by:
\begin{align}
    \text{Bias}\{\hat{f}_h(x)\} &\approx f'(m_j)(m_j - x), \nonumber \\
    & \text{where } m_j = x_0 + \left(j - \tfrac{1}{2}\right)h, \\
    \text{Var}\{\hat{f}_h(x)\} &\approx \frac{1}{nh} f(x).
\end{align}
As we can see, the bias increases with bin width \(h\), while the variance decreases with it, creating a trade-off. To balance this, one commonly uses the \textit{mean integrated squared error} (MISE) as a global measure of estimation accuracy:
\begin{eqnarray}
    \text{MISE}\{\hat{f}_h\} = \mathbb{E} \left\{ \int_{-\infty}^{\infty} \left( \hat{f}_h(x) - f(x) \right)^2 dx \right\}.
\end{eqnarray}
In the asymptotic limit, the MISE is approximated by the \textit{asymptotic mean integrated squared error} (AMISE):
\begin{eqnarray}
    \text{AMISE}\{\hat{f}_h\} = \frac{1}{nh} + \frac{h^2}{12} \| f' \|_2^2,
\end{eqnarray}
where \(\| f' \|_2^2 = \int_{-\infty}^\infty (f'(x))^2 dx\) is the \(L^2\)-norm of the derivative of the true density.

By minimizing the AMISE with respect to \(h\), one obtains the \textit{optimal bin width}:
\begin{eqnarray}
    h_0 \sim n^{-1/3},
\end{eqnarray}
and substituting back, the corresponding optimal AMISE scales as:
\begin{eqnarray}
    \text{AMISE}\{\hat{f}_{h_0}\} \sim n^{-2/3}.
\end{eqnarray}
This shows that as the sample size \(n\) increases, the histogram estimator \(\hat{f}_{h_0}\) converges to the true density \(f(x)\), albeit at a relatively slow rate compared to more advanced kernel estimators.
\subsection{Kernel Density Estimator}
As discussed in subsection \ref{KDE}, an alternative non-parametric estimator of the probability density function with superior asymptotic properties is the {\textit{kernel density estimator}} (KDE). Unlike the histogram, which produces a piecewise constant estimate that depends on arbitrary choices like bin edges and origin, KDE constructs a smooth estimate by centering a continuous kernel function at each data point. To discuss the bias-variance and {\textit{AMISE}} we reproduce Eq (\ref{eq:kde}) below. Formally, for a sample \(\{x_1, x_2, \cdots, x_n\}\) drawn from a density \(f(x)\), the Gaussian KDE, \(\hat{f}_h(x)\) is defined as:
\begin{eqnarray*}
    \hat{f}_h(x) = \frac{1}{n h \sqrt{2\pi}} \sum_{i=1}^n \exp\left(-\frac{(x - x_i)^2}{2h^2}\right).
\end{eqnarray*}
This estimator can be interpreted as a convolution of the empirical distribution function with a Gaussian kernel of variance \(h^2\).

\noindent{\bf{Bias-Variance Trade-off and AMISE}}:  
The KDE is a biased estimator of \(f(x)\), and the bias-variance trade-off is governed by the choice of the bandwidth \(h\). Under the assumptions that \(K\) is symmetric, integrable, and that \(f(x)\) is twice differentiable in a neighborhood of \(x\), the asymptotic bias and variance of \(\hat{f}_h(x)\) are given by:
\begin{eqnarray}
    \text{Bias}\{\hat{f}_h(x)\} &\approx& \frac{1}{2} h^2 f''(x) \mu_2(K), \\
    \text{Var}\{\hat{f}_h(x)\} &\approx& \frac{1}{n h} f(x) R(K),
\end{eqnarray}
where \(\mu_2(K) = \int u^2 K(u)\, du\) is the second moment of the kernel, and \(R(K) = \int K^2(u)\, du\). This leads to the asymptotic mean integrated squared error (AMISE),
\begin{eqnarray}
    \text{AMISE}\{\hat{f}_h\} = \frac{1}{n h} R(K) + \frac{h^4}{4} \mu_2(K)^2 \left\|f''\right\|_2^2,
\end{eqnarray}
where \(\left\|f''\right\|_2^2 = \int \left(f''(x)\right)^2 dx\). Minimizing the AMISE with respect to \(h\) yields the optimal bandwidth,
\begin{eqnarray}
    h_{\text{opt}} \sim n^{-1/5},
\end{eqnarray}
and the corresponding convergence rate of the estimator is \(\text{AMISE}\{\hat{f}_{h_{\text{opt}}}\} \sim n^{-4/5}\), which improves over the histogram estimator's rate of \(n^{-2/3}\). It can be shown that, under weak assumptions, there is no non-parametric estimator that converges at a faster rate than the kernel estimator \cite{10.1214/aos/1176342997}.

\section{Additional results}\label{app:additional_results}
\begin{figure*}[h!]
\includegraphics[width=0.48\textwidth, height=0.34\textwidth]{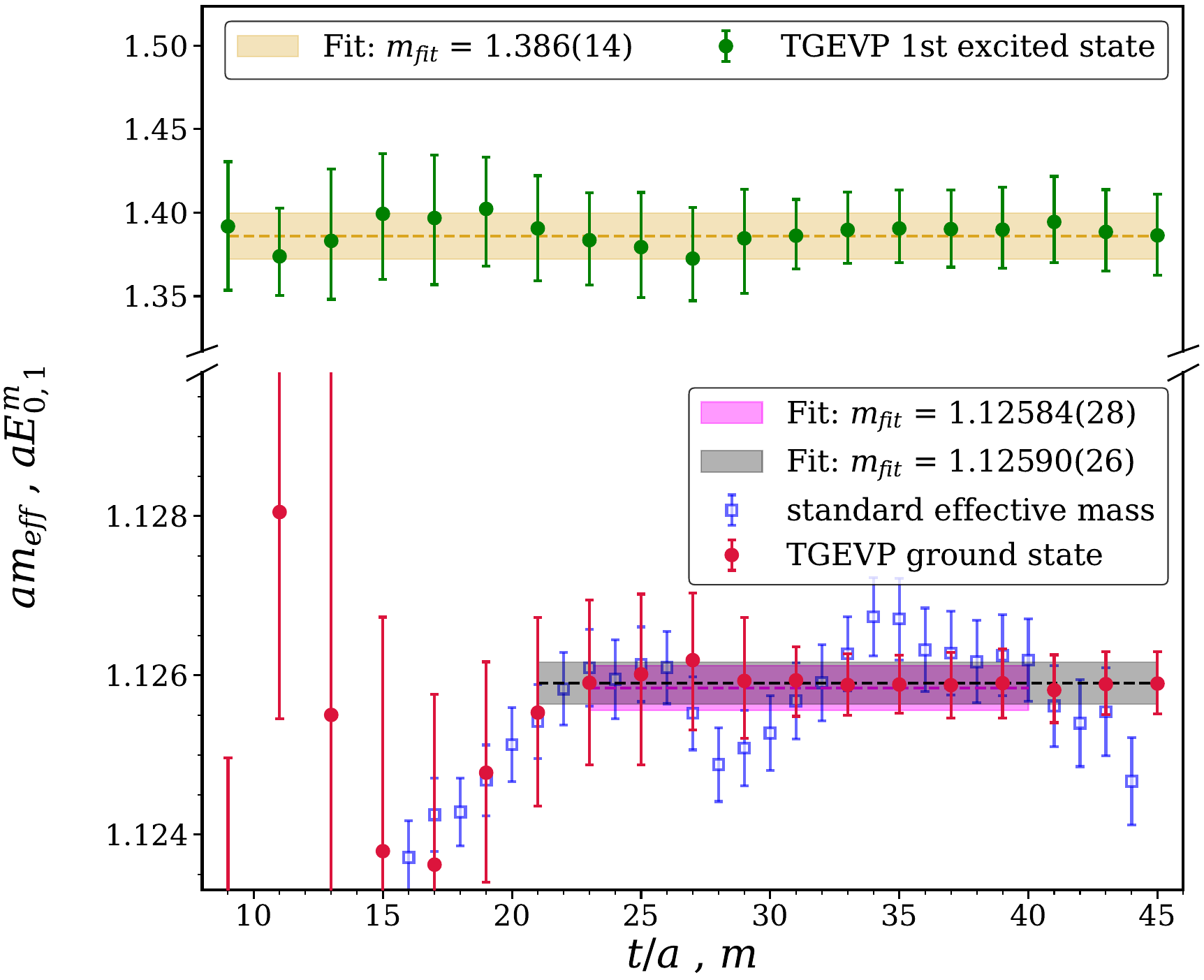}
\includegraphics[width=0.48\textwidth, height=0.34\textwidth]{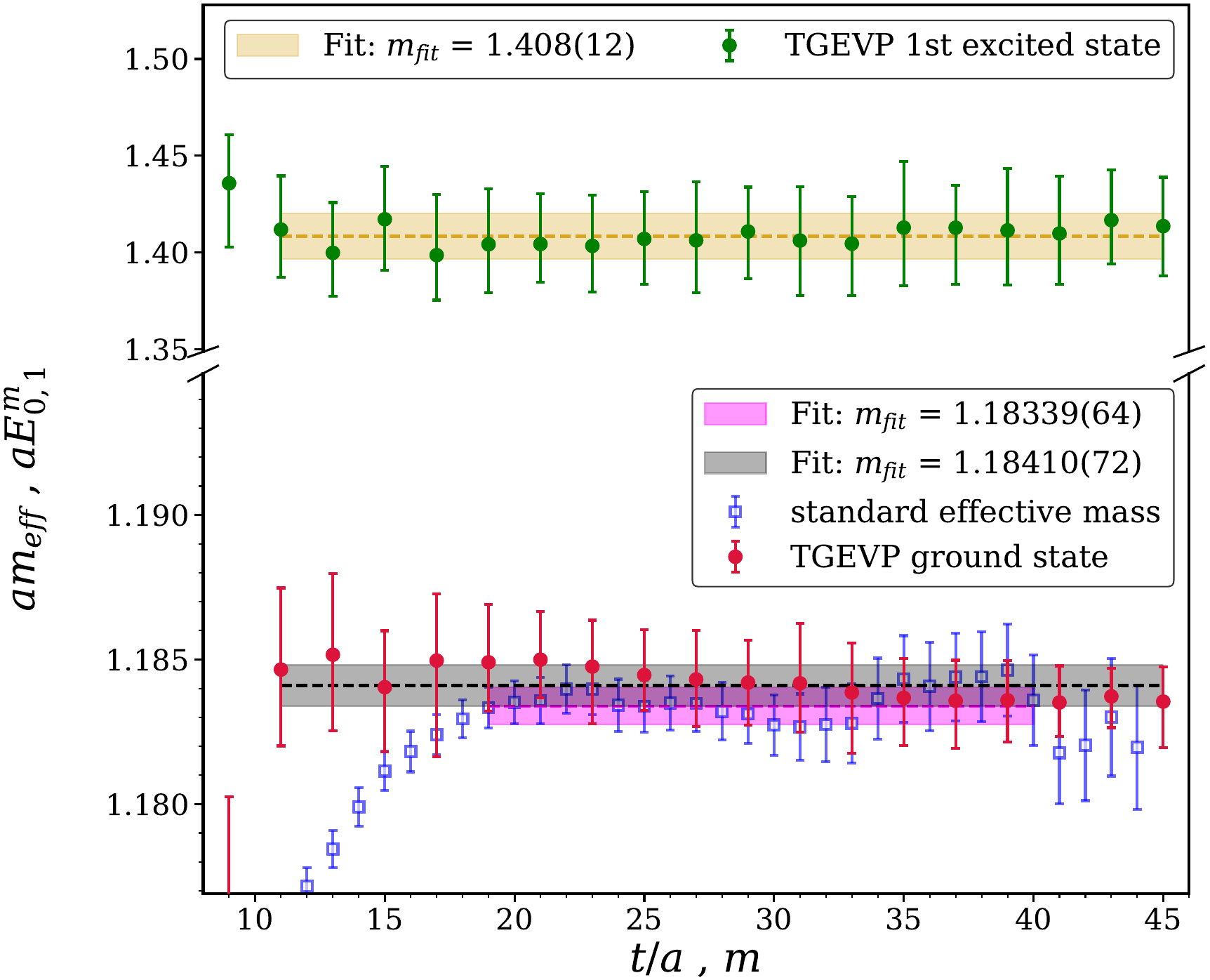}
\includegraphics[width=0.48\textwidth, height=0.34\textwidth]{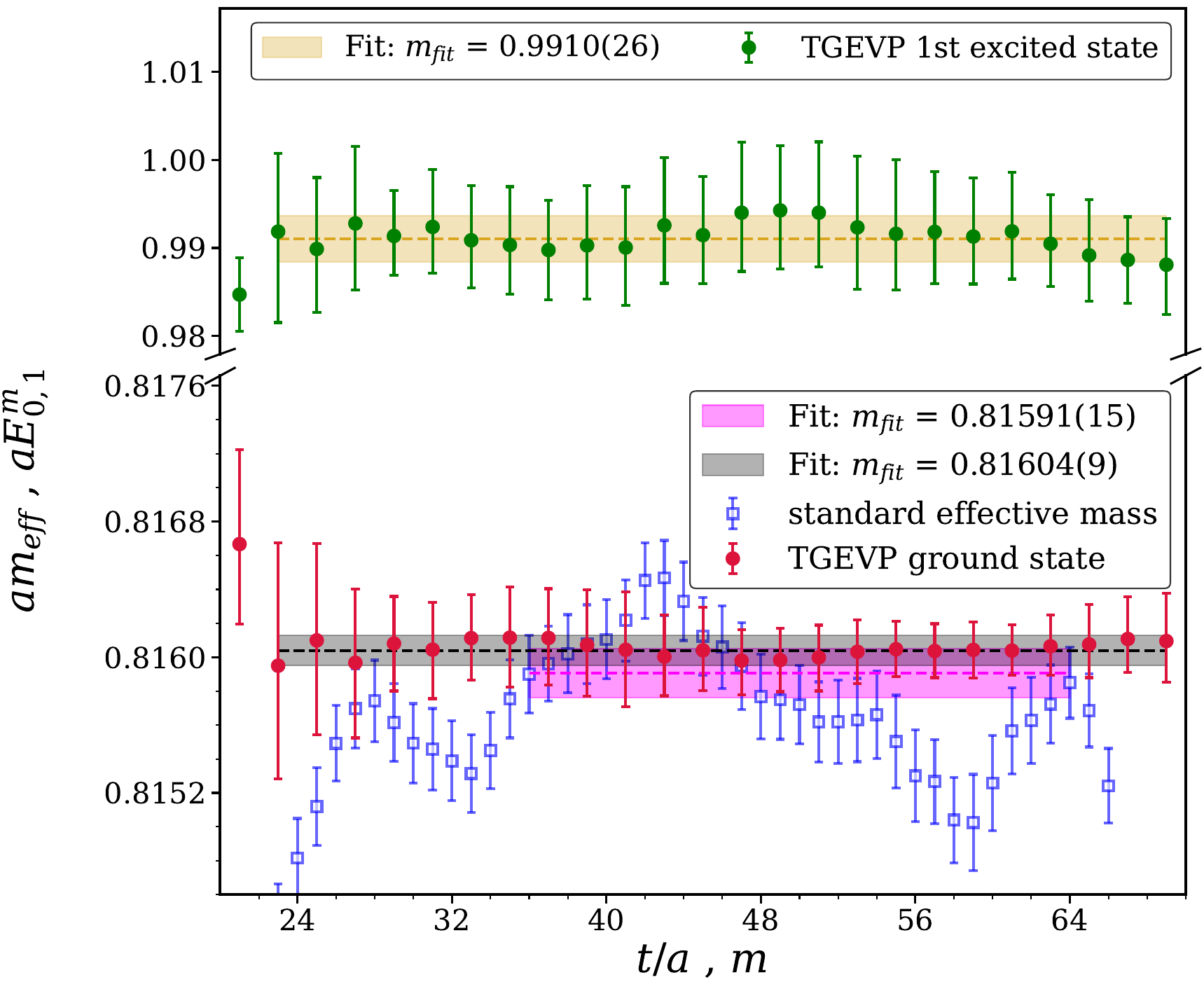}
\includegraphics[width=0.48\textwidth, height=0.34\textwidth]{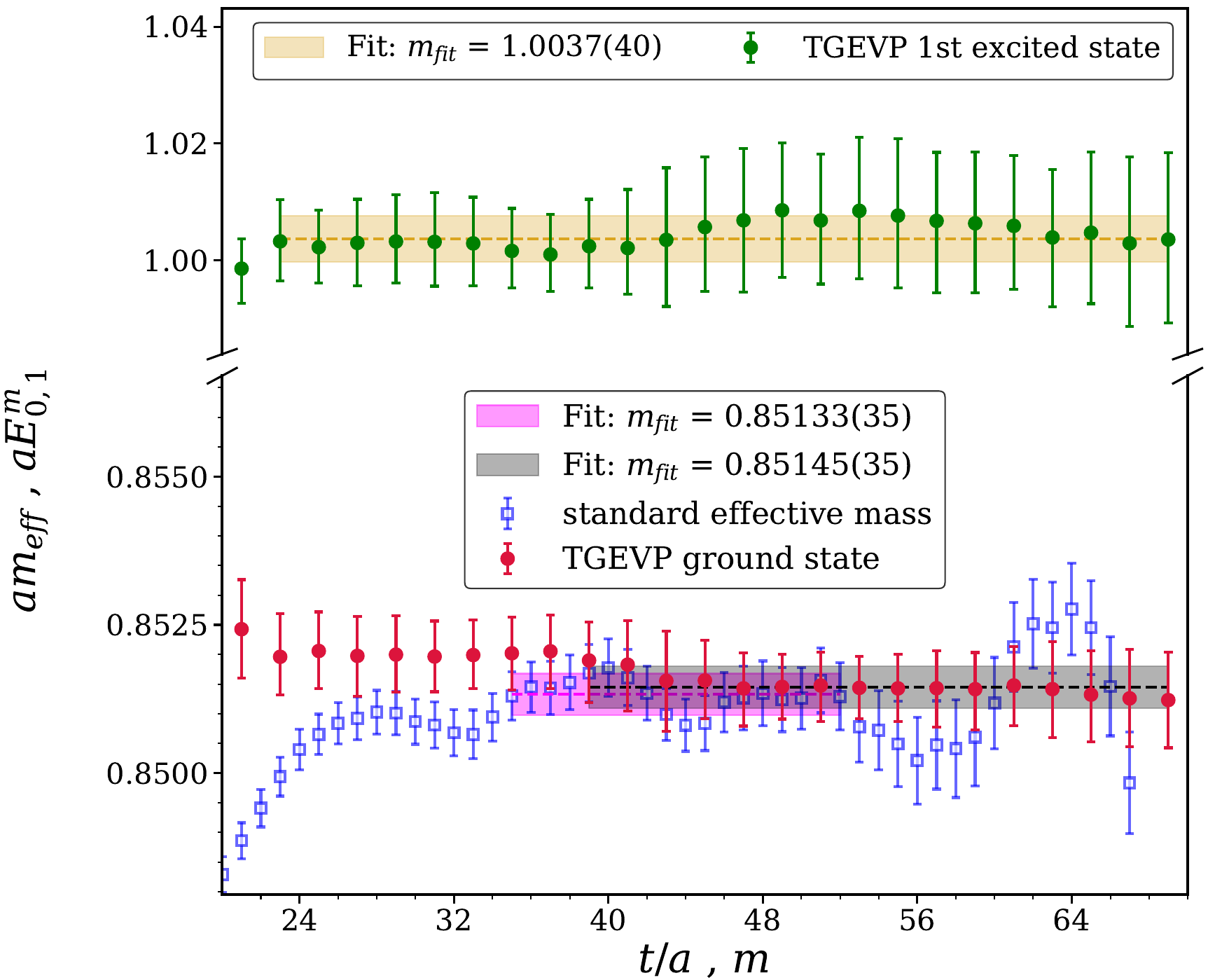}
\includegraphics[width=0.48\textwidth, height=0.34\textwidth]{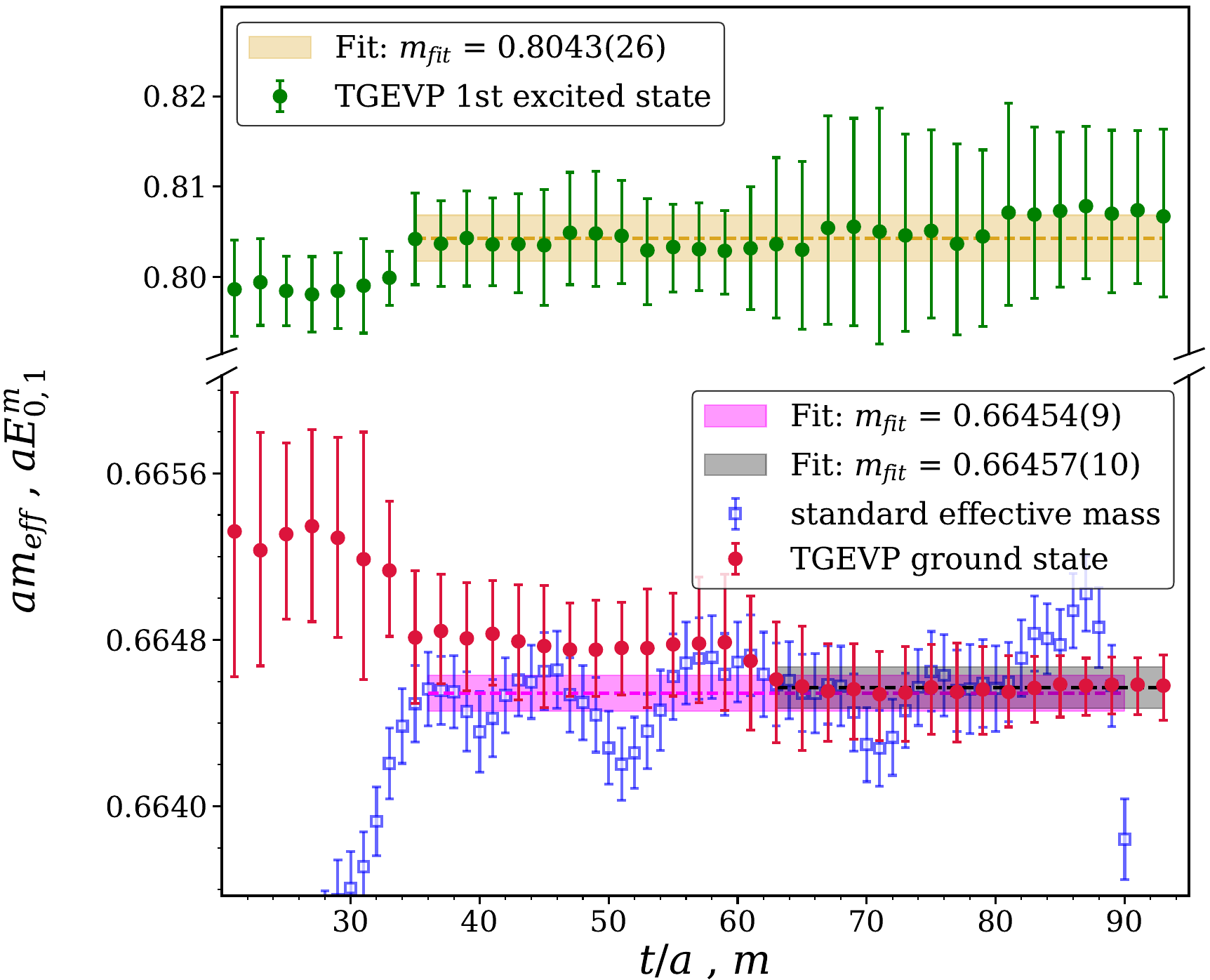}
\includegraphics[width=0.48\textwidth, height=0.34\textwidth]{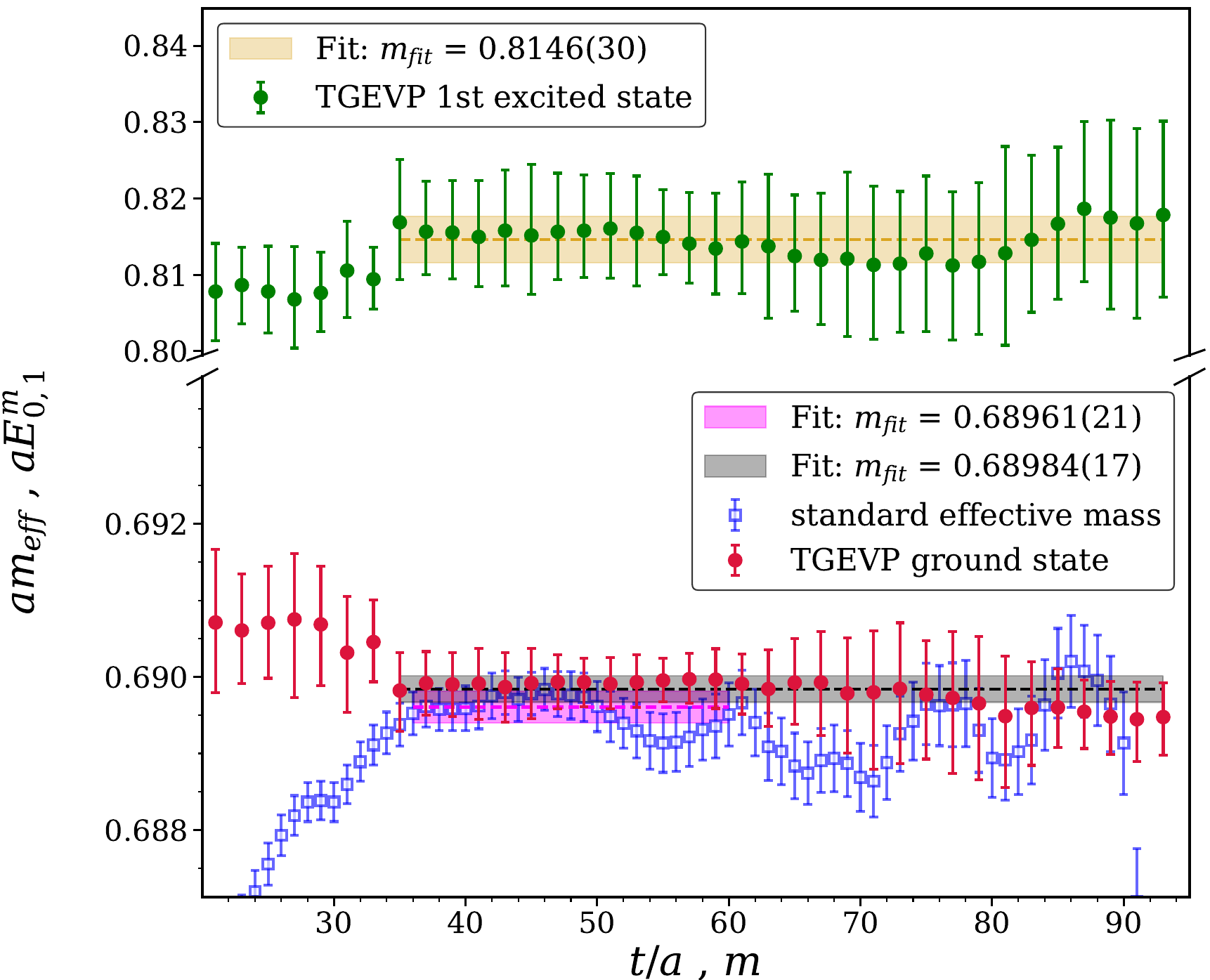}
        \caption{These plots represent the regular effective mass and the first two TGEVP-eigenvalues corresponding to charmonia states. Left panel is for $0^{-}$ and the right one corresponds to $1^{-}$. These results are obtained with overlap valence quarks on the $N_f = 2+1+1$ HISQ lattice ensembles with the specifications, $L^3\times T$ (lattice spacing $a$ in fm), $n_{meas}$  $\equiv$ Top: $32^3 \times 96 ~(0.0888)$, 169, Middle: $48^3 \times 144 ~ (0.0582)$, 180, and Bottom: $64^3 \times 192 ~ (0.0441)$, 142.}
        \label{fg:charm2}
  \end{figure*}

\begin{figure*}[htb!]
\includegraphics[width=0.48\textwidth, height=0.34\textwidth]{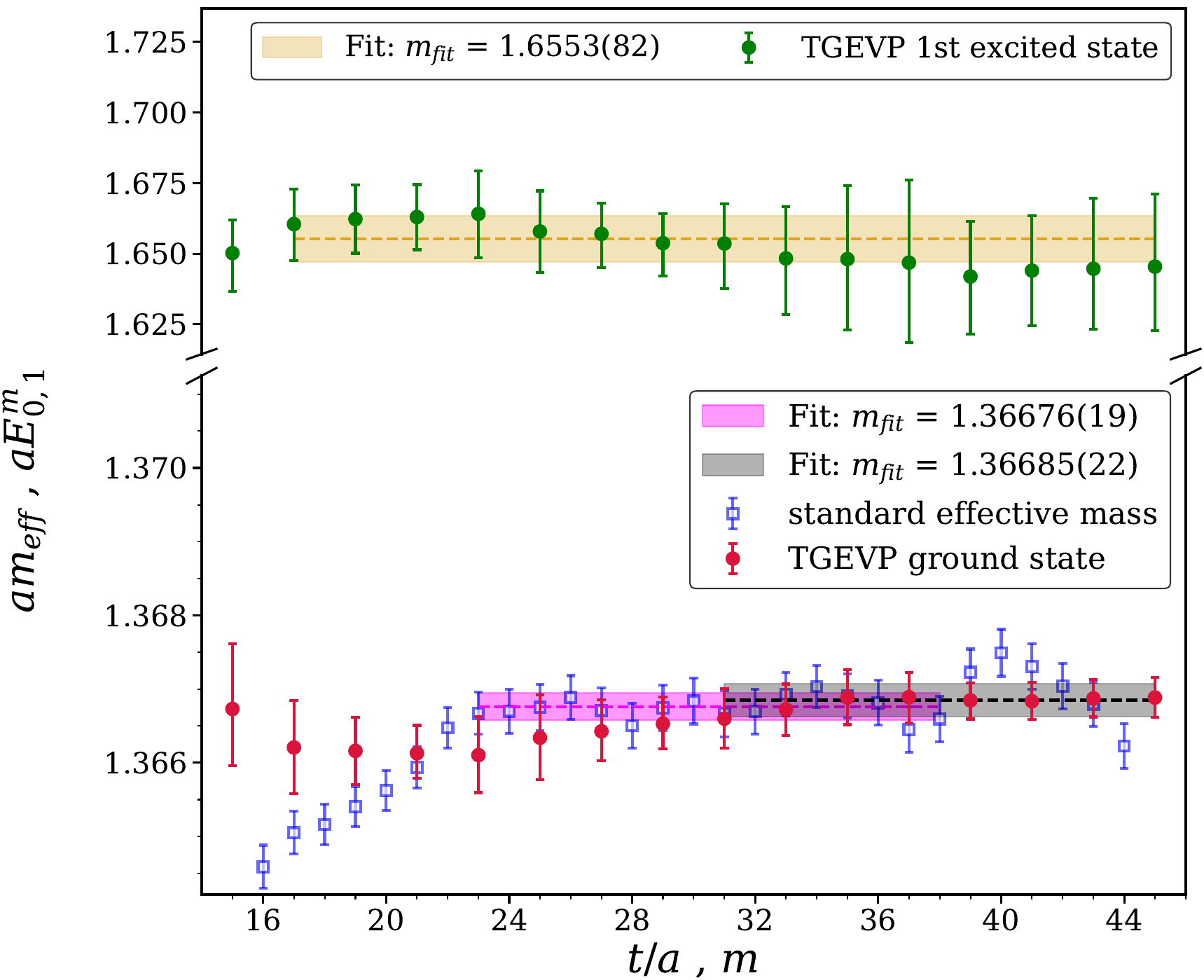}
\includegraphics[width=0.48\textwidth, height=0.34\textwidth]{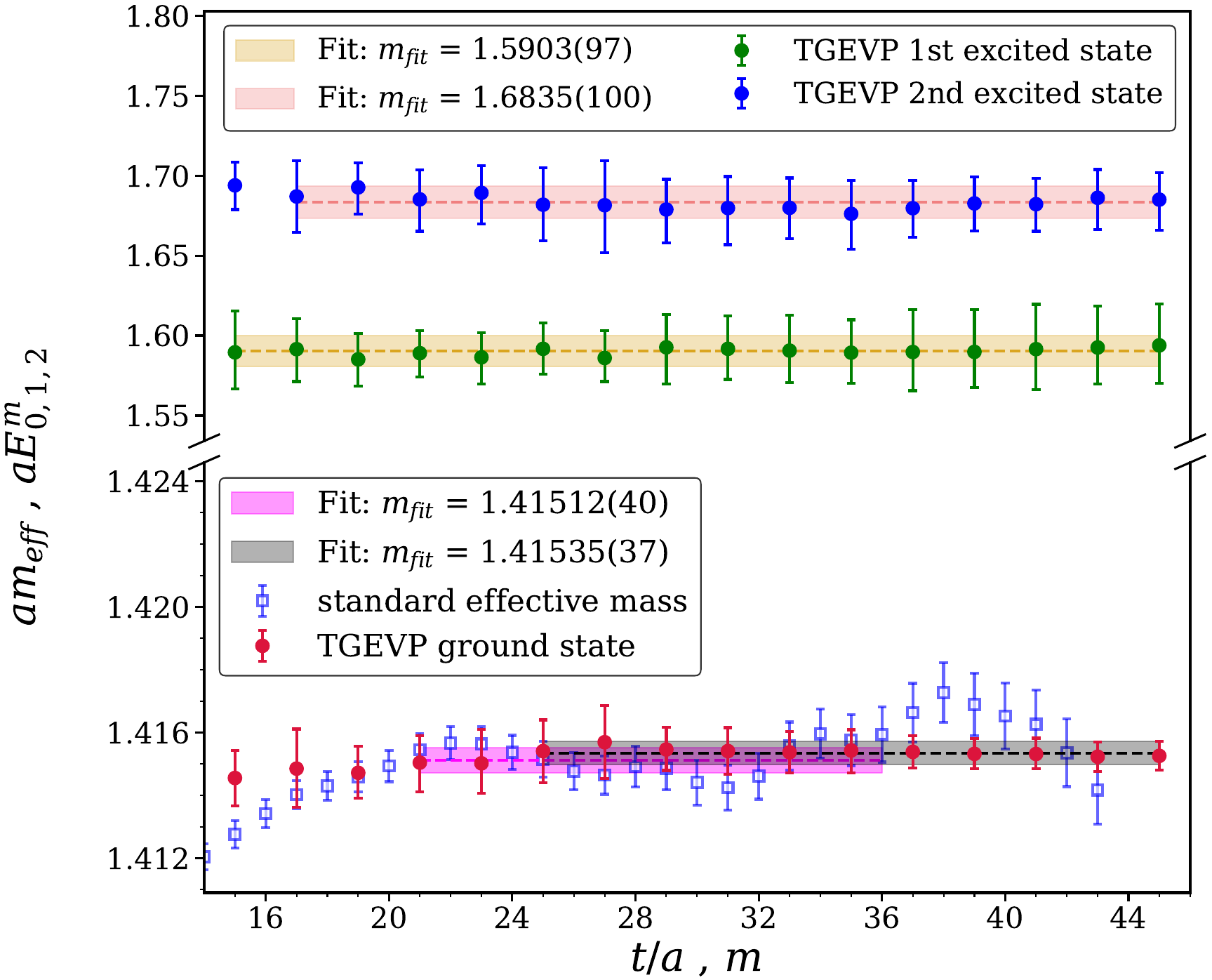}
\includegraphics[width=0.48\textwidth, height=0.34\textwidth]{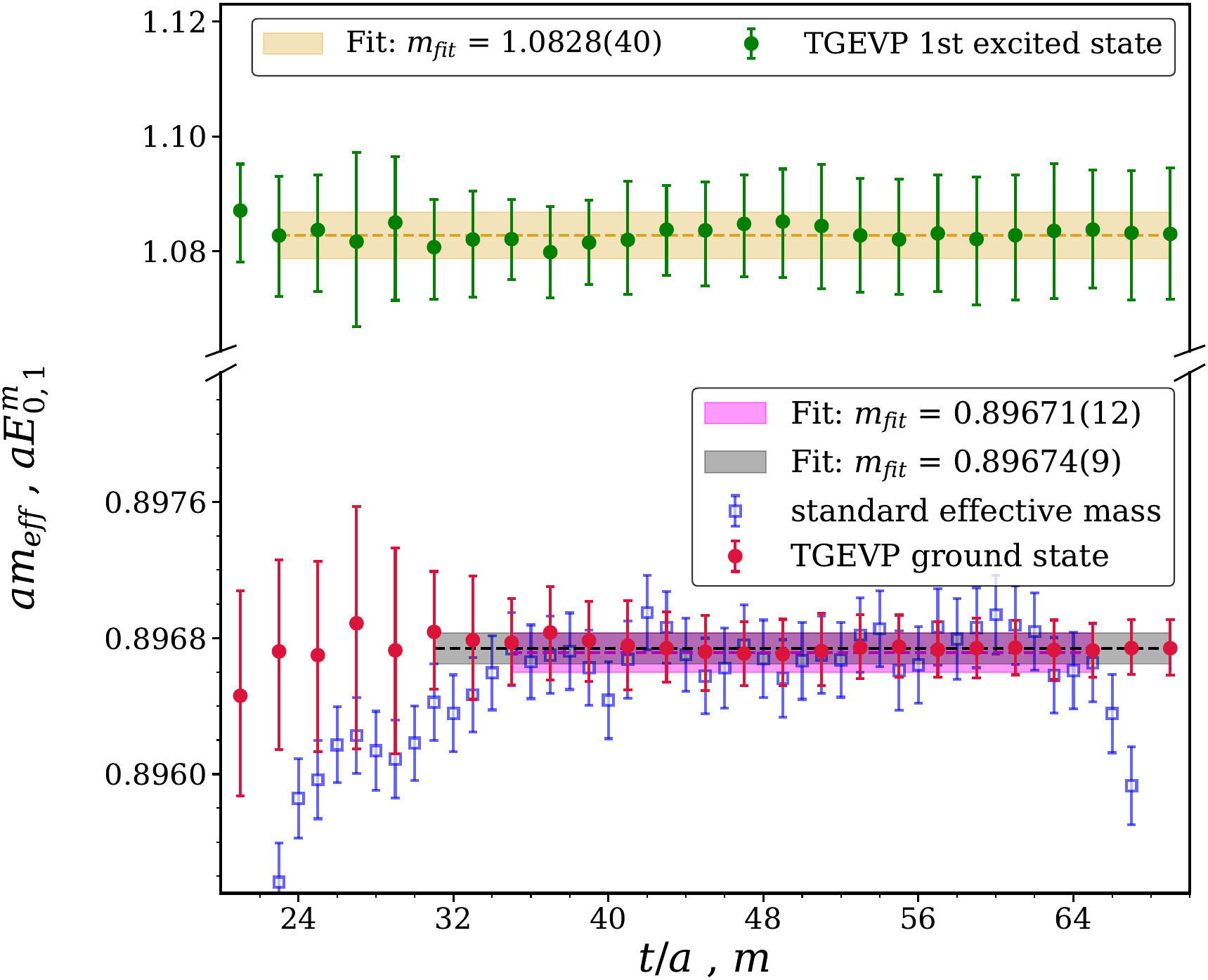}
\includegraphics[width=0.48\textwidth, height=0.34\textwidth]{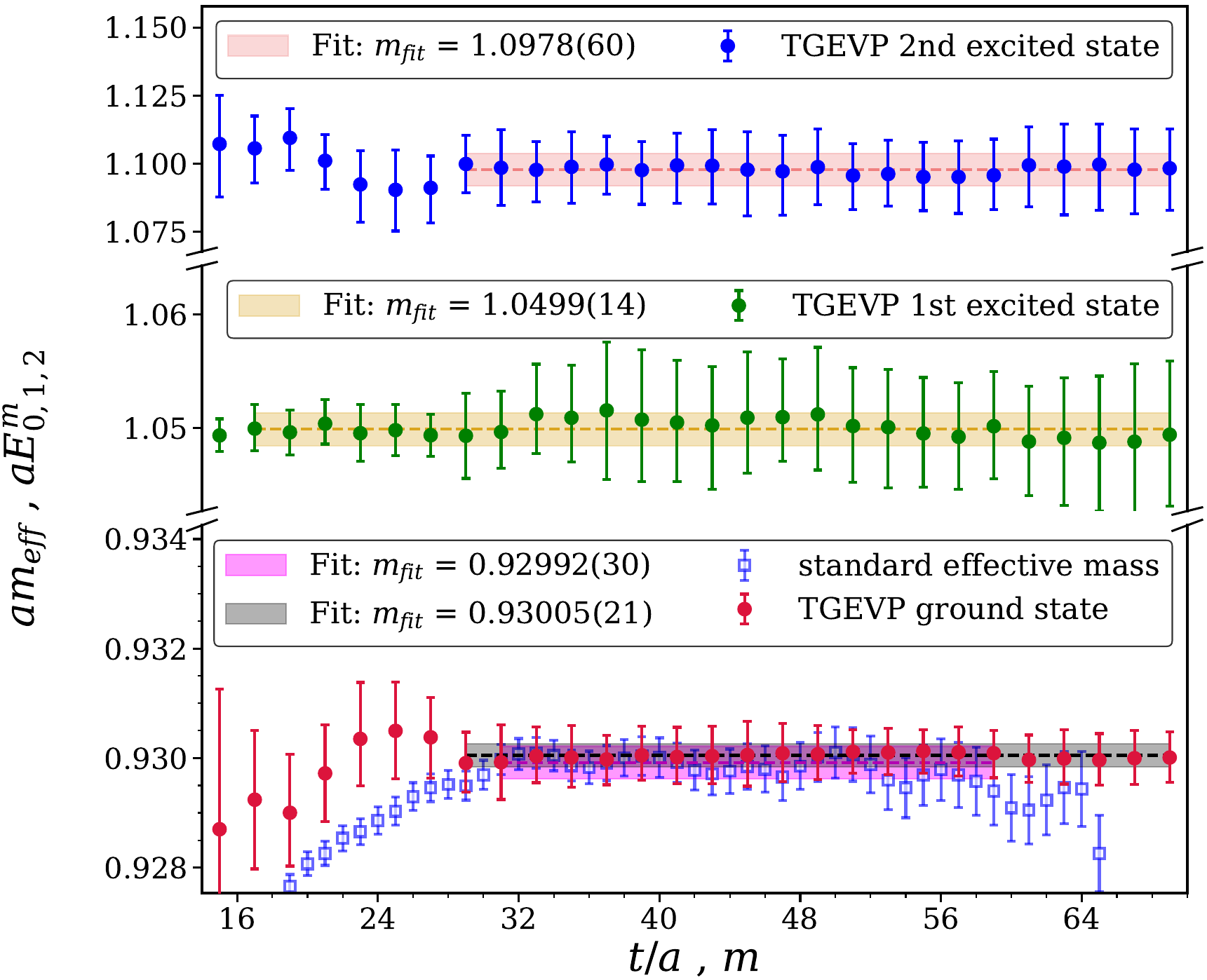}
        \caption{These plots represent the regular effective mass and the first two (left)/three (right) TGEVP-eigenvalues of charmonia states. Left panel is for $J^{P} ~\equiv ~0^-$ and the right one corresponds to $1^-$ together with $1^+$ states. These results are obtained with HISQ valence quarks on the $N_f = 2+1+1$ HISQ lattice ensembles; [Top: $32^3 \times 96 ~(a = 0.0888)$ ($n_{meas} = 396$),
        bottom: $48^3 \times 144 ~(a = 0.0582)$ fm ($n_{meas} = 386$).]}
        \label{fg:charm3}
  \end{figure*}

In addition to the results presented in the main section, here we present more such results for various hadrons including their excited states. In Fig. \ref{fg:charm2}, we show the first two eigenvalues as determined employing TGEVP method from a two point correlation functions .
These dataset corresponds to 1S-charmonia states, where the left panel is for $0^{-}$ and the right one corresponds to $1^{-}$ states. The regular effective mass, $m_{eff}(t)$ is also shown (blue circles) to compare that with the lowest eigenvalue (red circles). In addition to the lowest eigenvalue we find the second eigenvalue (green circles) is also stable across iterations, and one can get a good estimation of the first excited state.  These results are obtained from the two point correlation functions computed with overlap valence quarks on the $N_f = 2+1+1$ HISQ lattice ensembles with the specifications, $L^3\times T$ (lattice spacing $a$ in fm)  $\equiv$ $32^3 \times 96 ~(0.0888)$ (top), $48^3 \times 144 ~ (0.0582)$ (middle), and $64^3 \times 192 ~ (0.0441)$ (bottom) \cite{MILC:2012znn, Bazavov:2017lyh}, and was discussed in Refs. \cite{Mathur:2018epb, PhysRevD.99.031501, Dhindsa:2024erk} .  
In Fig.~\ref{fg:charm3}, we present similar results using the HISQ action for valence quarks on the same set of lattices.  We show the effective mass and the first two/three TGEVP-eigenvalues of charmonia states. Left panel represents the $J^{P} ~\equiv ~0^-$ state while the right one corresponds to $1^-$ and $1^+$ states.
In this case
$1^-$ requires special attention due to oscillating contributions in the temporal direction. To address that, we employ the smoothed effective mass technique for the ground state of $J/\psi$, as described in Ref. \cite{PhysRevD.91.034504}. 
From the non-oscillating part of the correlators we extract the lowest two eigenvalues which correspond to the masses of $J/\psi(1S)$ and $J/\psi(2S)$ states. The lowest eigenvalue corresponding to the oscillating part of the correlator is also extracted and  identified that with the mass of $\chi_{c1}(1P)$ state. For the eigenvalue of the oscillating state, only the negative eigenvalues, $\lambda_n^{m,b}$, of Eq. (\ref{GEVPeq1}) are retained and then we follow the same analysis procedure as in the other cases.


\begin{figure*}[h]
\includegraphics[width=0.48\textwidth, height=0.28\textwidth]{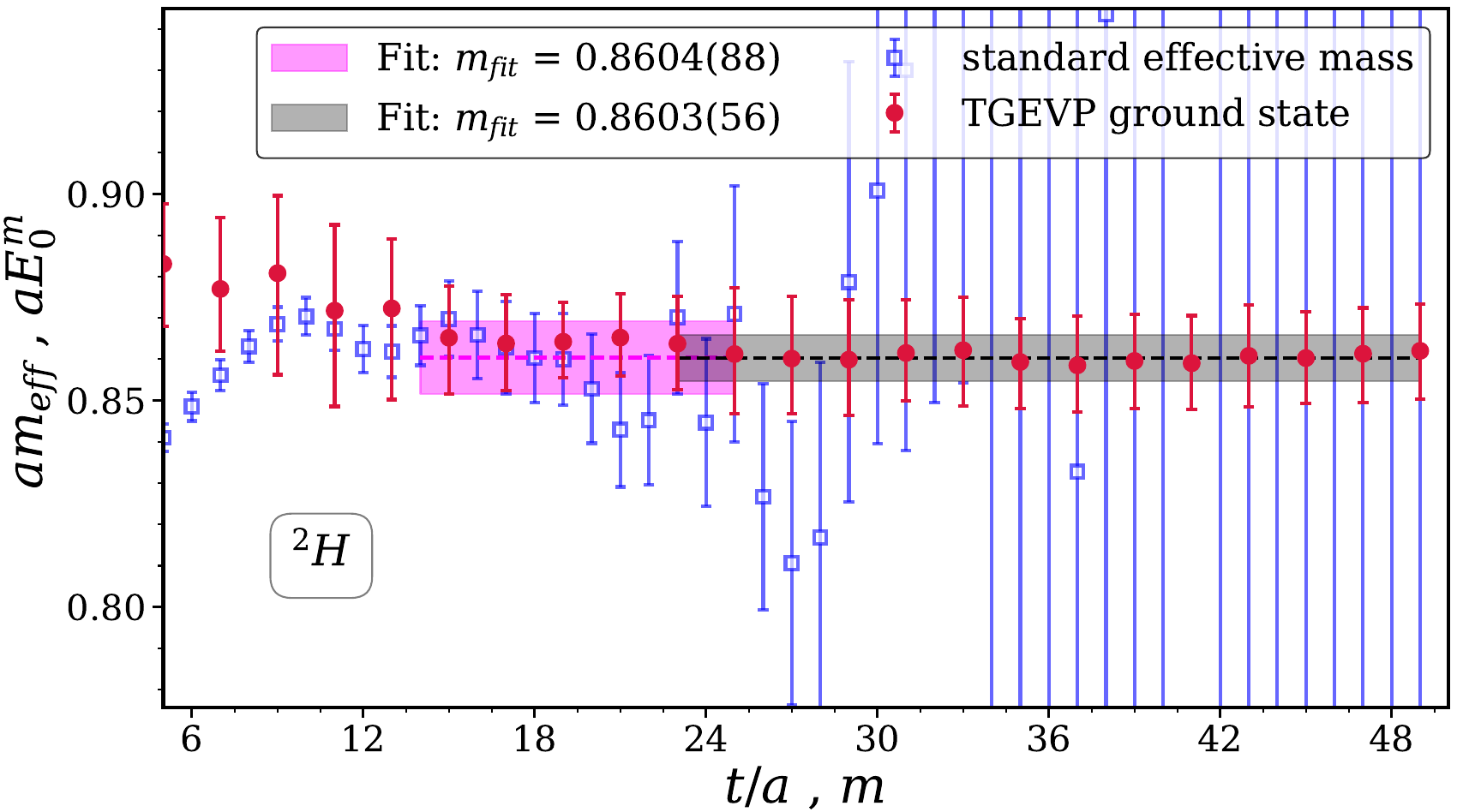}
\includegraphics[width=0.48\textwidth, height=0.28\textwidth]{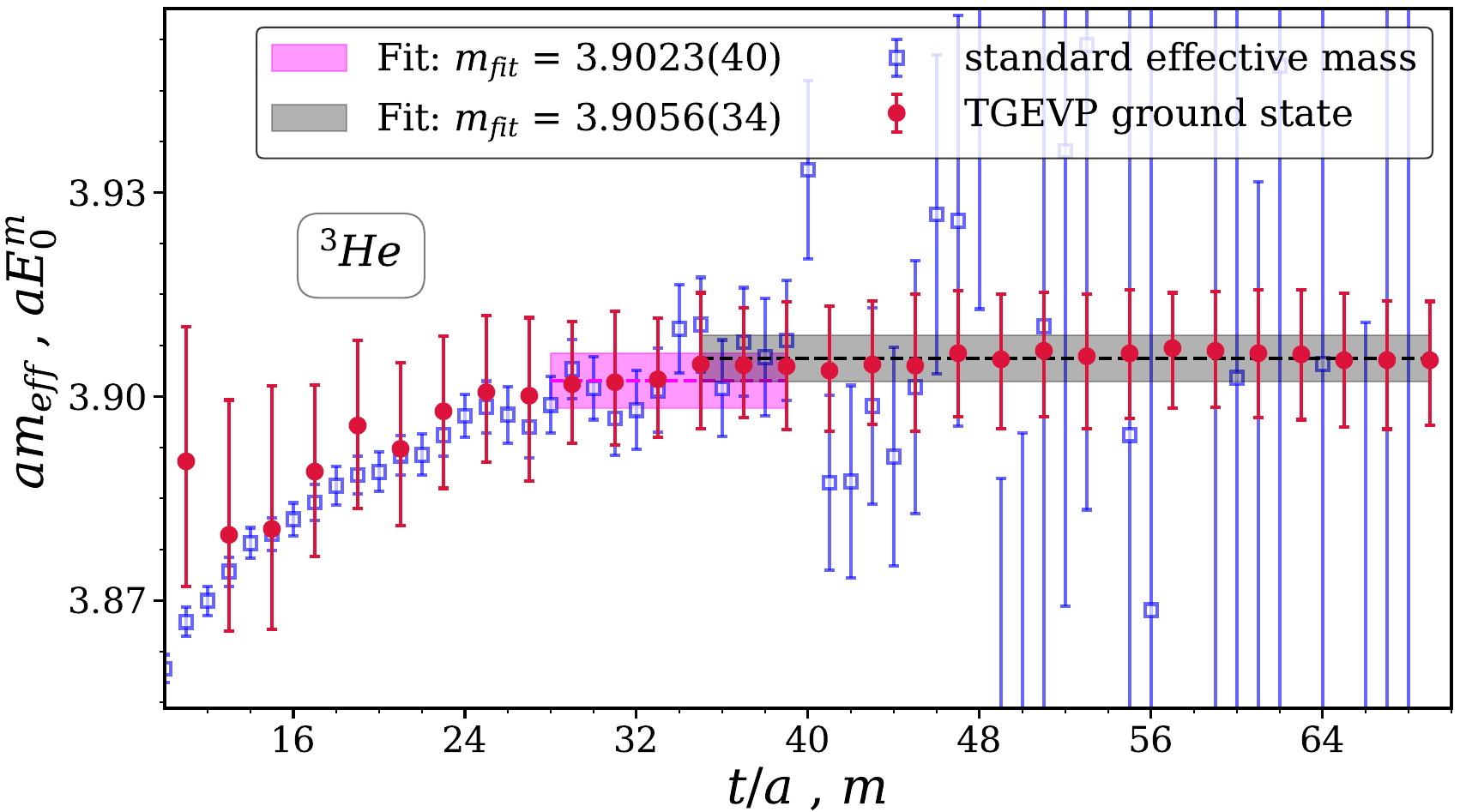}
\includegraphics[width=0.48\textwidth, height=0.28\textwidth]{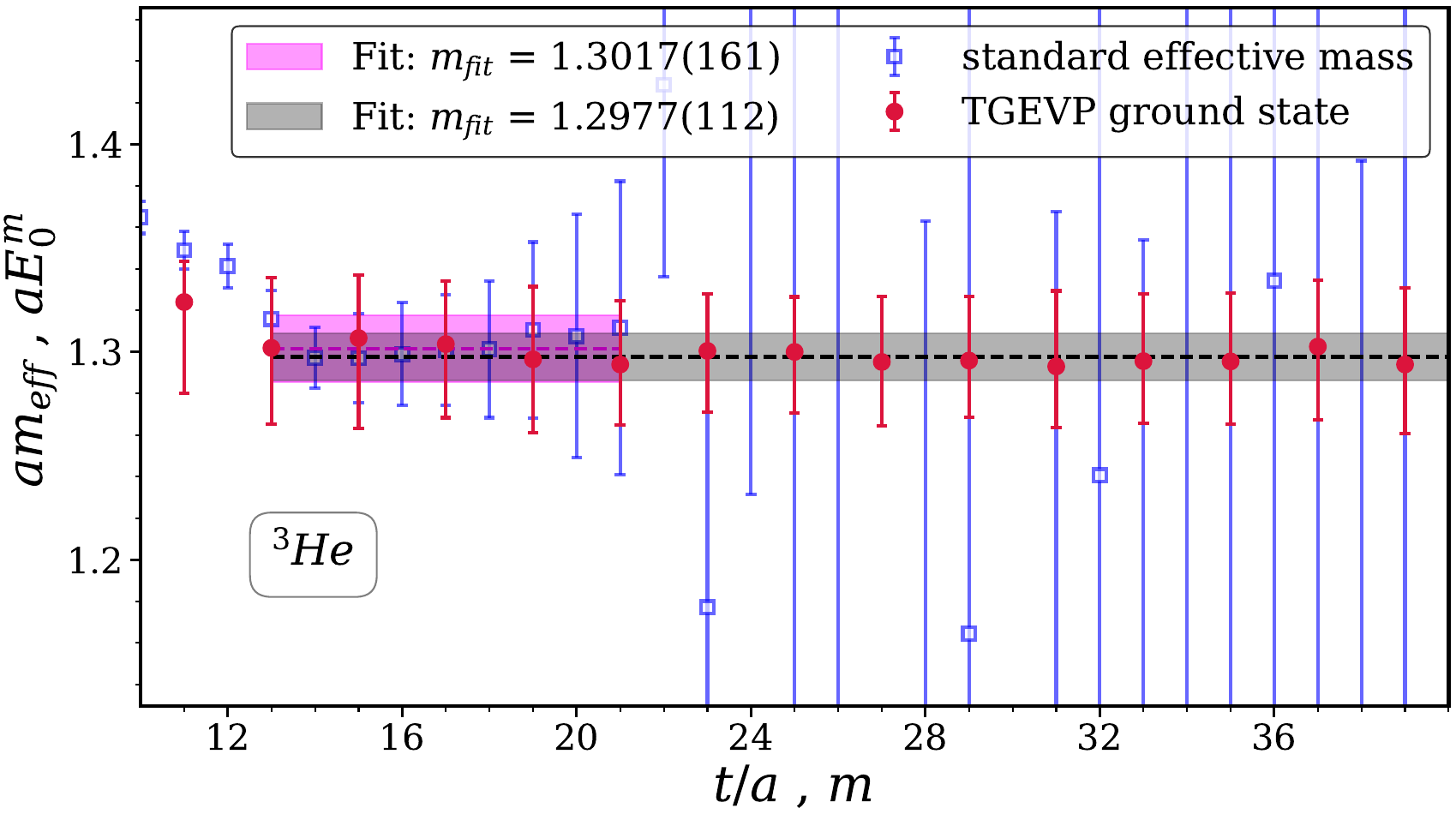}
\includegraphics[width=0.48\textwidth, height=0.28\textwidth]{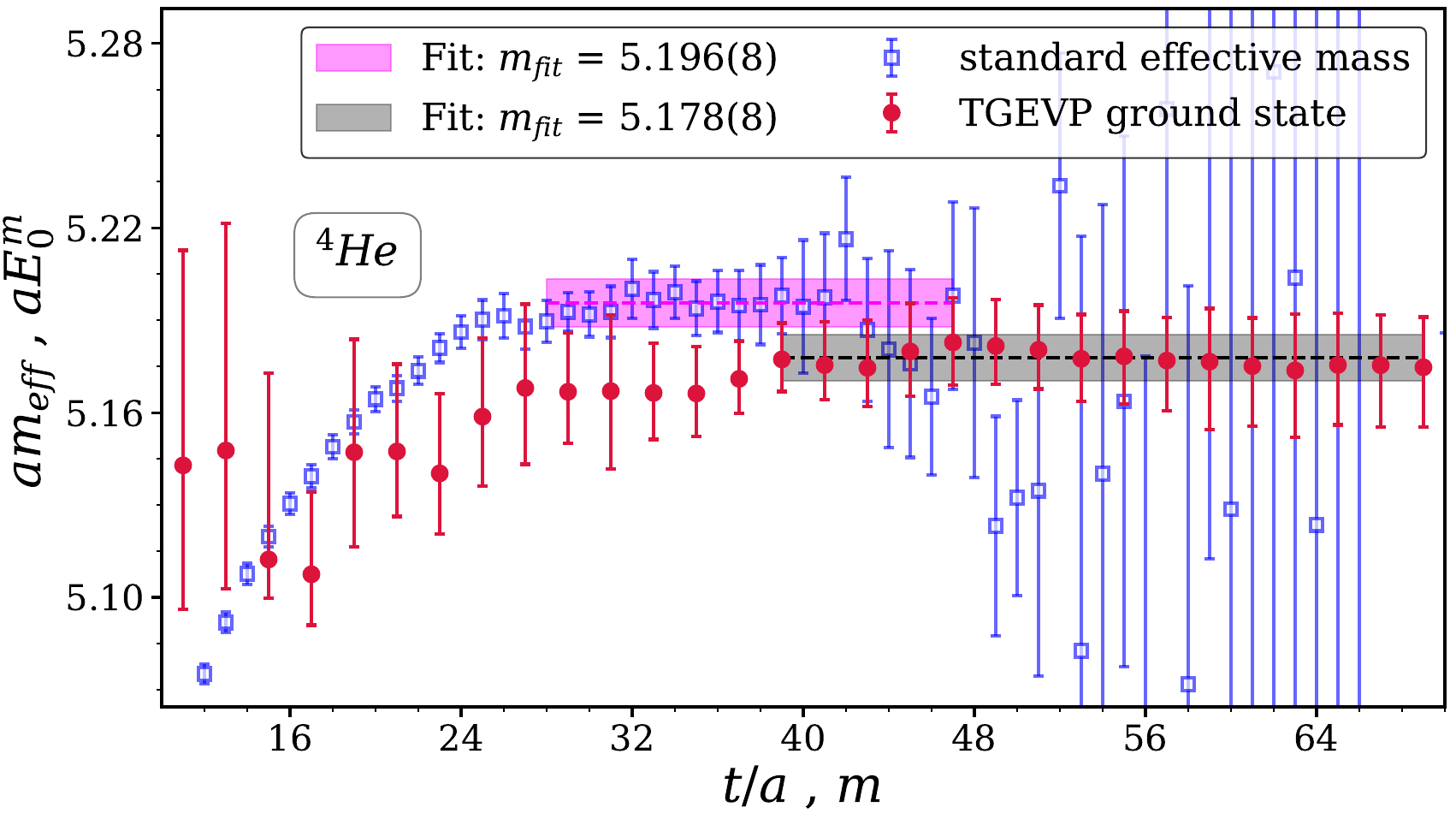}
\includegraphics[width=0.48\textwidth, height=0.28\textwidth]{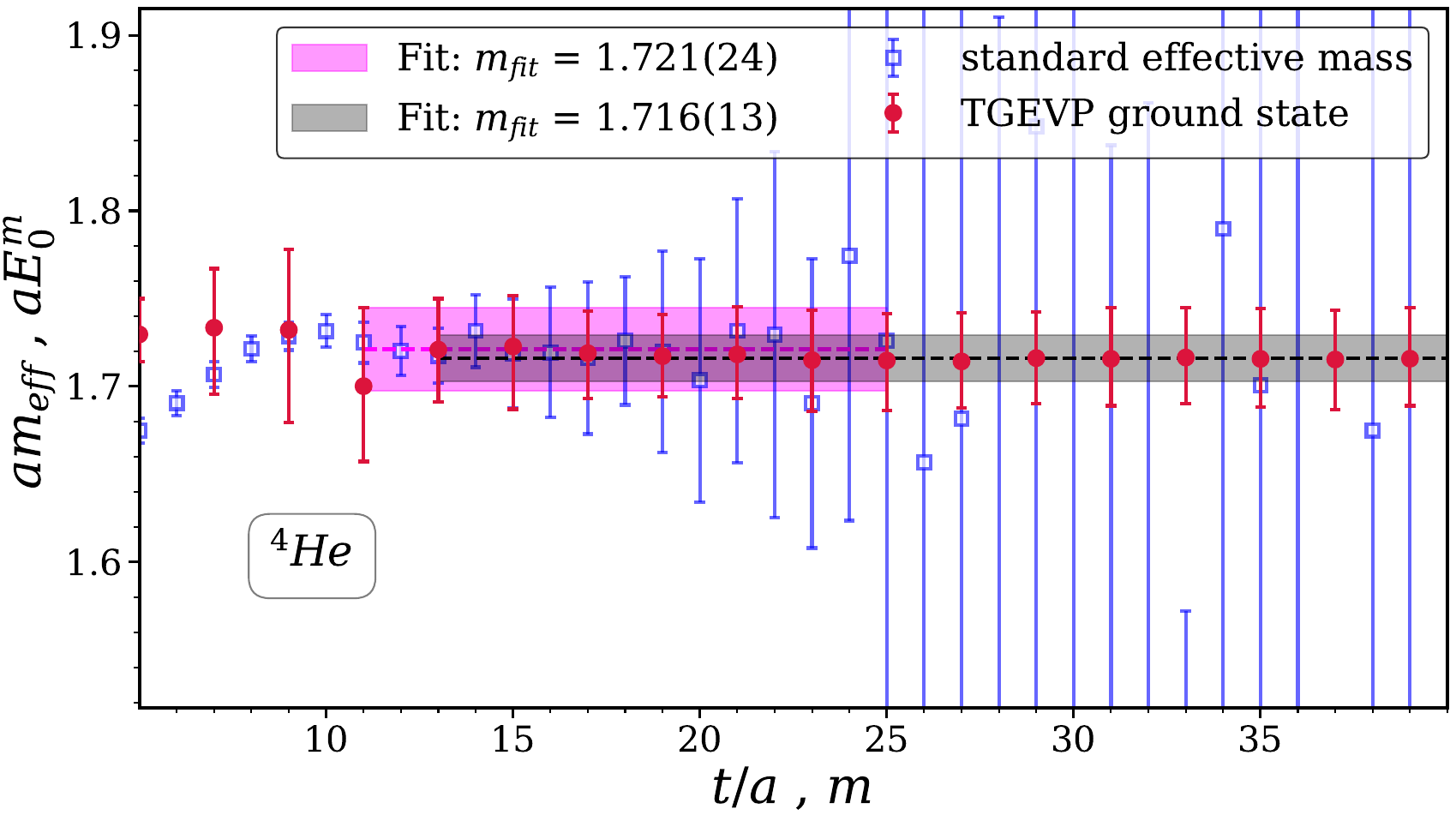}
\includegraphics[width=0.48\textwidth, height=0.28\textwidth]{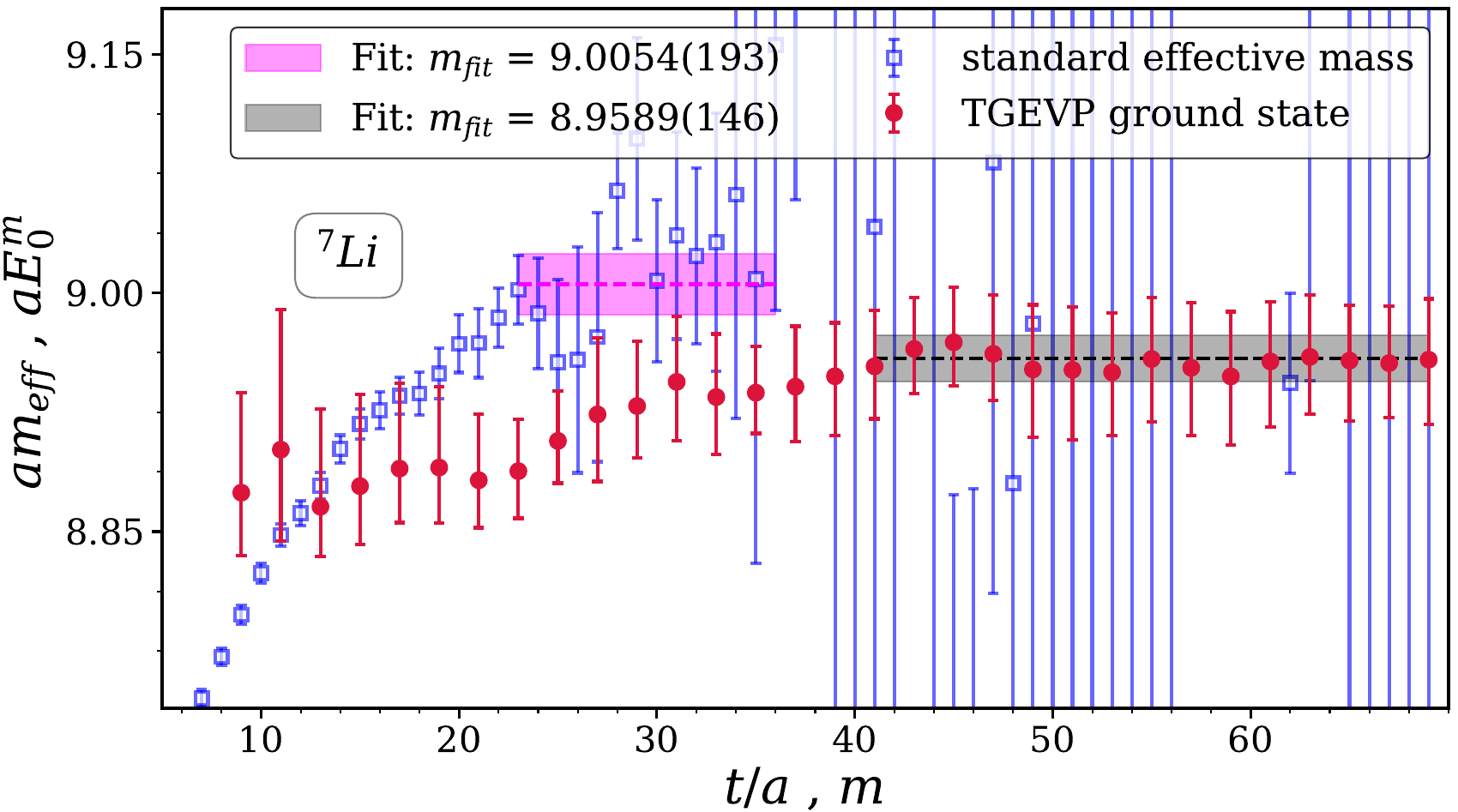}
        \caption{Comparison of the results between the regular effective mass (blue circles) and TGEVP ground state (red circles) corresponding to two-point functions of a few light nuclei. Top: $^2$H, middle: $^3$He, and bottom: $^4$He, with the quark masses, $m_u = m_d = m_s$ $(n_{meas} = 168)$.}
        \label{fg:mult_nu2}
  \end{figure*}

In Fig. \ref{fg:mult_nu2}, we present results for various nuclei, similar to Fig. \ref{fg:mult_nu1} of the main text, but
for a different set of ensembles. The two point correlation functions are generated with overlap valence quarks on the $N_f = 2+1+1$ HISQ lattice ensembles of size
 $48^3 \times 144$ ($a = 0.0582$ fm). The left panel are for the case with $m_u = m_d = m_s$, and the right panel show results of various nuclei, including $^7$Li for the case with $m_u = m_d = m_c$. The two point functions for these nuclei are generated using the algorithms described in Ref. \cite{Chakraborty:2024oym}.
  Again for each case we find the TGEVP eigenvalues are much more reliable compared to the fluctuating effective mass. For the cases of $^4$He and $^7$Li (right panel), we observe that the lowest GEVP eigenvalue plateau is somewhat lower than the probable effective mass plateau.  That leads to the following question: Whether this is because of the presence of a large number of close-by states, which becomes denser at the higher quark masses, TGEVP fails or the effective mass in this asymmetric source-sink setup becomes quite unreliable? This is an interesting observation and needs to be investigated when more statistics become available.

  Similar to Figs. \ref{fg:cor_mat1}, \ref{fg:cor_mat2}, \ref{fg:cor_mat3}, \ref{fg:cor_mat4}, \ref{fg:cor_mat5}, in the main text, in Fig. \ref{fg:cor_mat4}, we show additional correlation matrix plots for nuclei.
One can observe that the correlation over iterations increases with increasing  nuclear number (H$^2$ to He$^4$) and decreasing of quark masses (right to left column).
  The right column which are for the case of $m_u = m_d = m_c$ have comparatively low correlations compared to the left column for $m_u = m_d = m_c$ . For lattice calculations of nuclei with light quark masses, and with smaller statistics,  one thus need to address these correlations properly while fitting TGEVP results over iterations with a constant term.

\begin{figure*}[htb!]
\includegraphics[scale=0.33]{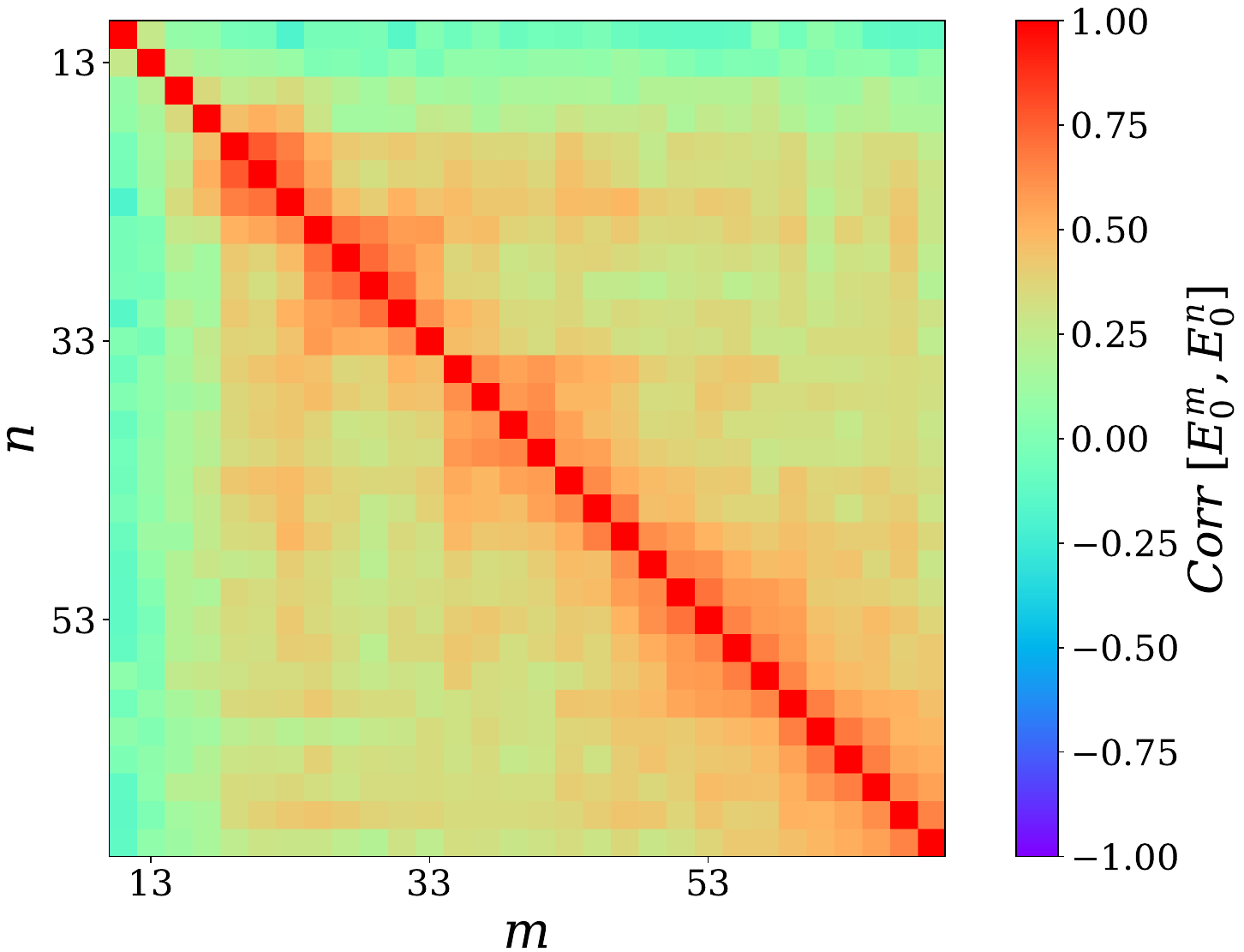}
\includegraphics[scale=0.33]{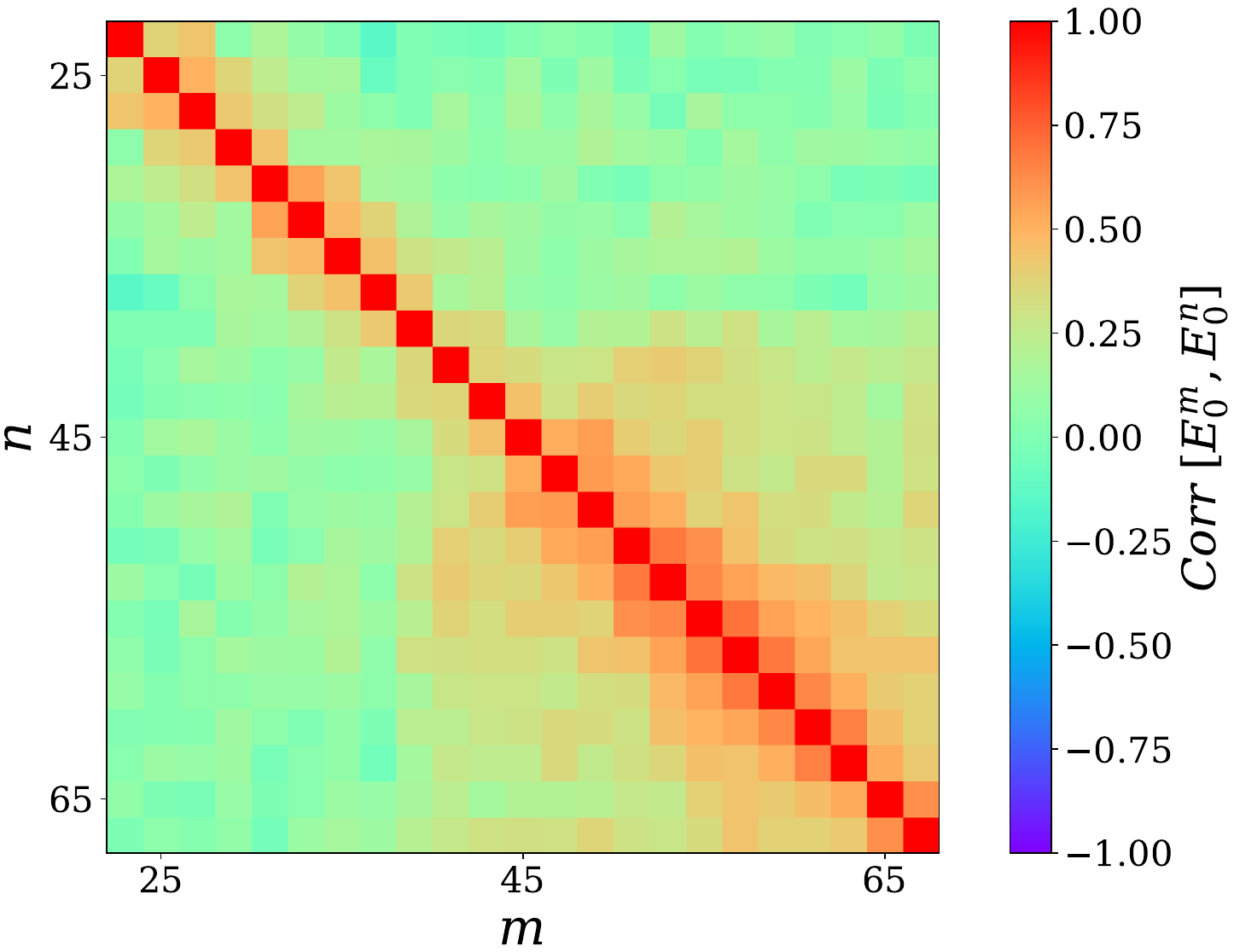}
\includegraphics[scale=0.33]{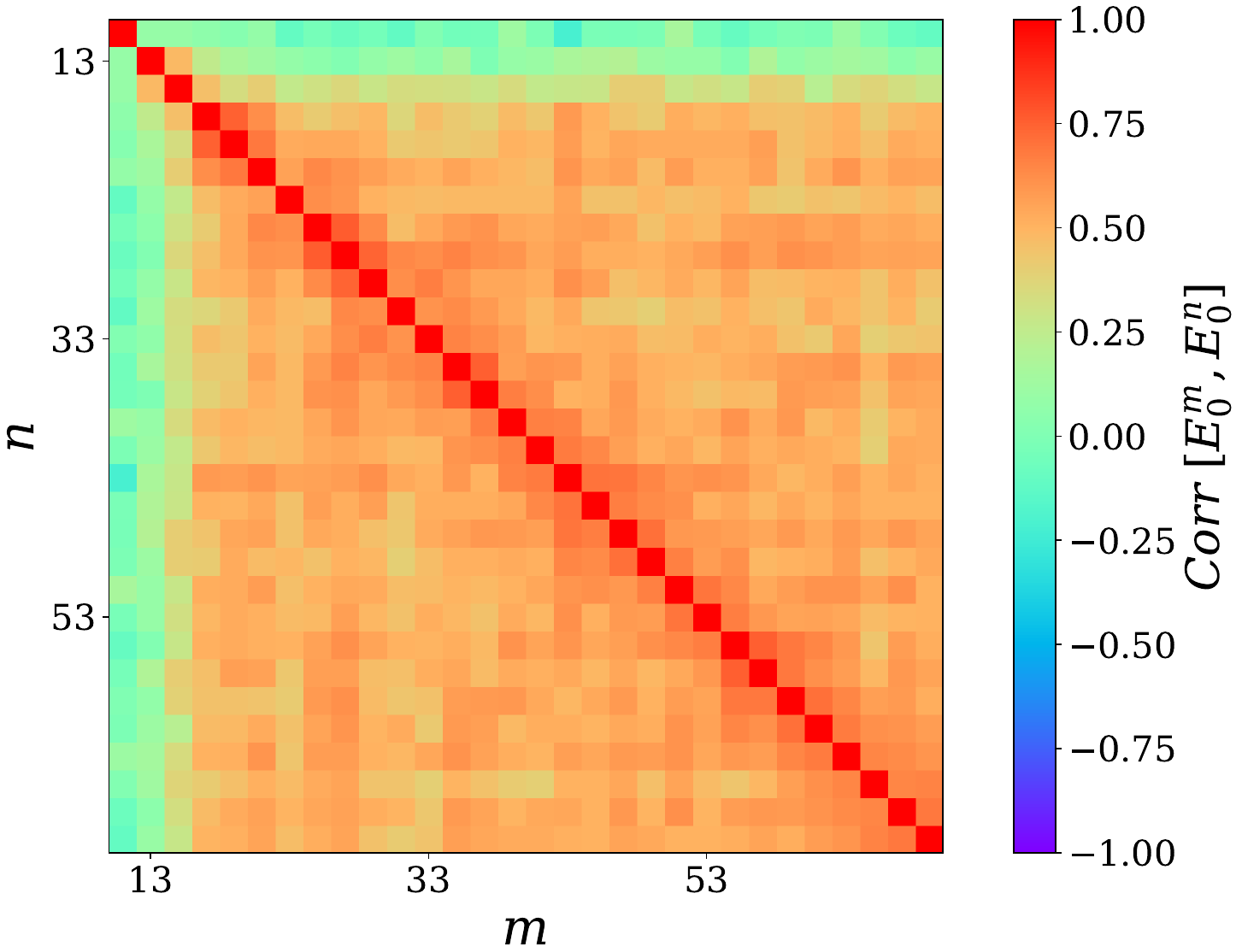}
\includegraphics[scale=0.33]{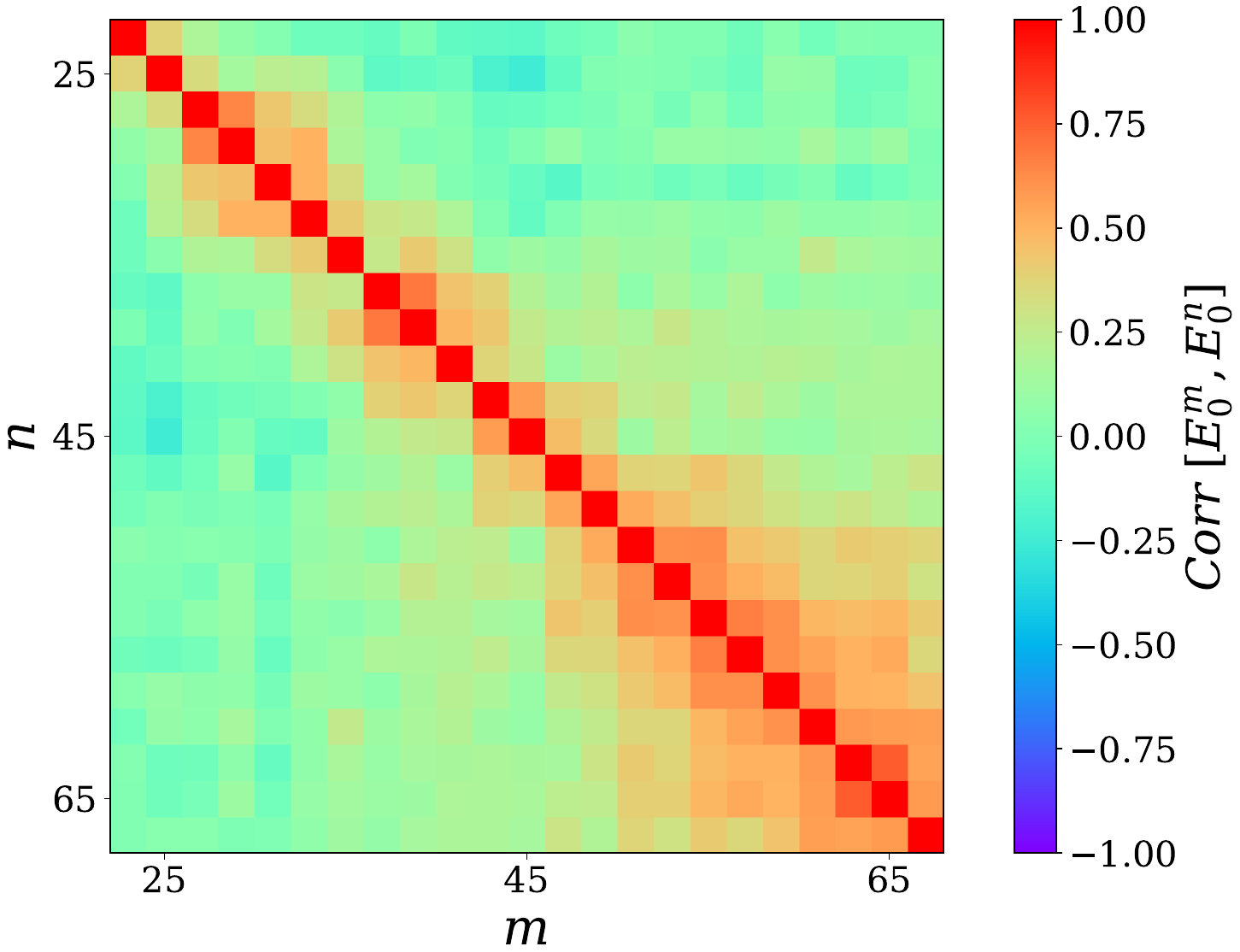}
        \caption{Correlation matrix plot corresponding to the lowest eigenvalues of  $^2$H and $^4$He (Fig. \ref{fg:mult_nu2}). The top row corresponds to $^2$H computed with the quark masses $m_u = m_d = m_s$ (left) and $m_u = m_d = m_c$ (right), respectively. The bottom row represents the same for $^4$He. }
        \label{fg:cor_mat4}
  \end{figure*}

\clearpage
\section{Impact of Complex Eigenvalues on Energy Estimates} \label{app:complx_eigs}
Here we examine whether the removal of complex or negative eigenvalues, $\lambda_n^{m,b}$, of the transfer matrix, evaluated through Eq.~(\ref{GEVPeq1}), introduces any bias in the final extracted energy estimates. There are two primary sources for the emergence of such eigenvalues, despite in principle, the transfer matrix is a positive operator.
The first source is numerical artifacts, including ill-conditioning, finite-precision limitations, or instabilities in the generalized eigenvalue problem (GEVP) solution procedure. These issues can give rise to spurious complex eigenvalues that do not reflect deviation from positivity of the stochastically determined transfer matrix ~\cite{Wagman:2024rid,CULLUM1981329}. The second source is statistical: the transfer matrix is estimated stochastically from finite Monte Carlo samples. As a result, statistical fluctuations can cause deviations from positivity, even though the exact transfer matrix is strictly positive. 
When complex or negative eigenvalues arise due to numerical artifacts, removing them does not introduce statistical bias. However, the second case is more subtle. If such eigenvalues originate from statistical fluctuations in a finite ensemble, their removal could potentially bias the extracted energy levels obtained from the TGEVP framework. We therefore investigate this possibility quantitatively below.
 
\subsection*{Robustness of Central Values to Complex Eigenvalue Removal}

To explicitly test whether the removal of complex eigenvalues biases the extracted energy levels, we compare the central values and bootstrap uncertainties obtained by including all eigenvalues (real and complex) against those obtained by discarding complex eigenvalues from the bootstrap ensemble. The results for several representative hadronic correlation functions are shown in Table~\ref{tab:energy_comparison}.

\begin{table}[h!]
\centering
\caption{Comparison of energy estimates $E_n$ (in lattice units) with and without discarding complex eigenvalues across bootstrap samples. Results are shown for the ground and first-excited states of various hadronic channels.}
\label{tab:energy_comparison}
\begin{tabular}{lcc}
\toprule
Channel / State & \shortstack{With Complex\\Eigenvalues} & \shortstack{Without Complex\\Eigenvalues} \\
\midrule
$N$ (Ground) & $0.5707(4)$ & $0.5706(4)$ \\
$NN$ (Ground) & $1.1319(9)$ & $1.1321(9)$ \\
$\eta_c$ (Ground) & $0.81612(15)$ & $0.81608(17)$ \\
$\eta_c$ (1st Excited) & $0.9915(24)$ & $0.9918(22)$ \\
\bottomrule
\end{tabular}
\end{table}

The near-identical central values and uncertainties confirm that the removal of complex eigenvalues has no statistically significant impact on the extracted energy levels for the hadronic channels considered. This indicates that such eigenvalues, while present at low frequency due to statistical fluctuations, do not bias the physical observables when discarded. However, it remains important to assess this case by case, since in scenarios with poor signal quality, the presence and removal of complex eigenvalues may lead to noticeable effects.
\vspace{0.2cm}

\subsection*{ Histogram Analysis of Bootstrap Energy Samples}

To assess the distribution of complex eigenvalues $E^{m,b}_n$ evaluated across bootstrap samples $b$, we construct histograms of the magnitudes of the extracted energy levels $|E_n^{m,b}|$. Each bin in the histogram is color-coded according to the fraction of complex-valued entries, quantified by the metric $\mathcal{C}$ defined as follows:
\begin{equation}
    \mathcal{C} = \frac{\text{Number of entries in the bin with }\arg(E_n^{m,b}) > \epsilon}{\text{Total number of entries in the bin}},
\end{equation}
where $\epsilon$ is a small numerical threshold used to distinguish non-zero imaginary parts. For this work, we have utilized $\epsilon = 10^{-12}$.

\begin{figure}[h]
    \centering
    \includegraphics[width=0.48\textwidth]{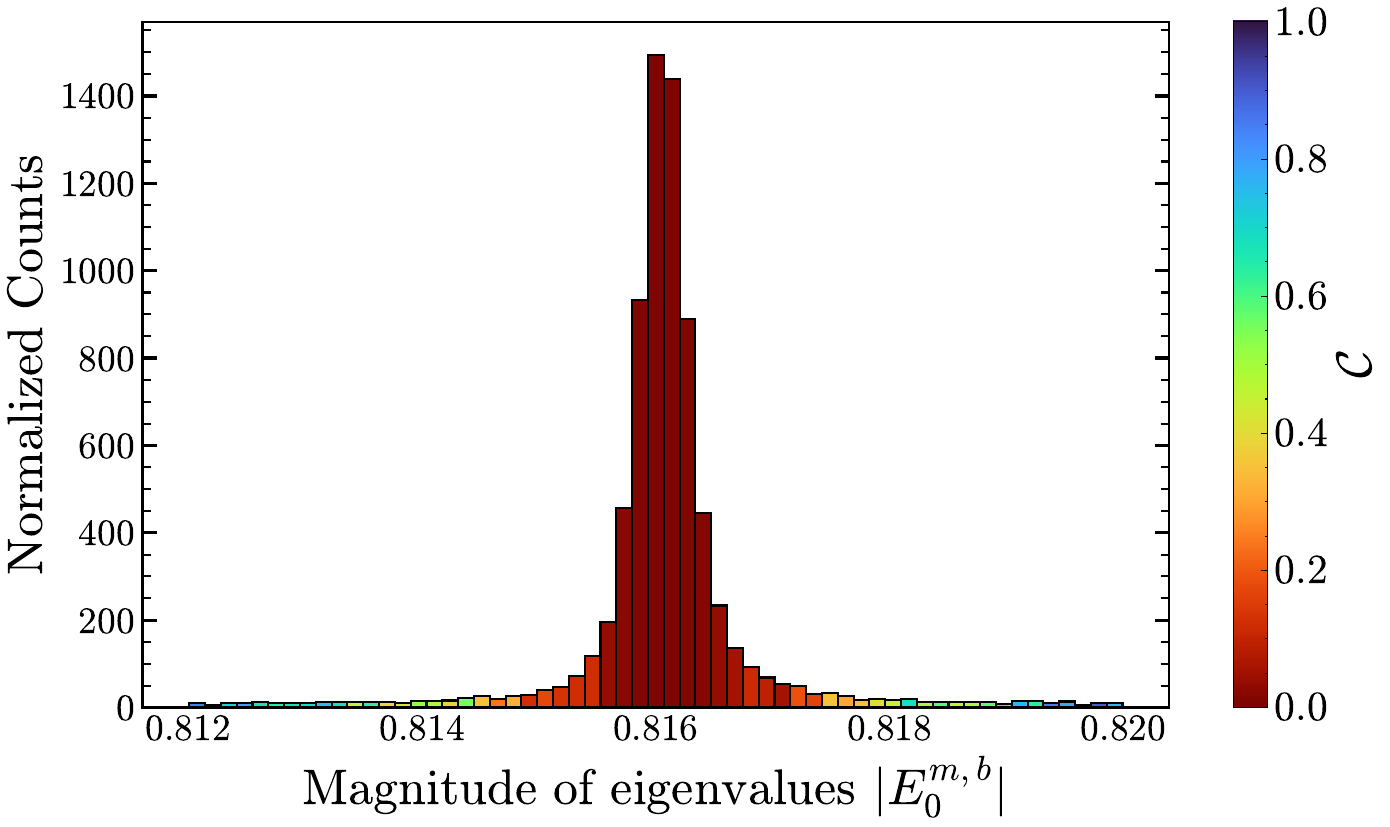} 
    \includegraphics[width=0.48\textwidth]{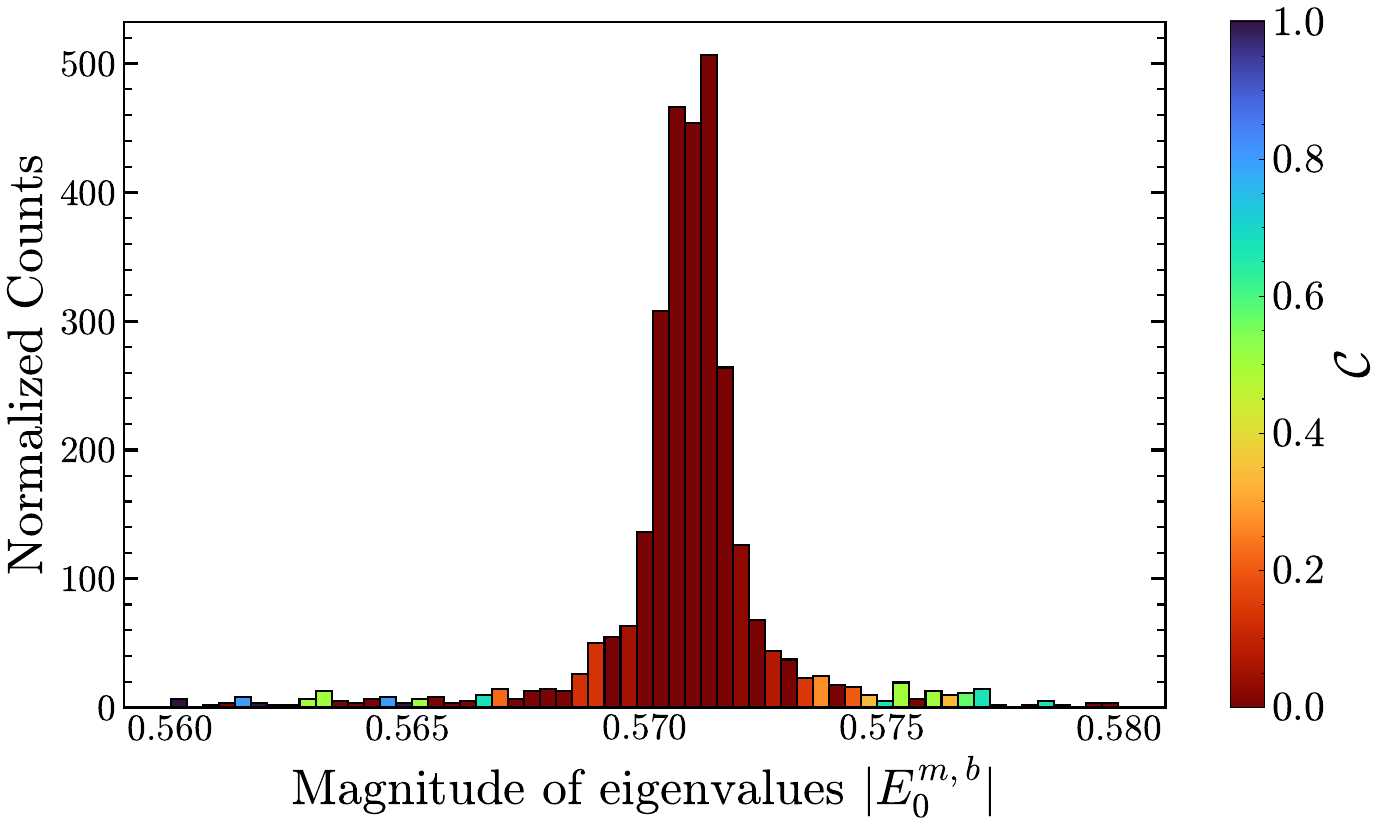}
    \includegraphics[width=0.48\textwidth]{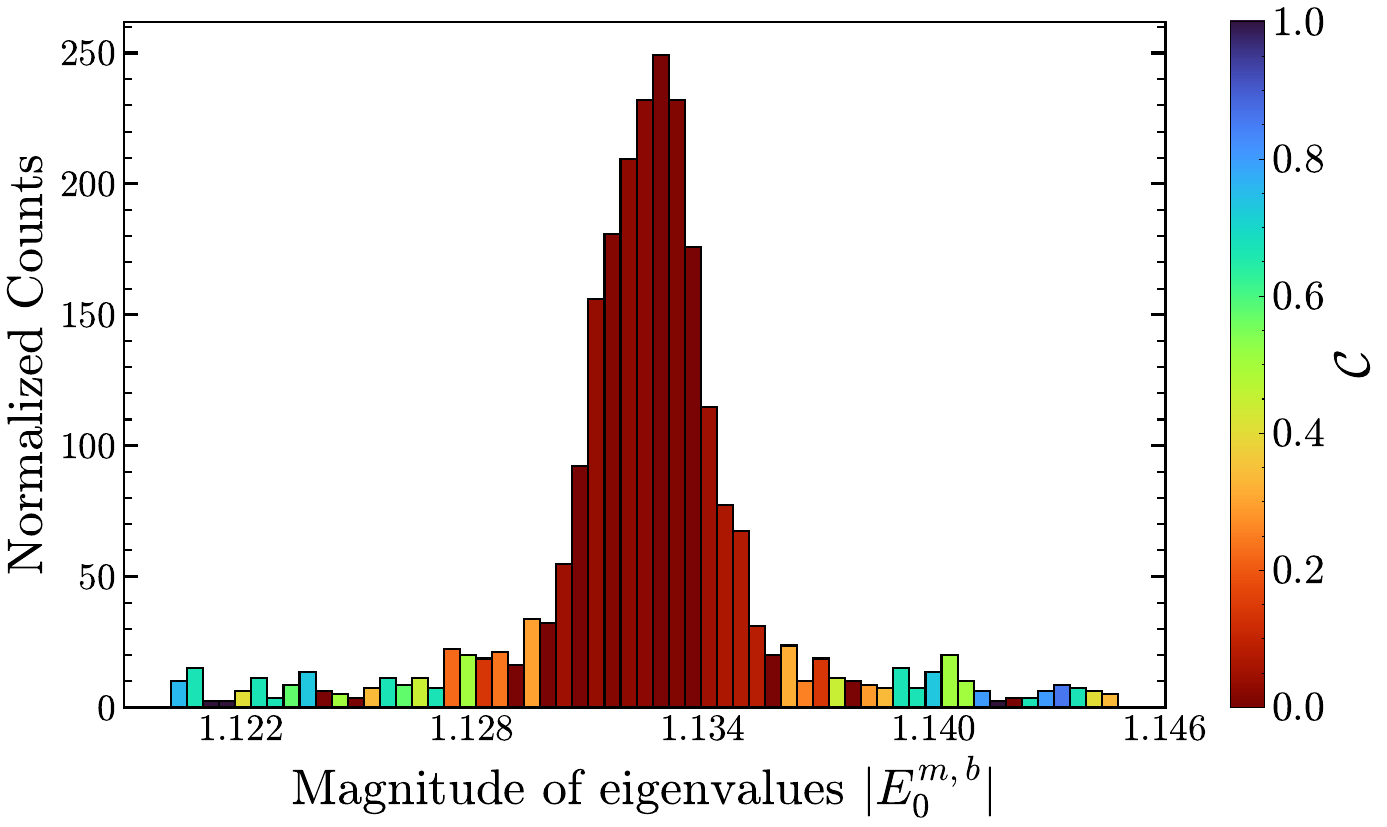}
    \caption{Histogram of bootstrap energy samples for $0^-$ charmonium (1S) (top panel), single nucleon (middle panel) and two-nucleon (bottom panel). Bins are color-coded by the fraction of complex eigenvalues present in the corresponding bin.}
    \label{fig:histogram}
\end{figure}

This visualization allows us to quantify the presence of complex eigenvalues in different regions of the energy spectrum. Since we employ Gaussian-smeared histogram and adaptive kernel density estimators (KDE) to identify peaks and estimate peak widths—used subsequently to define thresholds for outlier removal—the fraction $\mathcal{C}$ near the peak regions serves as a diagnostic for assessing whether complex eigenvalues influence the final extracted energy levels. In Fig. \ref{fig:histogram}, low value of $\mathcal{C}$ around the peaks clearly indicate that complex eigenvalues have negligible effect on the physical energy estimates for the ground state and first excited state (for charmonia). 

\begin{table}[h!]
    \centering
    \caption{Total fraction $\mathcal{C}_{\text{total}}$ of complex eigenvalues (i.e., with $\arg(E_n^{m,b}) > \epsilon$) across all bootstrap samples, for the ground and first-excited states of different hadronic channels. A small threshold $\epsilon = 10^{-12}$ is used to identify non-zero imaginary parts.}
    \label{tab:complex_fraction}
    \begin{tabular}{lcc}
        \toprule
        Channel & State & $\mathcal{C}_{\text{total}}$ \\
        \midrule
        Single Nucleon ($N$)       & Ground          & 0.015 \\
        Two-Nucleon ($NN$)         & Ground          & 0.028 \\
        $\eta_c$ Meson             & Ground          & 0.007 \\
                                   & First-Excited   & 0.018 \\
        \bottomrule
    \end{tabular}
\end{table}

In Table~\ref{tab:complex_fraction}, we present the total fraction of complex eigenvalues across all bootstrap samples for the single nucleon, two-nucleon ground state, and both the ground and first-excited states of the $\eta_c$ meson computed within $6\sigma$ from the central values. The negligible values of $\mathcal{C}_{\text{total}}$ support the conclusion that complex eigenvalues have an insignificant impact on the central values and uncertainties of the extracted energy levels. However, it is important to assess this diagnostic for each specific case of interest, as the presence of complex eigenvalues can increase substantially---particularly near the spectral peaks---if the underlying correlation functions suffer from poor statistical quality.

\bibliography{TGEVP}

\end{document}